\documentclass[fleqn,10pt]{wlscirep}
\usepackage[utf8]{inputenc}
\usepackage[T1]{fontenc}
\usepackage{subcaption}
\usepackage{multirow}
\usepackage{hhline}
\usepackage{graphicx}
\usepackage{colortbl,arydshln}

\ADLnullwidehline
\ADLnullwide
\title{Diverse Image Generation with Diffusion Models and Cross Class Label Learning for Polyp Classification}

\author[1,*]{Vanshali Sharma}
\author[2]{Debesh Jha}
\author[3]{M.K. Bhuyan}
\author[1]{Pradip K. Das}
\author[2]{Ulas Bagci}
\affil[1]{Department of Computer Science \& Engineering, Indian Institute of Technology Guwahati, Guwahati, 781039, India}
\affil[3]{Department of Electronics \& Electrical Engineering, Indian Institute of Technology Guwahati, Guwahati, 781039, India}
\affil[2]{Machine \& Hybrid Intelligence Lab, Department of Radiology, Northwestern University, Chicago, 60611, USA}

\affil[*]{vanshalisharma@iitg.ac.in}

\begin{abstract}
Pathologic diagnosis is a critical phase in deciding the optimal treatment procedure for dealing with colorectal cancer (CRC). Colonic polyps, precursors to CRC, can pathologically be classified into two major types: adenomatous (malignant potential) and hyperplastic (benign). For precise classification and early diagnosis of such polyps, the medical procedure of colonoscopy has been widely adopted paired with various imaging techniques, including narrow band imaging (NBI) and white light imaging (WLI). These imaging modalities have different advantages in capturing polyp-specific features for accurate classification. However, the existing classification techniques mainly rely on a single imaging modality and show limited performance due to data scarcity. Recently, generative artificial intelligence has been gaining prominence in overcoming such issues. Additionally, various generation-controlling mechanisms using text prompts and images have been introduced to obtain visually appealing and desired outcomes in a better-controlled manner. However, such mechanisms require class labels to make the model respond efficiently to the provided control input. In the colonoscopy domain, such controlling mechanisms are rarely explored; specifically, the text prompt is a completely uninvestigated area. Moreover, the unavailability of expensive class-wise labels for diverse sets of images limits such explorations. Therefore, in this work, we develop a novel model, \textbf{\textit{PathoPolyp-Diff}}, that generates text-controlled synthetic images with diverse characteristics in terms of pathology, imaging modalities, and quality. In the process, we introduce cross-class label learning to make the model learn features from other classes, reducing the burdensome task of data annotation. We validate the effectiveness of text-controlled synthesis and cross-class label learning by performing polyp classification (adenomatous/hyperplastic) with different imaging modalities (NBI/WLI) and text prompts. The experimental results report an improvement of up to 7.91\% in balanced accuracy using a publicly available dataset. Moreover, cross-class label learning achieves a statistically significant improvement of up to 18.33\% in balanced accuracy during video-level analysis. The code is available at \url{https://github.com/Vanshali/PathoPolyp-Diff}.               
\end{abstract}
\begin{document}

\flushbottom
\maketitle
\thispagestyle{empty}

%\noindent Please note: Abbreviations should be introduced at the first mention in the main text – no abbreviations lists. Suggested structure of main text (not enforced) is provided below.

\section*{Introduction}

Colorectal cancer (CRC) is the second most common cause of cancer-related mortality, accounting for about 1 million deaths per year~\cite{cancer-data}. The precursor lesions of CRC occur in the form of polyps, which in their initial stages are likely to be non-cancerous. However, they could turn into invasive cancer over time. The risk associated with such abnormal growth depends on the pathological characteristics of the lesion. For instance, a hyperplastic polyp is categorized as benign and generally not a cause of concern, whereas an adenomatous polyp might have the potential to become cancerous. Hence, early detection and categorization of polyps are necessary for optimal surgical procedures and effective treatment. Colonoscopy is a gold standard tool widely adopted for this purpose, in which a camera-mounted colonoscope is inserted into the patient's colon to investigate it for any abnormalities or lesions. This procedure encompasses various optical imaging technologies to capture polyp-specific features. Some such imaging techniques include white light imaging (WLI) and narrow band imaging (NBI). Among these imaging modalities, WLI is the most commonly used technique; however, it fails to capture topographic contrast, such as polyp elevation and pit patterns that increase the lesion miss rate. On the contrary, NBI uses electronically activated filters to limit the wavelengths of red, green and blue lights, which helps accentuate the superficial mucosa and vascular details~\cite{cheng2017narrow}. Despite its advantageous applicability in differentiating benign and neoplastic polyps, a standard colonoscopy procedure follows WLI. Besides clinical use, the research community is mainly dependent on images captured using a single imaging technique, i.e. WLI. The reason is the unavailability of publicly accessible image data encompassing different imaging modalities and pathological conditions. 

The recent breakthrough in generative artificial intelligence (AI) has sparked the potential to obtain realistic medical images spanning diverse sets of classes. Consequently, various tasks in the medical domain, including super-resolution~\cite{zhu2019can}, image reconstruction~\cite{bhadra2020medical}, segmentation~\cite{xun2022generative}, and image-to-image translation~\cite{sharma2023can}, have witnessed remarkable performance enhancements using generative adversarial network (GAN) based methodologies. GANs are one the prominent types of generative AI techniques; however, they suffer from convergence instability and show limited contributions and control over data generation. The advent of diffusion models marked a significant stride in dealing with such issues. These models facilitate a better control and manipulation ability to generate diverse, realistic images. This control can be achieved using conditioning in the form of text, images, or both. More recently, numerous works~\cite{wolleb2022diffusion, rahman2023ambiguous} in different fields focused on text-to-image synthesis have been proposed. Such approaches use random noise and text prompts as input to produce synthetic images. These images are expected to represent the conditions provided in the text. The literature utilizes existing text-to-image techniques by fine-tuning foundation models on the downstream tasks. This fine-tuning process requires text prompts that describe the desired synthetic images appropriately.

\begin{figure*}[t]
     \subfloat[\small \label{fig:y equals x}]{%
       \includegraphics[width=0.24\textwidth]{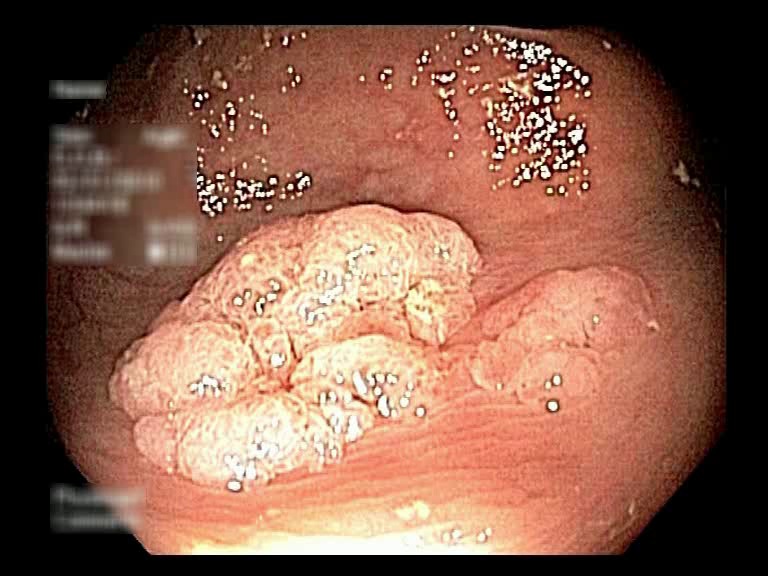}
        }  
      \hfill
     \subfloat[\small \label{fig:y equals x1}]{%
       \includegraphics[width=0.24\textwidth]{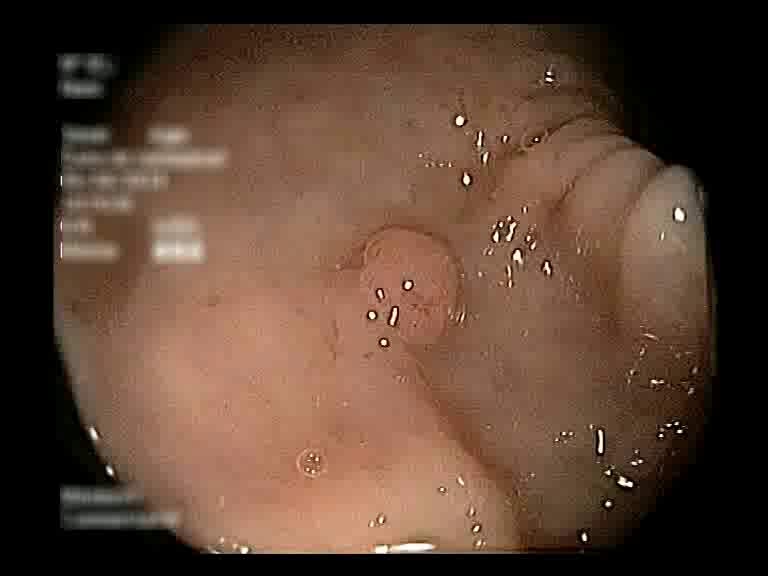}
        }
     \hfill
     \subfloat[\small \label{fig:y equals 2x}]{%
       \includegraphics[width=0.24\textwidth]{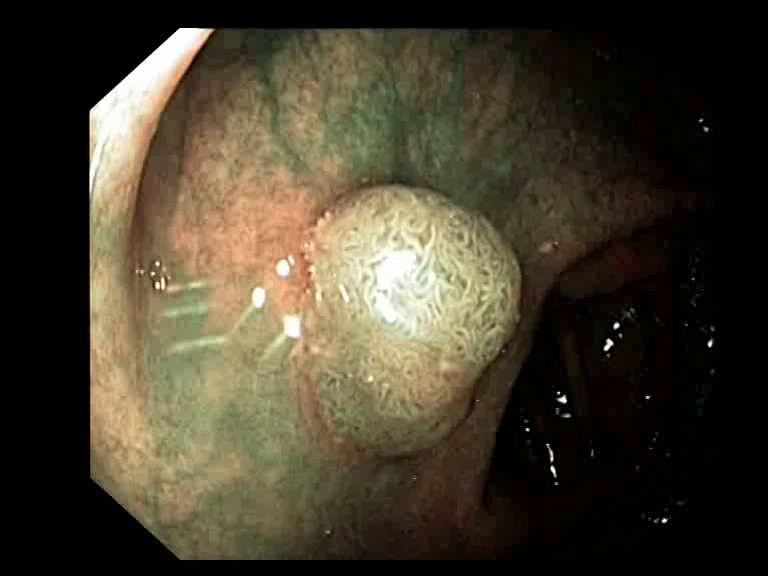}
        }
      \hfill
      \subfloat[\small \label{fig:y equals 3x}]{%
       \includegraphics[width=0.24\textwidth]{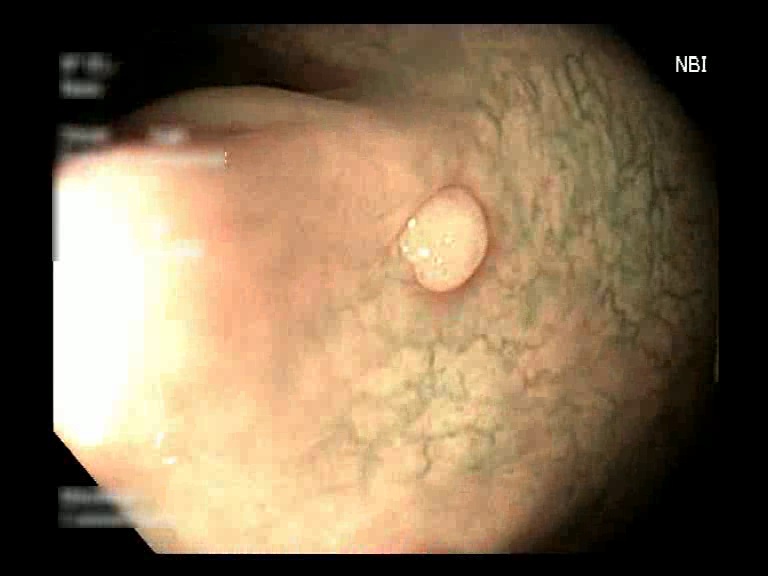}
        }
     \hfill
     \subfloat[\small \label{fig:y equals 4x}]{%
       \includegraphics[width=0.185\textwidth]{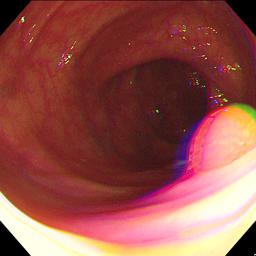}
        }    
      \hfill
       \subfloat[\small \label{fig:y equals 5x}]{%
       \includegraphics[width=0.185\textwidth]{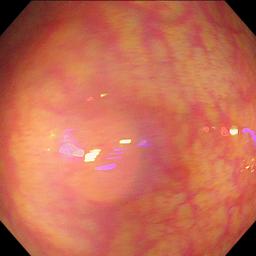}
        }
      \hfill
      \subfloat[\small \label{fig:y equals 6x}]{%
       \includegraphics[width=0.185\textwidth]{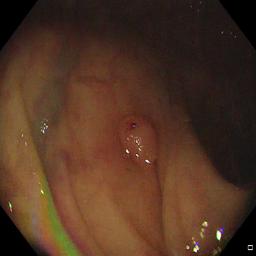}
        } 
      \hfill
      \subfloat[\small \label{fig:y equals 7x}]{%
       \includegraphics[width=0.185\textwidth]{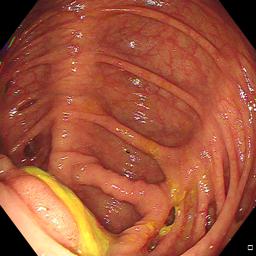}
        }
        \hfill
    \subfloat[\small \label{fig:y equals 8x}]{%
       \includegraphics[width=0.185\textwidth]{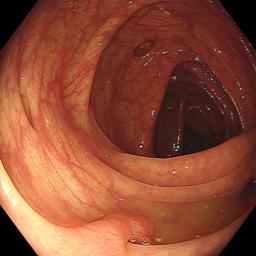}
        } 
      
      %\subfloat[\small Iteration-10K\label{fig:y equals 9x}]{%
      % \includegraphics[width=0.32\textwidth]{pic/t-SNE_plot_10k.png}
      %  }
      %  \hfill
      % \subfloat{%
      % \includegraphics[width=0.25\textwidth]%{pic/legend1.png}
       %} 
        \caption{\small Sample images depicting (a) adenomatous polyp in WLI (b) hyperplastic polyp in WLI, (c) adenomatous polyp in NBI, (d) hyperplastic polyp in NBI, (e)-(h) represent low-quality frames with artefacts, namely, ghost colors, motion blur, low illumination, and fecal depositions, respectively, and (i) shows a good-quality polyp image. }
        \label{fig:samples}
\end{figure*}

In the colonoscopy domain, diffusion models have not been explored much, especially with different conditioning possibilities. Some of the existing works \cite{machavcek2023mask, pishva2023repolyp} mainly focus on generating polyp images with binary mask conditioning. These limited explorations render the study of pathological conditions (adenoma/hyperplastic) of polyps and various imaging modalities unexplored. Additionally, the significance of controlling polyp generation using text prompts remains overlooked. Moreover, a well-designed dataset with comprehensive annotations is essential for training generative models. This ensures their ability to grasp intricate patterns linked to polyp-characterizing features and their associated pathological conditions. However, obtaining labels for each subtask, considering pathology, quality, and imaging modality, could be significantly expensive.  Therefore, in this framework, we develop a method to perform cross-class label learning that helps leverage annotations from other classes and produce synthetic images representing a combination of text prompts. This approach ensures obtaining synthetic polyp images with different pathological characteristics (adenoma/hyperplastic) captured with different imaging techniques (WLI/NBI). This diverse set is obtained while maintaining the quality, overcoming the general artifacts present in a colonoscopy video: ghost colors, motion blur, low illumination, and fecal deposition. Some sample images categorized under the aforementioned classes are shown in Fig. \ref{fig:samples}. 

Our contributions can be summarized as:
\begin{itemize}
    \item \textbf{Generated diverse set of colonoscopy images using text-controlled mechanism:} We develop PathoPolyp-Diff, a novel model to generate text-controlled synthetic images that cover a wide range of categories, including pathology, imaging modalities, and quality. It can generate adenomatous and hyperplastic polyps combined with the desired imaging modalities, including NBI and WLI. To the best of our knowledge, this is the first work to explore text prompts to generate synthetic colonoscopy images.
    \item \textbf{Introduced the concept of cross-class label learning:} We introduce the concept of cross-class label learning that allows the model to learn labels from different classes, hence expanding the diversity of data generation and reducing the cumbersome task of dataset annotation. Additionally, we integrate this learning with a weighted mechanism and study the impact of weights with text prompts.  
    \item \textbf{Improved polyp classification performance:} We report enhanced polyp classification outcomes when PathoPolyp-Diff's generated images are used to augment the real pathological dataset. This enhancement is achieved without additional manual annotations. We also performed an exhaustive analysis to study the impact of different text prompts on the quality of synthetic images generated for improved polyp classification. 
\end{itemize}

\section*{Related Works}
With the advent of generative AI, several works have been proposed to generate synthetic colonoscopy images. The main focus of existing related works has been to perform synthetic polyp generation irrespective of the pathological characteristics of polyps and the imaging modality. The techniques adopted so far can broadly be divided into two categories, namely, GAN, and Diffusion Models.  

\subsubsection*{Generative Adversarial Network (GAN) based Techniques}
The initial frameworks for polyp generation are based on the adversarial concept and adopt different variants of GANs. For example, Shin et al.~\cite{shin2018abnormal} used a conditional-GAN approach to translate normal colonoscopy images to polyp images. This translation is achieved using an input-conditioned image which is a combination of an edge map and a polyp binary mask. A similar concept of converting normal frames to polyp frames is proposed in~\cite{qadir2022simple}. They utilized a conditional GAN architecture to produce polyps with varied characteristics by controlling the input-conditioned binary mask values. Such conditional translation is also reported by Fagereng et al.~\cite{fagereng2022polypconnect}. They developed a framework called PolypConnect which uses an EdgeConnect model to convert clean colon images to polyps when given an edge map and a polyp mask. Sasmal et al.~\cite{sasmal2020improved} performed polyp generation using DCGAN and used the obtained synthetic polyps to enhance classifier performance for differentiating adenoma and hyperplastic polyps. An identical augmentation approach is followed by Adjei et al.~\cite{adjei2022examining} using synthetic polyps generated using a Pix2Pix model. Unlike the traditional GAN architecture, He et al.~\cite{he2021colonoscopic} introduced an attacker in the framework to obtain false negative images. Sams and Shomee~\cite{sams2022gan} utilized a StyleGAN2-ada to generate random binary masks, which are combined with colon images. This integrated image is used as an input for a conditional GAN to obtain synthetic polyp images. The above methods focused on polyp generation irrespective of the imaging modalities. However, a few works have used GAN-based approaches to transfer styles between different imaging modalities, such as WLI and NBI. Golhar et al. \cite{golhar2022gan} utilized the GAN inversion approach, which uses a latent representation of images to perform translation between NBI and WLI modalities. Following this technique, interpolation methods are used to change the polyp size. Similarly, Bhamre et al.~\cite{bhamre2022colonoscopy} used CycleGAN to convert WLI images to NBI images. Although these few existing works established the significance of NBI images over WLI images in polyp classification, the generation of new synthetic polyp images with different imaging modalities has not been explored in the literature.

\subsubsection*{Diffusion Model based Techniques}
The related literature involves only a few works focused on polyp image generation. Machacek et al.~\cite{machavcek2023mask} used a conditional diffusion probabilistic model to produce synthetic polyp images using synthetic masks. They validated the effectiveness of generated data by utilizing it for training polyp segmentation models. Pishva et al.~\cite{pishva2023repolyp} performed polyp generation using two diffusion models. The two models are fine-tuned on cropped-out polyps and clean colon images, respectively. This fine-tuning is followed by performing an inpainting using the latter model and cropped-out images. Du et al.~\cite{du2023arsdm} proposed an adaptive reﬁnement semantic diﬀusion model which considers the polyp and background ratio to adjust the diffusion loss. They also incorporated a pre-trained segmentation model that modifies the refinement loss depending on the difference between the predicted mask of the synthetic polyp and the binary mask used for its generation. The above-mentioned approaches followed a similar pattern of polyp generation using binary masks. The impact of text prompt based training and the generation of colonoscopy images with different imaging modalities still remains unexplored.

\section*{Methodology}
\begin{figure}[h!]
    \centering
    \includegraphics[width=\linewidth]{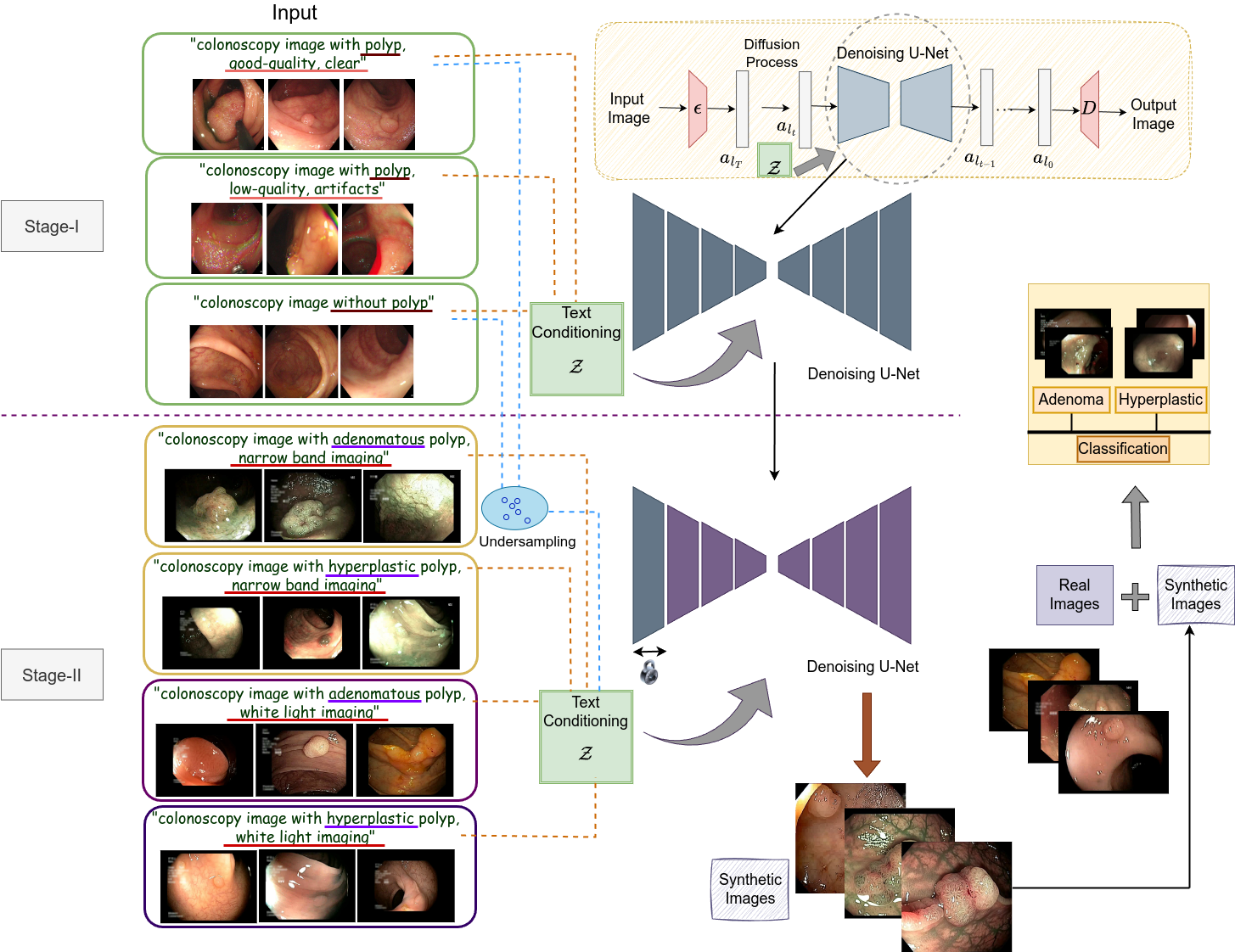}
    \caption{Overview of the proposed framework. It consists of two stages and uses various text conditioning to control the generation process. In Stage-II, some undersampled data from Stage-I is used for a smoother learning process. Also, the first block of U-Net is kept locked in the second stage. The performance of the proposed model is validated using a classification process which uses a combination of real and synthetic images in different proportions. }
    \label{fig:into}
\end{figure}

An overview of our approach is illustrated in Fig. \ref{fig:into}. Our novel framework, referred to as PathoPolyp-Diff, utilizes dual-stage training. The two stages, \textit{Stage-I} and \textit{Stage-II}, aim at generating colonoscopy images with diverse polyp types in different imaging modalities. They perform complementary tasks and the difference between the two lies in their training process. The \textit{Stage-I} network distils knowledge into the \textit{Stage-II} model in the form of a large set of features that enables it to differentiate between polyp and non-polyp characteristics and further helps to generate images for cross-class labels. Complementary to it, the training process in \textit{Stage-II} allows the model to learn pathological details and different imaging modality-related patterns.

%\subsubsection{Input}The input $I$ for training the first mode is defined by $\{e_i, a_i\}$ whereas that for the second mode is given by $\{e_i, a_i, c_i\}$ where $e$,$a$, and $c$ are the text prompt, target image and a control image, respectively.   

%Topical subheadings are allowed. Authors must ensure that their Methods section includes adequate experimental and characterization data necessary for others in the field to reproduce their work.

\begin{figure}[h!]
    \centering
    \includegraphics[width=\linewidth]{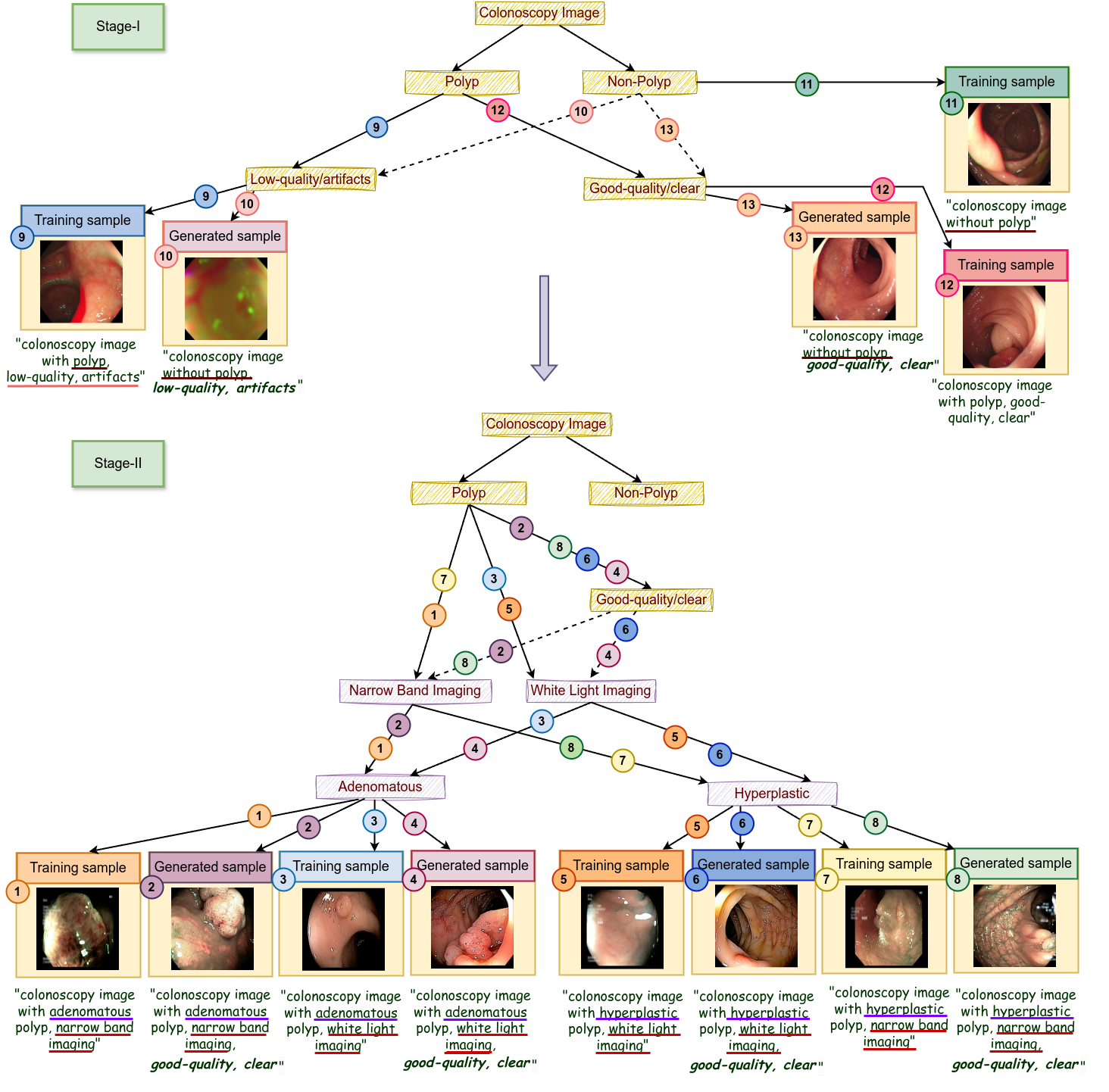}
    \caption{Flowchart depicting the different combinations of text prompt and cross-class labels used to generate images. The yellow nodes represent levels 1-3, while the nodes in another color represent levels 4-5. The solid arrows denote the labels already present in the dataset, whereas the dashed arrows represent the labels learnt from other classes (cross-class labels). Each number on a solid/dashed line represents the combination of strings used to form tokens for text prompts used in training/inference. For instance, following number \textit{`8'}, we obtain the text prompt \textit{``colonoscopy image with a hyperplastic polyp, narrow band imaging, good quality, clear''}, where \textit{``good quality, clear"} are part of indirectly inferred tokens and other are already present in the training annotations.}
    \label{fig:intro2}
\end{figure}

\subsection*{Architectural Details}

%\subsubsection*{Mode-I}
Our approach utilizes the stable diffusion (SD) model's \cite{rombach2022high} architecture with different settings and transfer learning approach. SD is a text-based image generation network that works on the principle of latent diffusion model (LDM) \cite{}. It is introduced to circumvent the issue of the high computational requirements of standard diffusion models (DM) \cite{ho2020denoising}. This improvement is achieved by executing the diffusion process on latent space instead of pixel level using an autoencoding procedure. To understand the advantages of using LDM over DM, some basic details about the two concepts are given below. 

\noindent \textit{Standard DMs:} The standard DMs follow a parameterized backward process of a fixed Markov Chain to gradually denoise a noisy image $a_t$.  It acts as a sequence of autoencoders $\epsilon_\theta(a_t,t)$ that serves as a denoising framework to predict the denoised version of $a_t$. Here, $t$ is uniformly sampled between $[1, T]$, and $T$ denotes the noise steps. The related objective can be defined in a refined form as:   
\vspace{-0.2cm}
\begin{equation}
    L_{DM} := \mathbb{E}_{a, \epsilon,t}[{\lVert \epsilon-\epsilon_\theta(a_t,t)\rVert}^2_2] 
\label{eq:dpm}
\end{equation}

where $\epsilon \sim \mathcal{N}(0,1)$. 

\noindent \textit{Standard LDMs:} These models leverage the ability of encoder and decoder architectures to represent significant information in compressed form and reconstruct it back in its original form. Such an attempt to use latent space enables the model to focus on important semantic details and perform efficiently with low computational resources. LDMs also keep track of time steps $t_i$ and are embedded with the U-Net architecture. However, instead of using an image $a$ directly, it is processed through an encoder $E$ to obtain $a_l$. 

\begin{equation}
    L_{LDM} := \mathbb{E}_{E(a), \epsilon,t}[{\lVert \epsilon-\epsilon_\theta({a_l}_t,t)\rVert}^2_2] 
\label{eq:ldm}
\end{equation}

SD has three main components, namely, a variational autoencoder (VAE) \cite{kingma2019introduction}, a U-Net \cite{ronneberger2015u}, and a text encoder, called CLIP \cite{radford2021learning}. The VAE comprises an encoder $E$ that converts the input image $a \in \mathbb{R}^{H\times W\times 3}$  in a latent space for low resource overhead during complete processing and a decoder $D$ that translates the encoded representation into a reconstructed image $a$. SD follows two processes similar to a standard LDM, i.e., forward diffusion and reverse diffusion. During forward diffusion, Gaussian noise is iteratively introduced in the image and during the reverse diffusion step, a U-Net architecture with ResNet blocks is employed to obtain denoised outcomes. This step is supported by text embeddings which represent the textual information about the image in a numeric form. These text embeddings are obtained using a pre-trained CLIP model that acts as conditioning on the image reconstruction.

%\textit{Standard LDM:} The base of a LDM is a standard diffusion model which learns the underlying distribution of the training data by denoising noisy image $a_t$. This denoising is achieved using a equally weighted series of autoencoders $\epsilon_{\theta(a_t,t)}$ 

%\begin{equation}
%    L_{DPM} := \mathbb{E}_{a, \epsilon,t}[{\lVert \epsilon-\epsilon_\theta(a_t,t)\rVert}^2_2] 
%\label{eq:dpm}
%\end{equation}

%where $\epsilon \sim \mathcal{N}(0,1)$.

\noindent \textit{Conditioning Mechanism:} DMs, like any other generative model, are capable of conditioning the sampling to guide the generation process. It is achieved by modeling $p(a|b)$ where b denotes the condition that could be in the form of text or image embeddings. The SD model incorporates this concept in a more effective way by using a cross-attention mechanism. This mechanism is used to fuse text embeddings with the U-Net structure after pre-processing $b$ through a text encoder $\mathcal{Z}$ (in case of text prompts). It encodes $b$ into an intermediate representation, which is then mapped to U-Net architecture using cross-attention layers. This conditioning is integrated with the overall objective as given below:

\begin{equation}
    L_{LDM_b} := \mathbb{E}_{E(a), b, \epsilon,t}[{\lVert \epsilon-\epsilon_\theta({a_l}_t,t,\mathcal{Z}_\theta(b))\rVert}^2_2] 
\label{eq:pmi}
\end{equation}

The conditioning mechanism can be represented as:

\begin{equation}
    Attention(Q,K,V) = softmax(\frac{QK^T}{\sqrt{h}}).V
\end{equation}

where $h$ denotes the dimension of vector embedding, $Q$, $K$, and $V$ are the query, key, and value sequences, respectively. These sequences can further be computed as:

\begin{equation}
    Q = W_Q^{(i)}.F_i({a_l}_t), K = W_K^{(i)}.\mathcal{Z}_\theta(b), V = W_V^{(i)}.\mathcal{Z}_\theta(b)
\end{equation}
where $F_i(a_{l_t})$ signifies the flattened intermediate U-Net representation of the denoising process and $W_Q^{(i)}$, $W_K^{(i)}$ and $W_V^{(i)}$ are the learnable projection matrices with shape $h \times h_{\mathcal{Z}}$, $h \times h_{\mathcal{Z}}$, and $h \times h_{\epsilon}^i$, respectively.

\noindent \textit{Text Prompts:} In our work, we have used text prompts as the conditioning to control the generation process. To construct the prompts, quality information was incorporated in Stage-I prompts whereas pathological and modality-related information was used in Stage-II. The stage-wise prompt construction process is explained below:

\textbf{Stage-I:} \textit{(a) Polyp/non-polyp}: The polyp frames are fed into the model with a prompt incorporating \textit{`with polyp'} substring, whereas non-polyp frames are provided with \textit{`without polyp'} substring. \textit{(b) Good-quality/low-quality}: The polyp frames have additional information about low-quality (frames with artifacts like ghost colors, fecal depositions, low-illumination, etc.) and good-quality (free from artifacts). The text prompts in the former case involve \textit{'low-quality, artifacts'} keywords, whereas \textit{'good-quality, clear'} is part of the latter case. 

\textbf{Stage-II:} \textit{(a) Adenomatous/hyperplastic}: To make the model learn and focus on the pathological features, text prompts depicting appropriate condition, \textit{`adenomatous polyp'} or \textit{`hyperplastic polyp'}, were provided. \textit{(b) WLI/NBI}: To generate diverse polyp images with both WLI and NBI modality, the text prompts at this stage also included \textit{`white light imaging'} or \textit{`narrow band imaging'} substrings.

\subsection*{Training Details}
\textbf{Stage-I:} The Stage-I uses a pre-trained Stable Diffusion v1-4\cite{rombach2022high} model and further fine-tunes it with some text conditions to generate desired colonoscopy images. This stage is focused on developing a model that learns the basic features to differentiate polyp and non-polyp characteristics. In the process, text prompts are used as the conditioning mechanism to control the model output. These text prompts comprise the embeddings pertaining to the strings presented at the first to third levels in Fig. \ref{fig:intro2}, starting from the top. During the fine-tuning of the model, a relatively large-scale dataset is used with polyp/non-polyp classes wherein the polyps have an additional annotation of low-quality/artifacts (uninformative) and good-quality/clear (informative). This process allows the model to generate polyp and non-polyp images with specific quality criteria. 

\noindent \textbf{Stage-II:} In this stage, the pre-trained model of Stage-I is used with the first block locked. The other blocks are further fine-tuned on our desired text conditioning. These conditions are shown in the fourth and fifth levels of Fig. \ref{fig:intro2}. For a successful implementation of cross-class label learning, we undersampled the non-polyp set and informative set of polyps and included them in the training iterations of Stage-II. Presenting these undersampled images allows the model to retain the features pertaining to Stage-I without undergoing overfitting with the new dataset.

\noindent{The different deep learning models used to validate the outcomes during the above two stages are given in Table \ref{tab:desc} with their corresponding objectives. The dataset details and the training settings are given in the next section. The source code developed in this work is available at \url{https://github.com/Vanshali/PathoPolyp-Diff}.              

\begin{table}[]
\resizebox{\linewidth}{!}{
\begin{tabular}{ccc}
\cellcolor[HTML]{C0C0C0} Model  & \cellcolor[HTML]{C0C0C0} Stage & \cellcolor[HTML]{C0C0C0}Description                                                     \\
DenseNet-201    & I              & \begin{tabular}[c]{@{}l@{}} Used to choose the best iteration of PathoPolyp-Diff training during Stage-I. Detects polyps, supports t-SNE plot visualization and validates the \\  polyp-characterizing features existing in the generated polyp images. Results are shown in Table \ref{tab:step1_res} and Fig. \ref{fig:step1}. \end{tabular}                \\
DenseNet-121    & II             & \begin{tabular}[c]{@{}l@{}}Used to choose the best iteration of PathoPolyp-Diff training during Stage-II. Supports t-SNE plot visualization and performs iteration-wise \\ adenomatous and hyperplastic classification with WLI and NBI modalities. Related results are given in Table \ref{tab:step2}, Fig.  \ref{fig:conf} and Fig. \ref{fig:step2}. \end{tabular}                                                        \\
EfficientNet-B0 & II             & \begin{tabular}[c]{@{}l@{}}Validates the effectiveness of using synthetic images by classifying adenomatous and hyperplastic classes with WLI and NBI modalities with \\ different proportions of real and synthetic image count in the training set. Corresponding results are shown in Table \ref{tab:framecomp}, Table \ref{tab:side_by_side_tables}, Fig. \ref{fig:patient_wli} and Fig. \ref{fig:patient_nbi}. \end{tabular}
\end{tabular}
}
\caption{Description of different models used to validate the outcomes during Stage-I and Stage-II.}
\label{tab:desc}
\end{table}

\begin{table}[]
\resizebox{\linewidth}{!}{
\begin{tabular}{lcccccccc}
\multicolumn{1}{c}{\cellcolor[HTML]{C0C0C0}}   & {\cellcolor[HTML]{C0C0C0}}         & \multicolumn{3}{c}{ \cellcolor[HTML]{C0C0C0} Videos} &  \multicolumn{4}{c}{\cellcolor[HTML]{C0C0C0} Frames}                                                                                                                                                                                     \\
\rowcolor[HTML]{EFEFEF} \multicolumn{1}{c}{\multirow{-2}{*}{\cellcolor[HTML]{C0C0C0} Stage/ Model}}   & \multirow{-2}{*}{\cellcolor[HTML]{C0C0C0} Dataset}                    & Total                                & Train                      & Test                 & Total                                                                                    & Train                                                                              & Validation & Test              \\
 \begin{tabular}[c]{@{}l@{}} Stage-I ((a) Pathopolyp-Diff, \\ (b) DenseNet-201) \end{tabular}                     &  SUN Database             &  100 cases                            & -                          & -                    &  \begin{tabular}[c]{@{}l@{}} 49,136 polyp / \\ 109,554 non-polyp \\ (random split) \end{tabular}                                         &  80\%                                                                               &  10\%       &  10\%              \\
 \begin{tabular}[c]{@{}l@{}} Stage-I ((a) Pathopolyp-Diff, \\ (b) FFT (only Test split applicable)) \end{tabular} &  SUN Database             &  100 cases                            & -                          & -                    &  \begin{tabular}[c]{@{}l@{}} 49,136 polyps \\ (31\% uninformative \\ frames, remaining are \\ informative frames, \\random split) \end{tabular} &  80\%                                                                               &  10\%       & 10\%              \\
 Stage-II (PathoPolyp-Diff)                                          &  ISIT-UMR                 &  \begin{tabular}[c]{@{}l@{}} 40 AD / 21 HP \\ (for each \\ WLI and NBI) \end{tabular} &  29 AD / 15 HP              & -                    &  -                                              &  \begin{tabular}[c]{@{}l@{}} 12,883 AD /  12,987 HP \\ (both WLI and NBI) \end{tabular}                                            & -          & -                 \\
 \begin{tabular}[c]{@{}l@{}} Stage-II ((a) DenseNet-121, \\ (b) EfficientNet-B0) \end{tabular}                    &  ISIT-UMR                 &  \begin{tabular}[c]{@{}l@{}} 40 AD / 21 HP \\ (for each \\ WLI and NBI) \end{tabular}&  15 AD / 15 HP              &  6 AD / 6 HP          & -                                                                                        &  \begin{tabular}[c]{@{}l@{}} Variable number of \\ frames from each \\ video (16, 23 and 64). \\ More details in Table \ref{tab:framecomp}. \end{tabular} & -          &  \begin{tabular}[c]{@{}l@{}} 4,376 AD / \\ 2,597 HP \\ (both WLI \\ and NBI)\end{tabular}
\end{tabular}
}
\caption{ Dataset details with video-wise and frame-wise split for different models.  AD and HP refer to adenoma and hyperplastic, respectively.}
\label{tab:dataset}
\end{table}

\section*{Results}

\subsection*{Dataset Details and Training Settings}
We used two publicly available datasets, namely, SUN Database~\cite{misawa2021development,sundatabase} and ISIT-UMR Colonoscopy Dataset~\cite{mesejo2016computer}, to evaluate the performance of our proposed model. SUN Database comprises 49,136 polyp frames and 109,554 non-polyp frames.  We further used additional annotations of polyp frames for informative/uninformative classes provided by \cite{sharma2023multi} (not provided in the original SUN Database). These additional annotations comprise 31\% uninformative polyp frames and are used to train our model in Stage-I. We used the same dataset split as used in \cite{sharma2023multi}. ISIT-UMR Colonoscopy Dataset is an NBI and WLI video dataset and consists of labels for hyperplastic, adenoma and sessile classes. We used the first two of these classes to train our model in Stage-II. The same dataset is used for validation purposes during classification. We converted 40 adenoma and 21 hyperplastic video streams of each NBI and WLI modalities into frames. We performed a train and test split (video level split) for each class for both modalities, thus creating a split of ratio 29:11 and 15:6 for adenoma and hyperplastic classes, respectively. 29 adenoma and 15 hyperplastic videos of both modalities are used for training PathoPolyp-Diff. The ISIT-UMR dataset has a notable class imbalance between adenoma and hyperplastic classes. These are pathology-based classes, and differentiating them is a complex task, which could unnecessarily impact the training and validation of our proposed diffusion model. Therefore, we undersampled video frames of adenoma class during PathoPolyp-Diff training. Further, during the classification task, we selected an equal number of variable frames (16, 32, or 64) from each video, therefore, to avoid class imbalance, an equal number of videos of both classes were randomly selected.  15 videos (a subset of the previous 29 adenomas) and 6 videos of each pathology and modality were randomly selected to form the train set and test set, respectively. Note that the test set is mutually exclusive to the training sets of both the diffusion model and the classifier. The dataset split details are summarized in Table \ref{tab:dataset}.

During PathoPolyp-Diff training in Stage-I, a pre-trained Stable Diffusion v1-4 model was loaded, which was further fine-tuned on the SUN Database with a learning rate, batch size and resolution set to $1e^{-06}$, 15, and 512, respectively. A lower learning rate allows stable training as mentioned in the literature\cite{rombach2022high} and a resolution of 512 maintains compatibility with the resolution on which the Stable Diffusion v1-4 model was originally trained. The best model chosen for subsequent fine-tuning in Stage-II was identified after 8,000 iterations (based on t-SNE plots and Table \ref{tab:step1_res}). In Stage-II, ISIT-UMR Colonoscopy Dataset is used for training with the same learning rate and one block of U-Net locked. The training of EfficientNet-B0 classifier involves 50 epochs and 0.00005 learning rate. All the implementations are carried out using the PyTorch framework and experiments are executed on NVIDIA A100 and NVIDIA Titan-Xp GPU for PathoPolyp-Diff and classification tasks, respectively.

\noindent \textbf{Evaluation Metrics:} In this work, we adopted some standard metrics used for classification. It includes precision, recall, F1-score, and balanced accuracy. The first three are commonly used to evaluate any classification model. The last metric, i.e. balanced accuracy, is generally used when an imbalance is encountered in data distribution, and the objective focuses on both minority and majority classes with equal importance during evaluation. This scenario aligns with our case, where we consider both adenomatous and hyperplastic classes to be equally important, as the clinical treatments depend on the diagnosed pathology class. The balanced accuracy can be defined as an arithmetic mean of specificity (true negative rate) and sensitivity (true positive rate). In addition, we used Kernel Inception Distance (KID)~\cite{binkowski2018demystifying} for quality assessment of the generated images. It quantifies the dissimilarity between the generated and real data distributions. Unlike the standard KID metric, which uses Inception-v3 as the feature extractor, we used the DenseNet model inspired by the literature \cite{sharma2023multi,jha2023gastrovision} where it achieved remarkable performance in a similar domain.

\begin{table}
\centering
\resizebox{\linewidth}{!}{
\begin{tabular}{ccccccccccccc}
\rowcolor[HTML]{C0C0C0} 
\multicolumn{1}{c}{\cellcolor[HTML]{C0C0C0}}                                                                  & \multicolumn{1}{c}{\cellcolor[HTML]{C0C0C0}}                          & \multicolumn{1}{c}{\cellcolor[HTML]{C0C0C0}}                           & \multicolumn{10}{c}{\cellcolor[HTML]{C0C0C0}Iterations} \\
\rowcolor[HTML]{EFEFEF} 
\multicolumn{1}{c}{\multirow{-2}{*}{\cellcolor[HTML]{C0C0C0}Prompt}}                                          & \multicolumn{1}{c}{\multirow{-2}{*}{\cellcolor[HTML]{C0C0C0}Metrics}} & \multicolumn{1}{c}{\multirow{-2}{*}{\cellcolor[HTML]{C0C0C0}Behavior}} & 1K        & 2K        & 3K       & 4K       & 5K  & 6K & 7K & 8K & 9K & 10K     \\     & Precision               &    $\uparrow$                                                                    &    0.8671       &    0.9085       &  0.9617        &    0.9720      & 0.9850   & 0.9858 & \textbf{0.9961} & 0.9859 & 0.9761 &   0.9899   \\
 & Recall               &    $\uparrow$                                                                    & 0.9567          &   \textbf{0.9933}        &  0.8367        &  0.9267        & 0.8733   & 0.9267 & 0.8467 & 0.9300 & 0.9533 &  0.9800    \\
\multirow{-3}{*}{\begin{tabular}[c]{@{}l@{}}colonoscopy image with\\ polyp, good-quality, clear\end{tabular}} & F1-score   & $\uparrow$          &   0.9097        &    0.9490       &  0.8948        &    0.9488      &   0.9258 & 0.9553 & 0.9153 & 0.9571 & 0.9646 & \textbf{0.9849}       \\  
  & KID          &     $\downarrow$                                                                        &  \textbf{0.045}          &    0.059       &  0.055        &  0.075        &  0.053   & 0.071 & 0.048 & 0.051 & 0.049 &  0.064   \\ 
\hline 
\begin{tabular}[c]{@{}l@{}}colonoscopy image \\ without polyp\end{tabular} & KID      & $\downarrow$           &  \textbf{0.027}         &  0.041        & 0.037    & 0.034  & 0.067 & 0.051 & 0.051 & 0.045 & 0.055 & 0.057     \\
\hline
& Average KID      & $\downarrow$           &     \textbf{0.036}      & 0.05         & 0.046         &  0.0545 & 0.06 & 0.061 & 0.0495 & 0.048 & 0.052 & 0.0605       \\
\end{tabular} }
\caption{Iteration-wise quality assessment of generated images in Stage-I. This assessment is done using KID (similarity with real images), precision, recall, and F1-score (polyp/non-polyp characterizing features). $\downarrow$ and $\uparrow$ denote ‘lower is
best’ and ‘higher is best’, respectively.}
\label{tab:step1_res}
\end{table}

\begin{figure*}[h!]
     \subfloat[\small Iteration-1K\label{fig:y equals 10x1}]{%
       \includegraphics[width=0.32\textwidth]{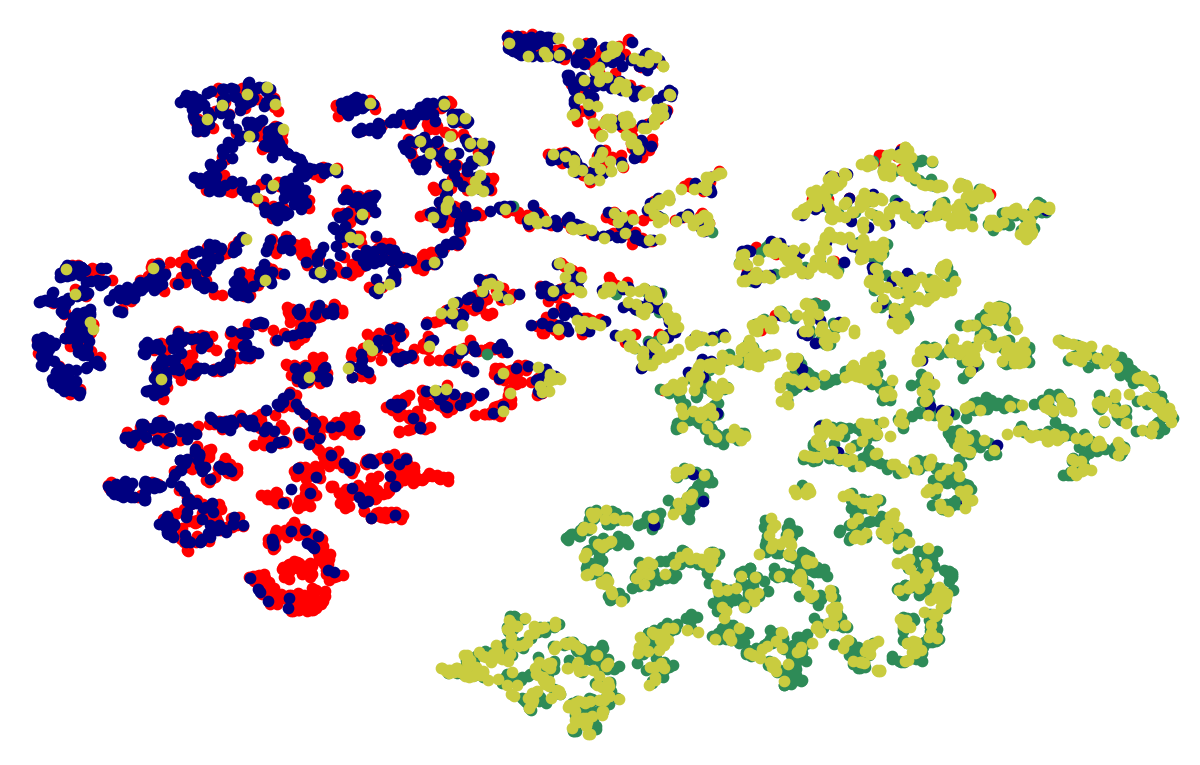}
        }  
      \hfill
     \subfloat[\small Iteration-2K\label{fig:y equals 11x2}]{%
       \includegraphics[width=0.32\textwidth]{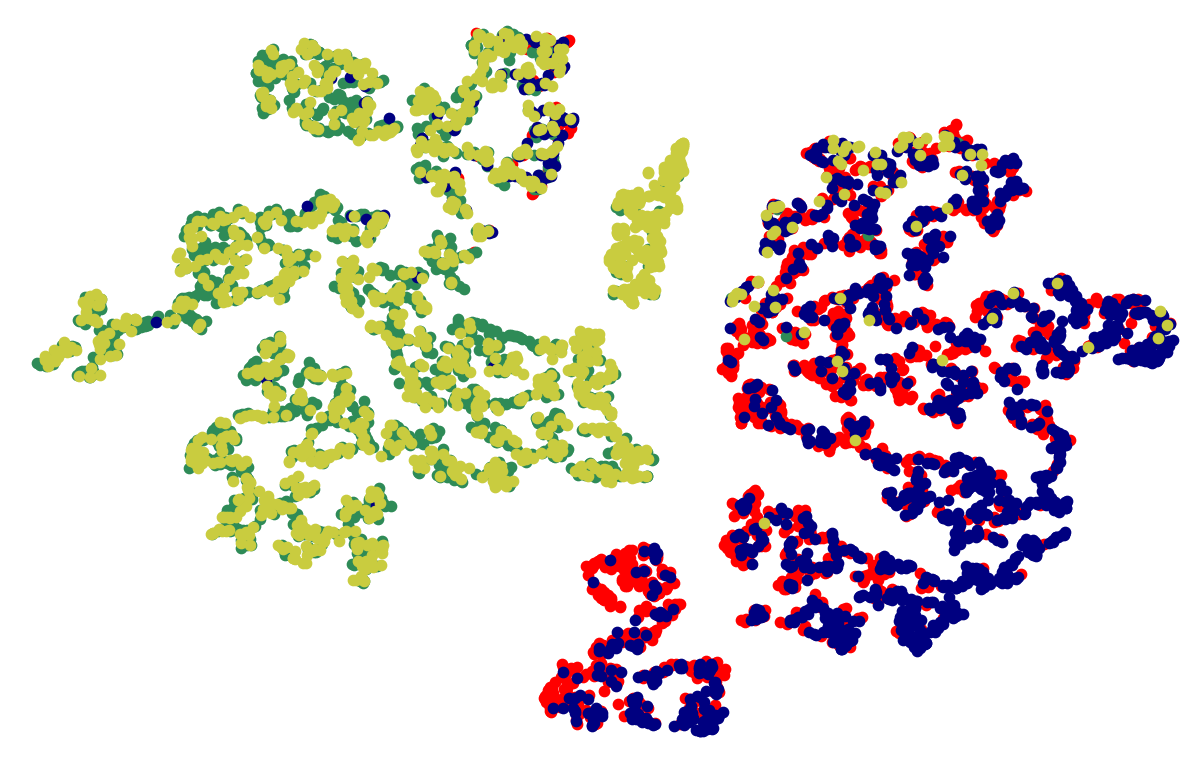}
        }
     \hfill
     \subfloat[\small Iteration-3K\label{fig:y equals 12x3}]{%
       \includegraphics[width=0.32\textwidth]{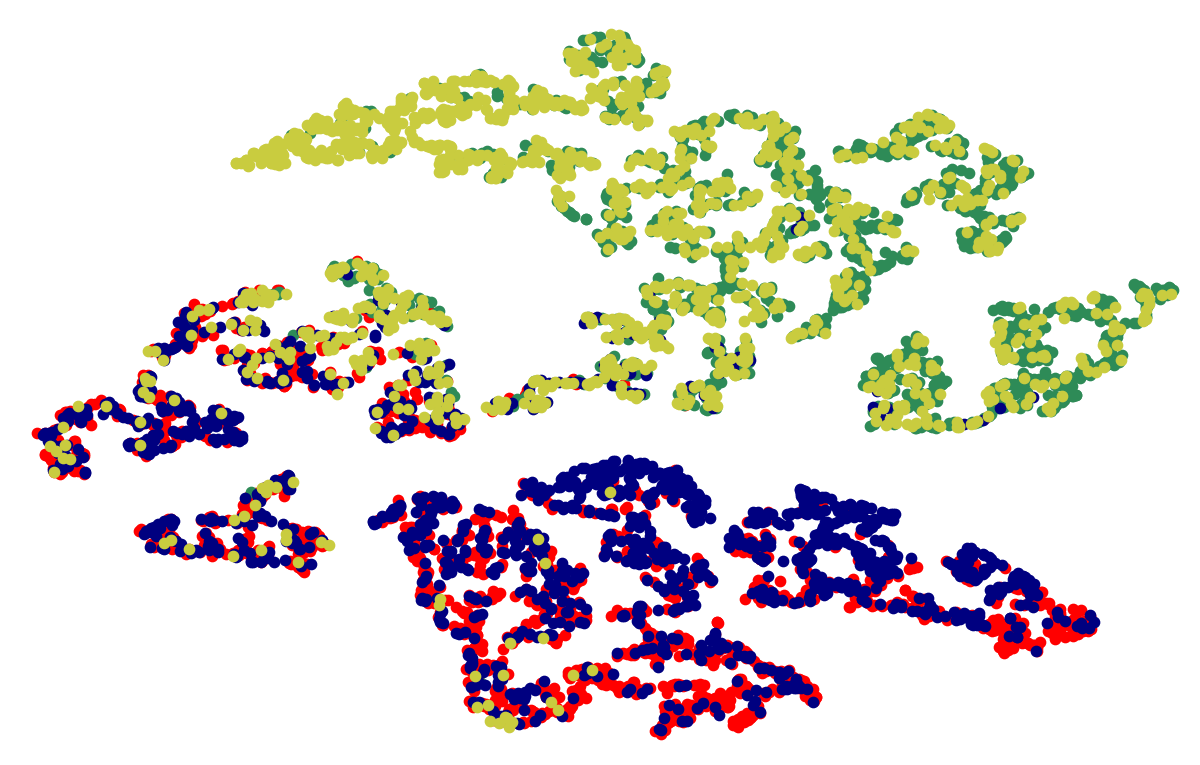}
        }
      
      \subfloat[\small Iteration-4K\label{fig:y equals 13x4}]{%
       \includegraphics[width=0.32\textwidth]{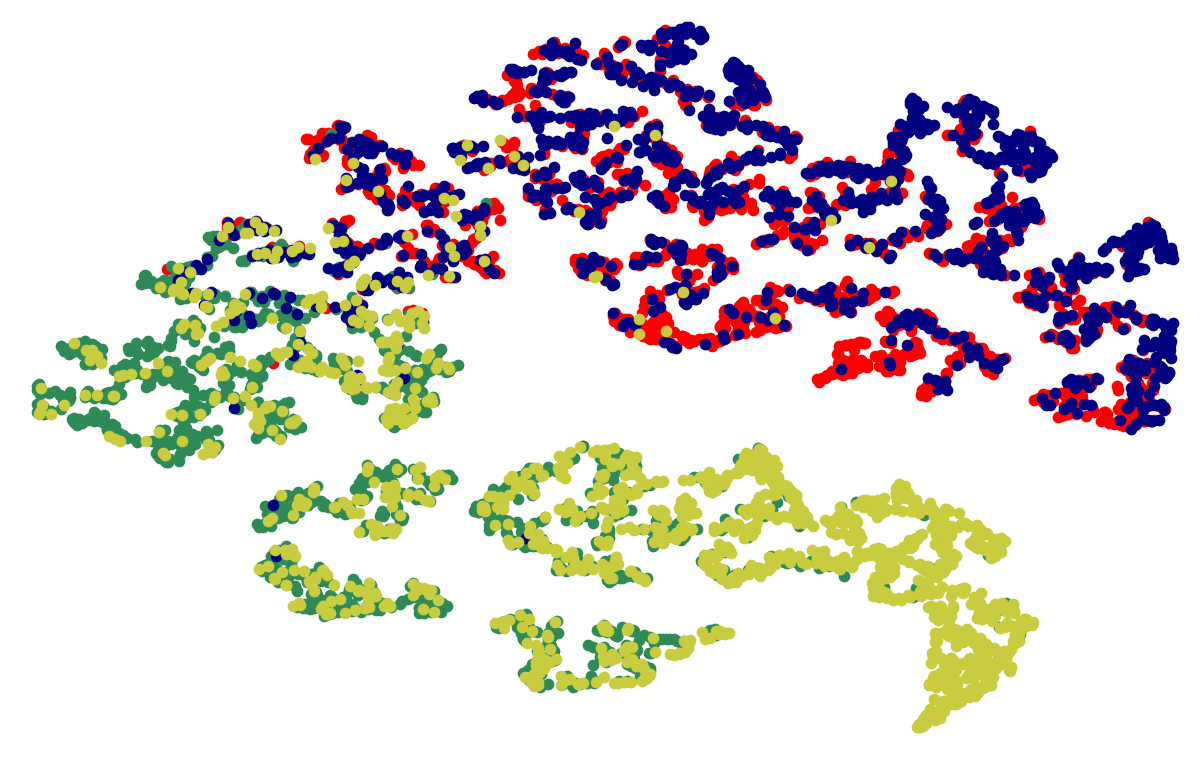}
        }
     \hfill
     \subfloat[\small Iteration-5K\label{fig:y equals x5}]{%
       \includegraphics[width=0.32\textwidth]{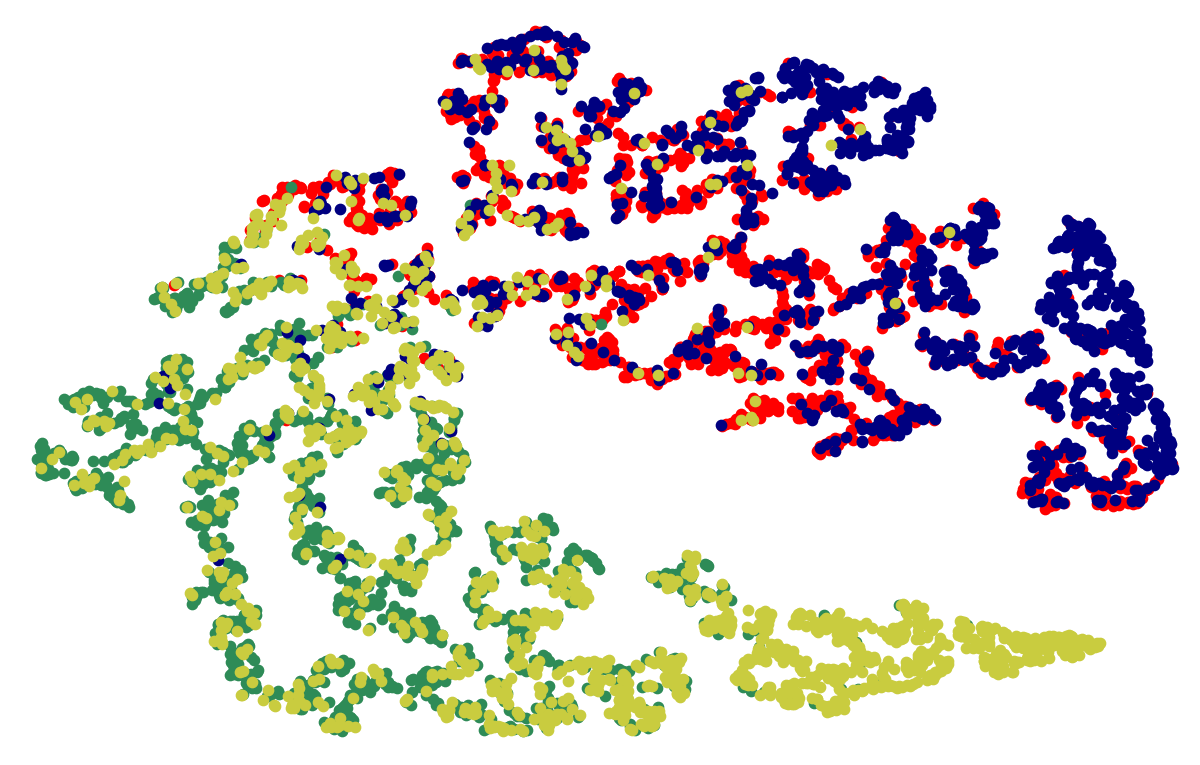}
        }    
      \hfill
       \subfloat[\small Iteration-6K\label{fig:y equals x6}]{%
       \includegraphics[width=0.32\textwidth]{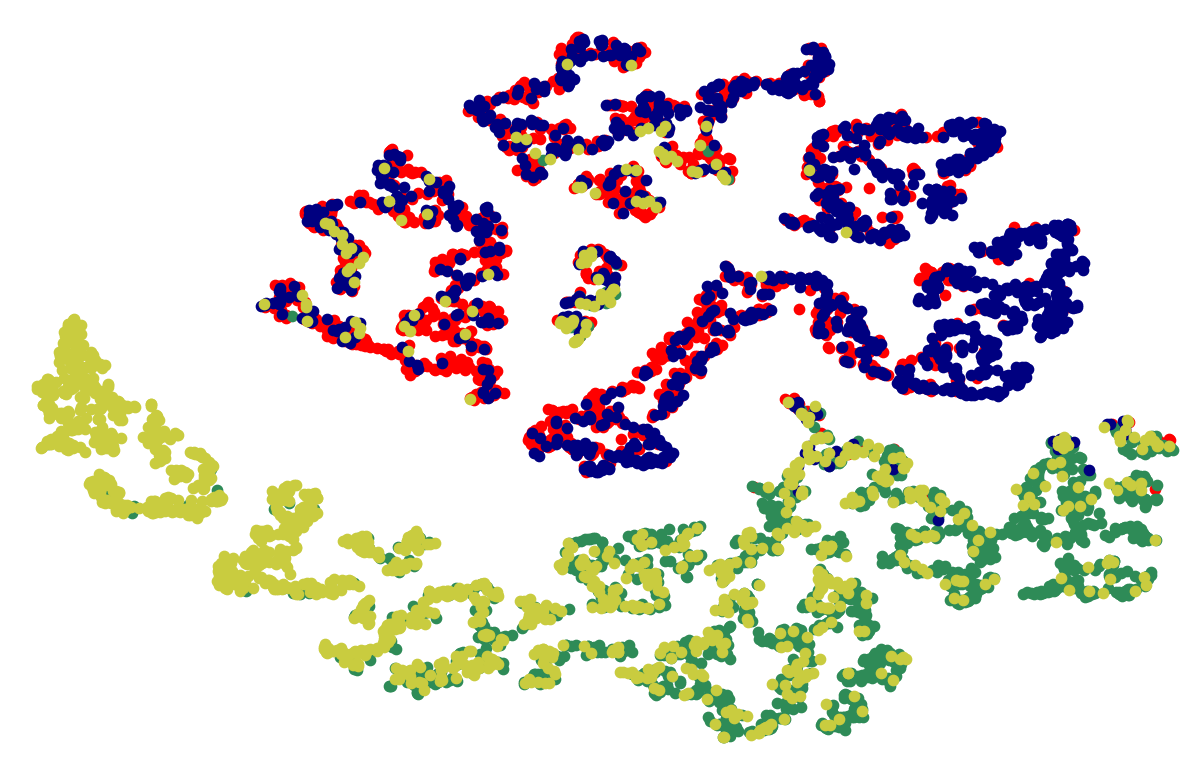}
        }
     
      \subfloat[\small Iteration-7K\label{fig:y equals x7}]{%
       \includegraphics[width=0.32\textwidth]{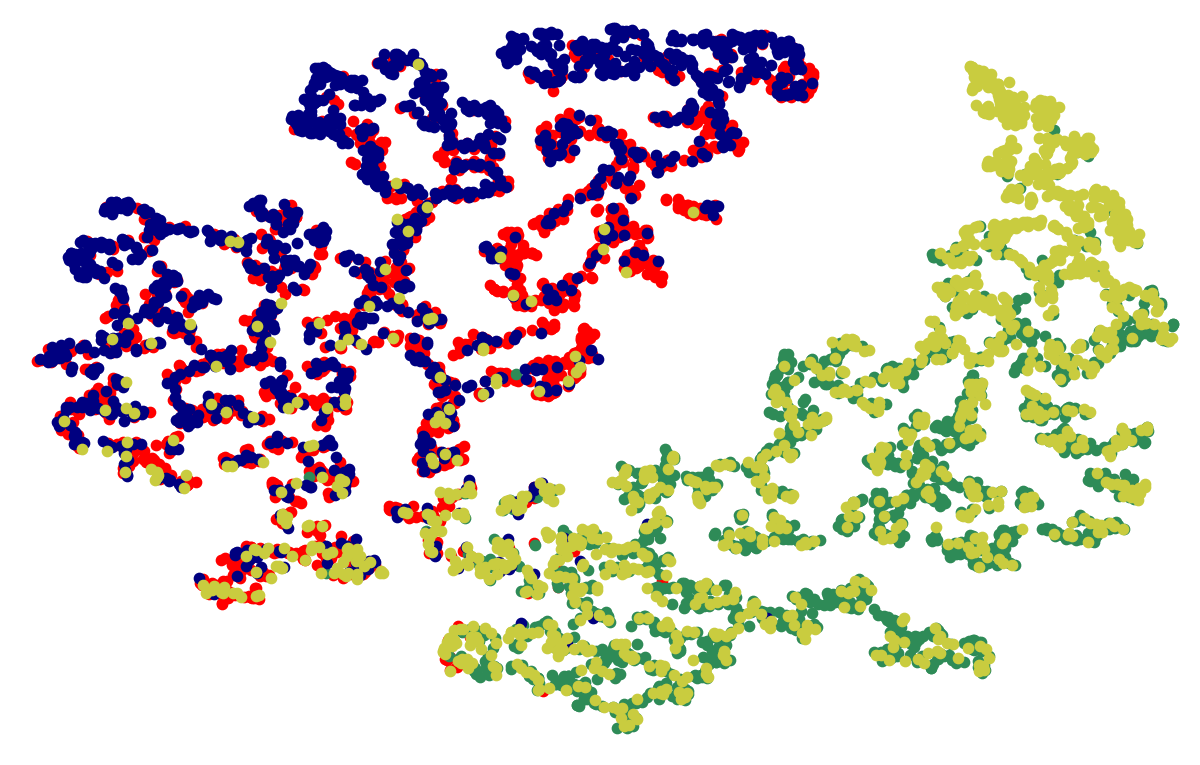}
        } 
      \hfill
      \subfloat[\small Iteration-8K\label{fig:y equals x8}]{%
       \includegraphics[width=0.32\textwidth]{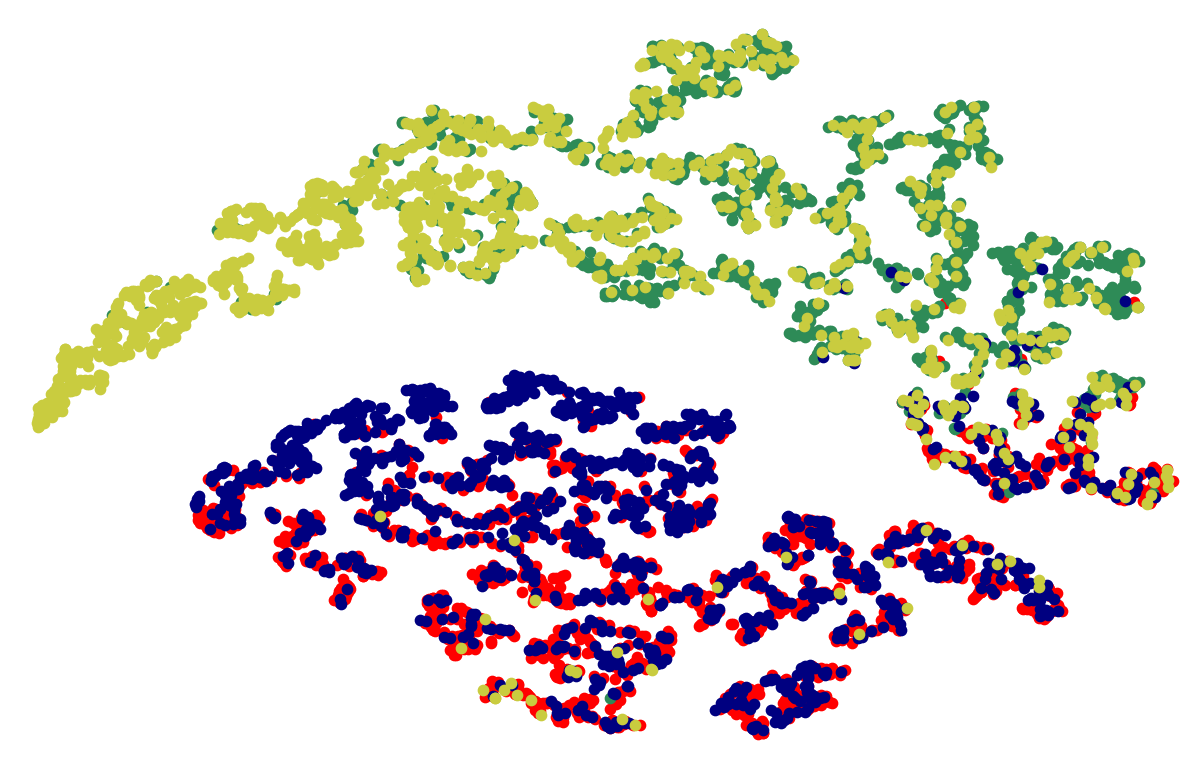}
        }
        \hfill
    \subfloat[\small Iteration-9K\label{fig:y equals x9}]{%
       \includegraphics[width=0.32\textwidth]{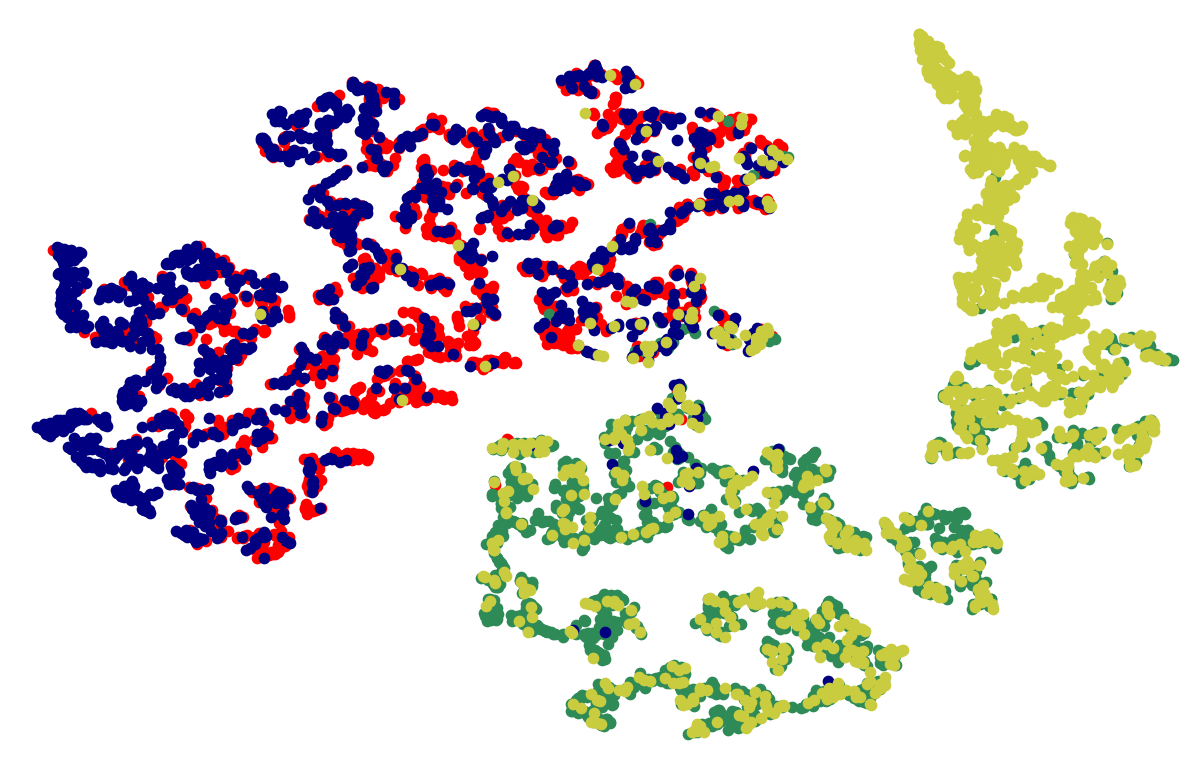}
        } 
      
      \subfloat[\small Iteration-10K\label{fig:y equals x10}]{%
       \includegraphics[width=0.32\textwidth]{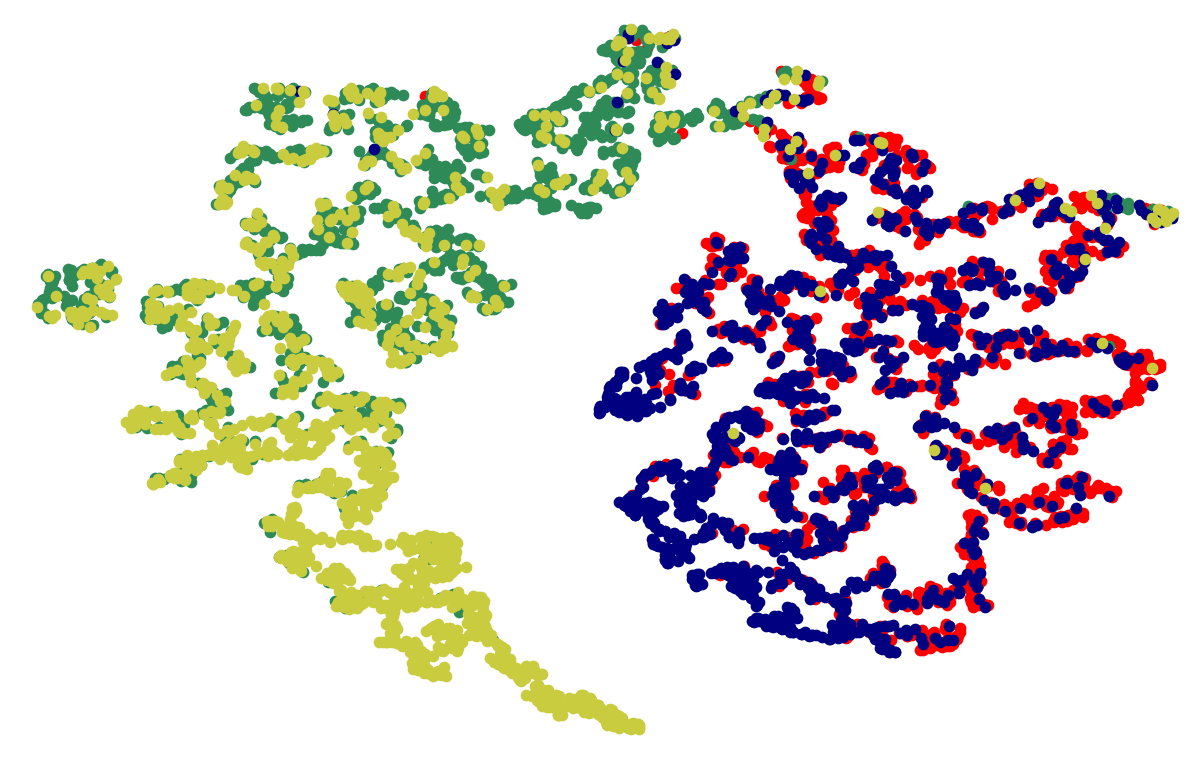}
        }
        \hfill
       \subfloat{%
       \includegraphics[width=0.25\textwidth]{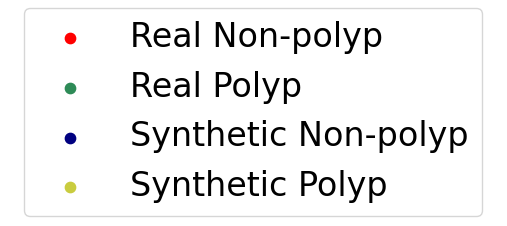}
       } 
        \caption{\small Iteration-wise two-dimensional t-SNE embeddings to visualize the data points pertaining to synthetic and real polyp/non-polyp images. }
        \label{fig:step1}
\end{figure*}

\subsection*{Model Performance}
\textbf{Stage-I:} To evaluate the performance of our model in generating polyp and non-polyp images and to select the best model for the subsequent training process, we used four assessment metrics, namely, KID, precision, recall, and F1-score. Initially, we trained the model for 10,000 iterations and selected the model after every 1,000 iterations for testing. The corresponding assessment results are given in Table \ref{tab:step1_res}. As our main focus is to obtain good-quality polyp images, we generated and validated images with substring \textit{``good-quality, clear''}. Additionally, we used a DenseNet-201~\cite{huang2017densely}, pre-trained with the same training settings as used by~\cite{sharma2023multi} to detect polyps. This model validates the polyp-characterizing features existing in the generated polyp images. For this, we tested generated polyp and non-polyp images using DenseNet-201. It can be observed that the lowest (also the best) average KID of 0.036 is obtained at $1,000^{th}$ iteration; however, it reported a low F1-score. Similar outcomes are achieved with the next lowest KID. Contrarily, the highest F1-score in the $10,000^{th}$ iteration is obtained with 0.0605 KID, the second least favorable among all iterations.

To study the reasoning behind the contradictory results, we plotted t-SNE embeddings, shown in Fig. \ref{fig:step1}. It can be observed that in the initial iterations (1,000 to 5,000), the polyp and non-polyp features of synthetic data are not entirely distinct; therefore, the associated F1-scores are relatively low. At the same time, they are finely overlapping with their real counterparts, therefore, resulting in lower KID. To establish a trade-off between the KID and F1-score, we leveraged the visualization capabilities of t-SNE plots. Finally, the model at $8,000^{th}$ iteration is selected as its KID score (in terms of all categories, i.e., polyp, non-polyp and average) is higher than the average score computed over each corresponding KID category. For instance, the average $KID_{polyp}$ is calculated as $KID_{{polyp}_{1k}} + KID_{{polyp}_{2k}} + KID_{{polyp}_{3k}}... + KID_{{polyp}_{10k}}$, which comes out to be 0.57. It can be noted that $KID_{{polyp}_{8k}} < KID_{polyp}$, and a similar observation can be identified in other categories. Moreover, the t-SNE plots signify that after $8,000^{th}$ iteration, the synthetic polyp and non-polyp features start to deviate from the feature space, representing their real counterparts. As one class of synthetic features is still far apart from the other class of synthetic features, the F1-score is higher at the last iterations. The above analysis establishes a relationship between the F1-score and the KID score. While the F1-score effectively measures the distinction between the two classes, the KID score assesses their similarity to real counterparts. Notably, these metrics may conflict in scenarios where the classes exhibit a fair overlap with the real counterparts yet remain poorly separated, or when they are well-separated but lack alignment with real data distributions. To further support these results, we included some sample images from the two extreme cases, i.e. iterations $1,000^{th}$ and $10,000^{th}$, and the best iteration $8,000^{th}$ in Fig. \ref{fig:gensamples_table1}. In the initial iteration, the model demonstrates an understanding of the fundamental structure; however, it lacks refinement in texture and overall appearance. Similarly, in the last iteration, the background structure appears distorted, exhibiting unusual color patterns. The samples corresponding to $8,000^{th}$ iteration show visually appealing outcomes, retaining texture and structural details.

\begin{figure*}[h!]
     \centering
     \subfloat[\small \label{fig:1k_polyp}]{%
       \includegraphics[width=0.15\textwidth]{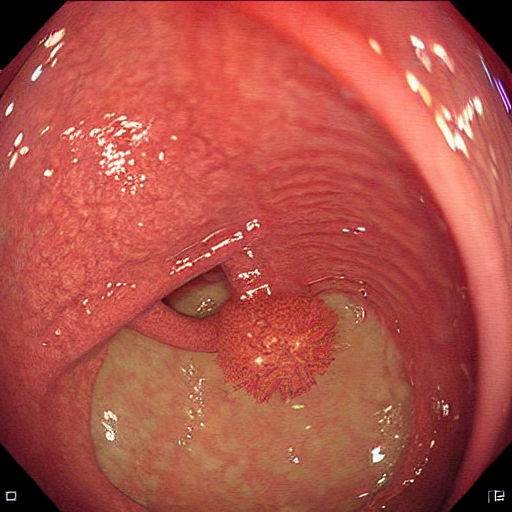}
        }  
      \hfill
     \subfloat[\small \label{fig:1k_nonpolyp}]{%
       \includegraphics[width=0.15\textwidth]{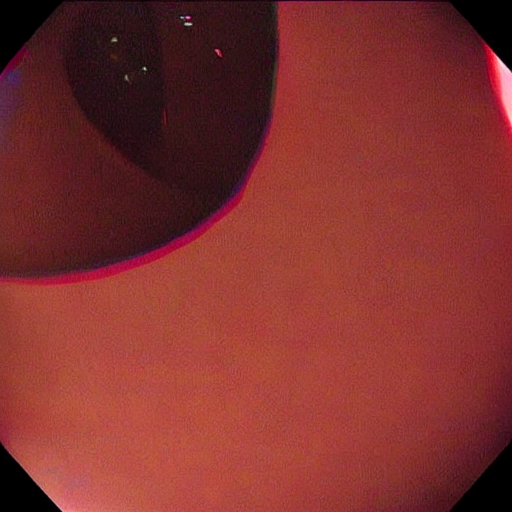}
        }
     \hfill
     \subfloat[\small \label{fig:8k_polyp}]{%
       \includegraphics[width=0.15\textwidth]{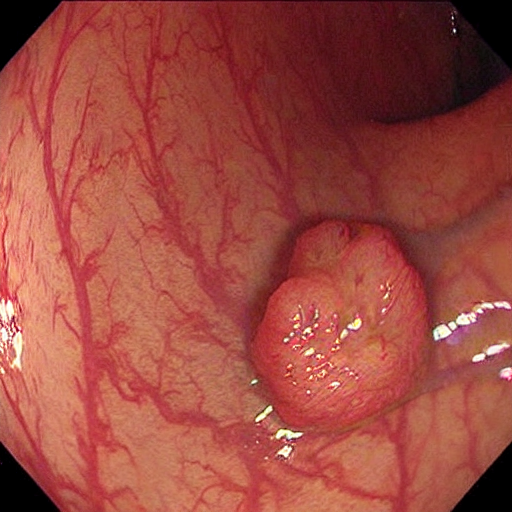}
        }
      \hfill
      \subfloat[\small \label{fig:8k_nonpolyp}]{%
       \includegraphics[width=0.15\textwidth]{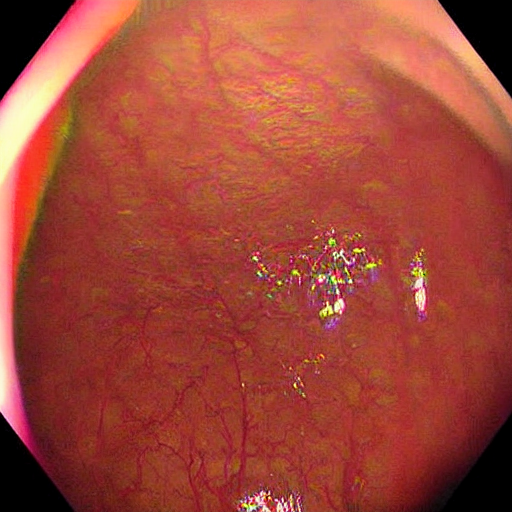}
        }
     \hfill
     \subfloat[\small \label{fig:10k_polyp}]{%
       \includegraphics[width=0.15\textwidth]{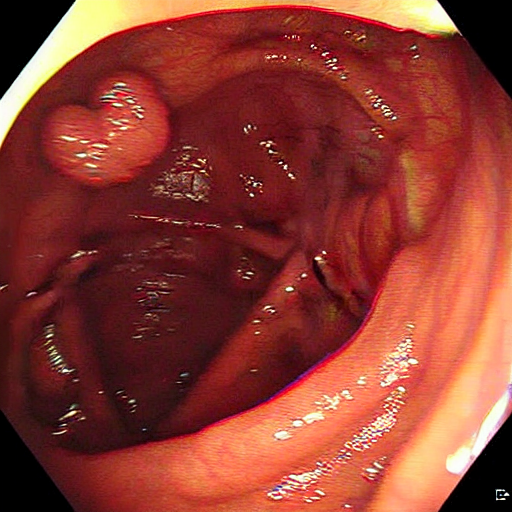}
        }
      \hfill
      \subfloat[\small \label{fig:10k_nonpolyp}]{%
       \includegraphics[width=0.15\textwidth]{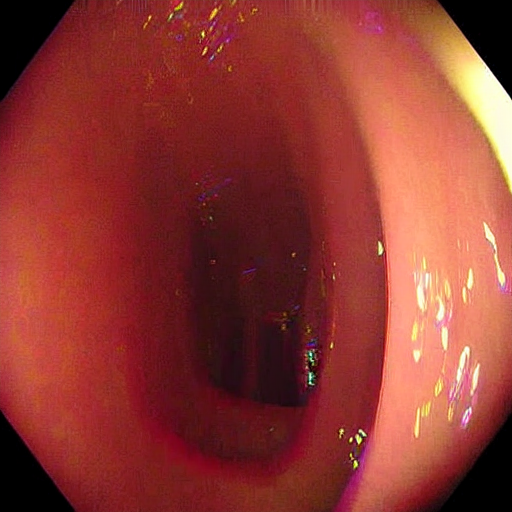}
        }
        \caption{\small Sample generated images from iterations (a) 1K polyp, (b) 1K non-polyp, (c) 8K polyp, (d) 8K non-polyp, (e) 10K polyp, and (f) 10K non-polyp.} 
        \label{fig:gensamples_table1}
\end{figure*}

\textbf{Impact of Negative Prompt:} Negative prompt is an additional parameter which guides the image generation process \textit{not} to include some specific objects or characteristics. It can assist in eliminating unwanted elements from the synthetic images. Hence, to further improve the quality of generated images, we evaluated our model on various negative prompts, including \textit{``low-quality'', ``blur''} and \textit{``low-quality, blur''}. This approach is experimented with both polyp and non-polyp frames, and quality assessment is performed using the fast Fourier transform (FFT). This metric quantifies the blurriness content of the given image. The results shown in Fig. \ref{fig:fft} present a significant improvement when a normal text prompt is combined with our specific quality-based negative prompt. The best combination is achieved when we club text prompt with \textit{``blur, low-quality''} negative prompt. Therefore, the subsequent experiments include this specific negative prompt to enhance quality, as it helps the model avoid generating blurred or low-quality images.

\begin{table}[]
\begin{tabular}{cccllllllllll}
\rowcolor[HTML]{C0C0C0} 
\multicolumn{1}{c}{\cellcolor[HTML]{C0C0C0}}                        & \multicolumn{1}{c}{\cellcolor[HTML]{C0C0C0}}  & \multicolumn{1}{c}{\cellcolor[HTML]{C0C0C0}}                          & \multicolumn{10}{c}{\cellcolor[HTML]{C0C0C0}Iterations}             \\
\rowcolor[HTML]{EFEFEF} 
\multicolumn{1}{c}{\multirow{-2}{*}{\cellcolor[HTML]{C0C0C0}Class}} & \multicolumn{1}{c}{\multirow{-2}{*}{\cellcolor[HTML]{C0C0C0}Metrics}} & \multicolumn{1}{c}{\multirow{-2}{*}{\cellcolor[HTML]{C0C0C0}Behavior}} & 1K   & 2K   & 3K   & 4K   & 5K   & 6K   & 7K   & 8K   & 9K   & 10K  \\
    & Precision    & $\uparrow$                                                      & 0.50 & 0.54 & 0.57 & 0.65 & 0.69 & \textbf{0.71} & 0.65 & 0.59 & 0.61 & 0.58 \\
  & Recall  & $\uparrow$                                                              & \textbf{0.99} & 0.98 & \textbf{0.99} & \textbf{0.99} & 0.98 & 0.97 & 0.98 & \textbf{0.99} & \textbf{0.99} & \textbf{0.99} \\
\multirow{-3}{*}{Adenoma}                                           & F1-score & $\uparrow$            & 0.67 & 0.69 & 0.73 & 0.79 & 0.81 & \textbf{0.82} & 0.78 & 0.74 & 0.76 & 0.73 \\ \hline  
& Precision &$\uparrow$        & 0.75 & 0.90 & \textbf{0.98} & 0.97 & 0.97 & 0.96 & 0.95 & 0.96 & \textbf{0.98} & 0.95 \\
& Recall   & $\uparrow$     & 0.02 & 0.15 & 0.26 & 0.48 & 0.55 & \textbf{0.61} & 0.48 & 0.33 & 0.37 & 0.29 \\
\multirow{-3}{*}{Hyperplastic}                                      & F1-score & $\uparrow$    & 0.04 & 0.26 & 0.42 & 0.64 & 0.70 & \textbf{0.74} & 0.64 & 0.49 & 0.54 & 0.45 
%& Precision  & $\uparrow$       & 0.63 & 0.72 & 0.77 & 0.81 & 0.83 & 0.83 & 0.80 & 0.78 & 0.79 & 0.77 
%& Recall & $\uparrow$           & 0.51 & 0.57 & 0.63 & 0.73 & 0.77 & 0.79 & 0.73 & 0.66 & 0.68 & 0.64 \\
%\multirow{-3}{*}{Weighted Average}    & F1-score  & $\uparrow$    & 0.35 & 0.48 & 0.57 & 0.71 & 0.76 & 0.78 & 0.71 & 0.61 & 0.65 & 0.59
\end{tabular}
\caption{Class-wise quality assessment of generated images after every 1000 iterations during Stage-II. $\uparrow$ denotes ‘higher is best’.} 
\label{tab:step2}
\end{table}

\begin{figure}[h!]
    \centering
    \includegraphics[width=\linewidth]{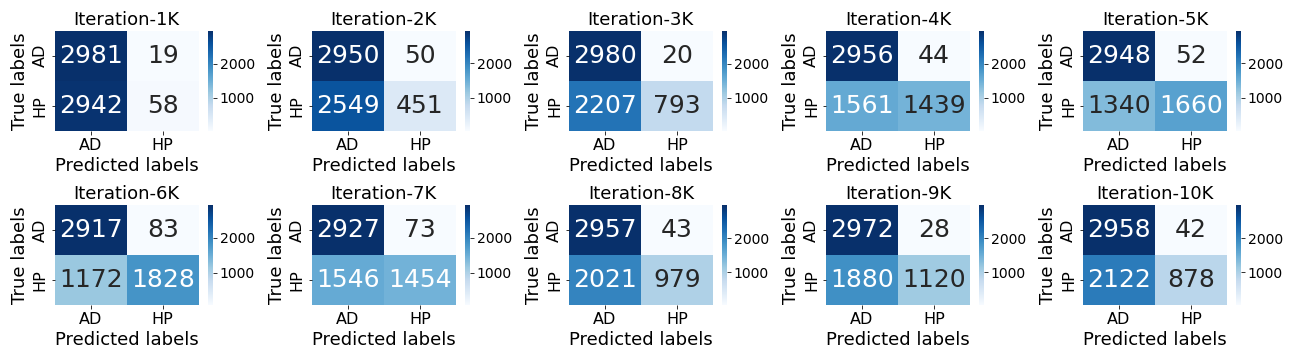}
    \caption{Confusion matrices to validate the iteration-wise performance of our model in generating adenomatous/hyperplastic polyp images with NBI/WLI imaging modalities. AD and HP refer to adenoma and hyperplastic, respectively.}
    \label{fig:conf}
\end{figure}

\begin{figure*}[h!]
     \subfloat[\small Iteration-1K\label{fig:y equals x11}]{%
       \includegraphics[width=0.3\textwidth]{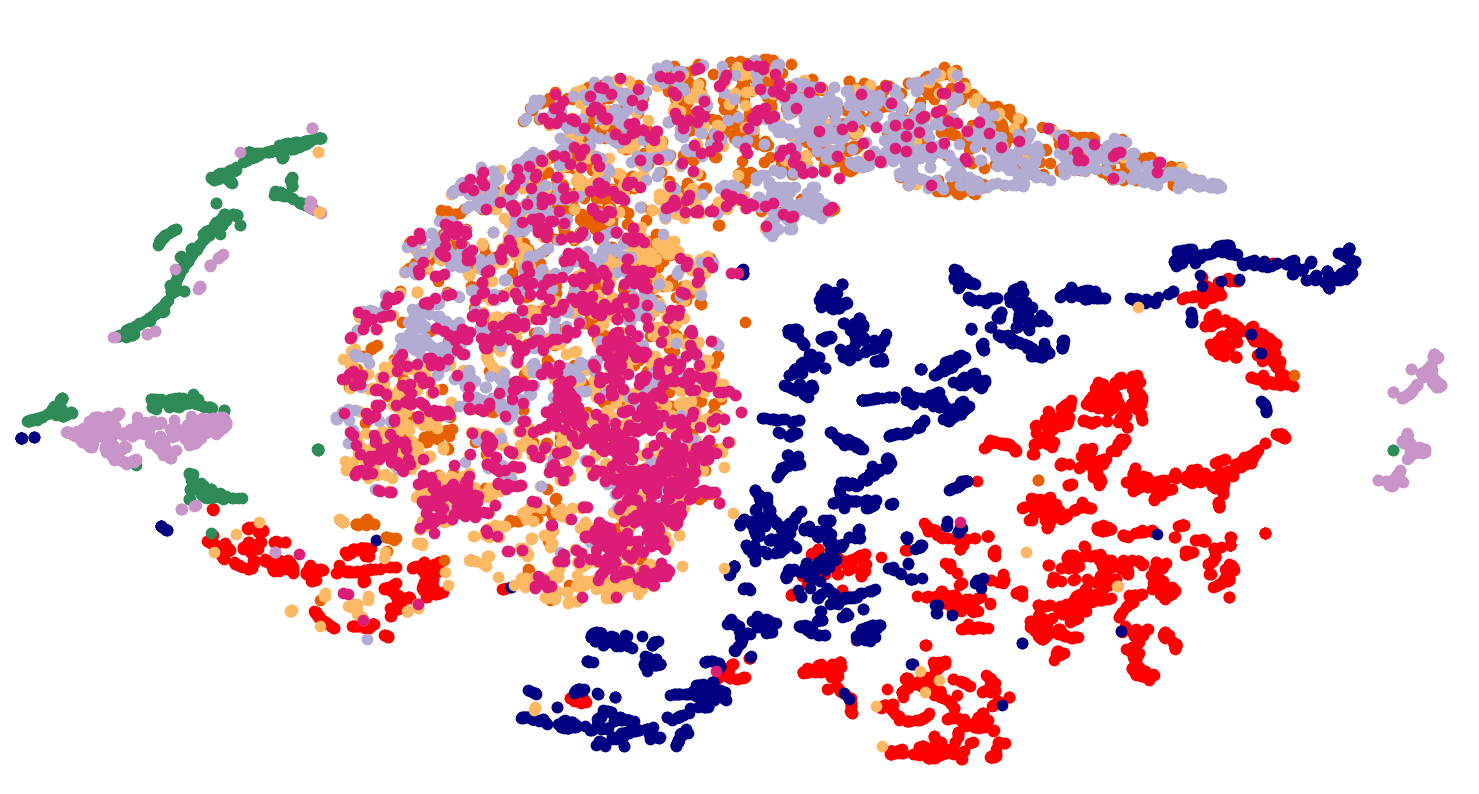}
        }  
      \hfill
     \subfloat[\small Iteration-2K\label{fig:y equals x12}]{%
       \includegraphics[width=0.3\textwidth]{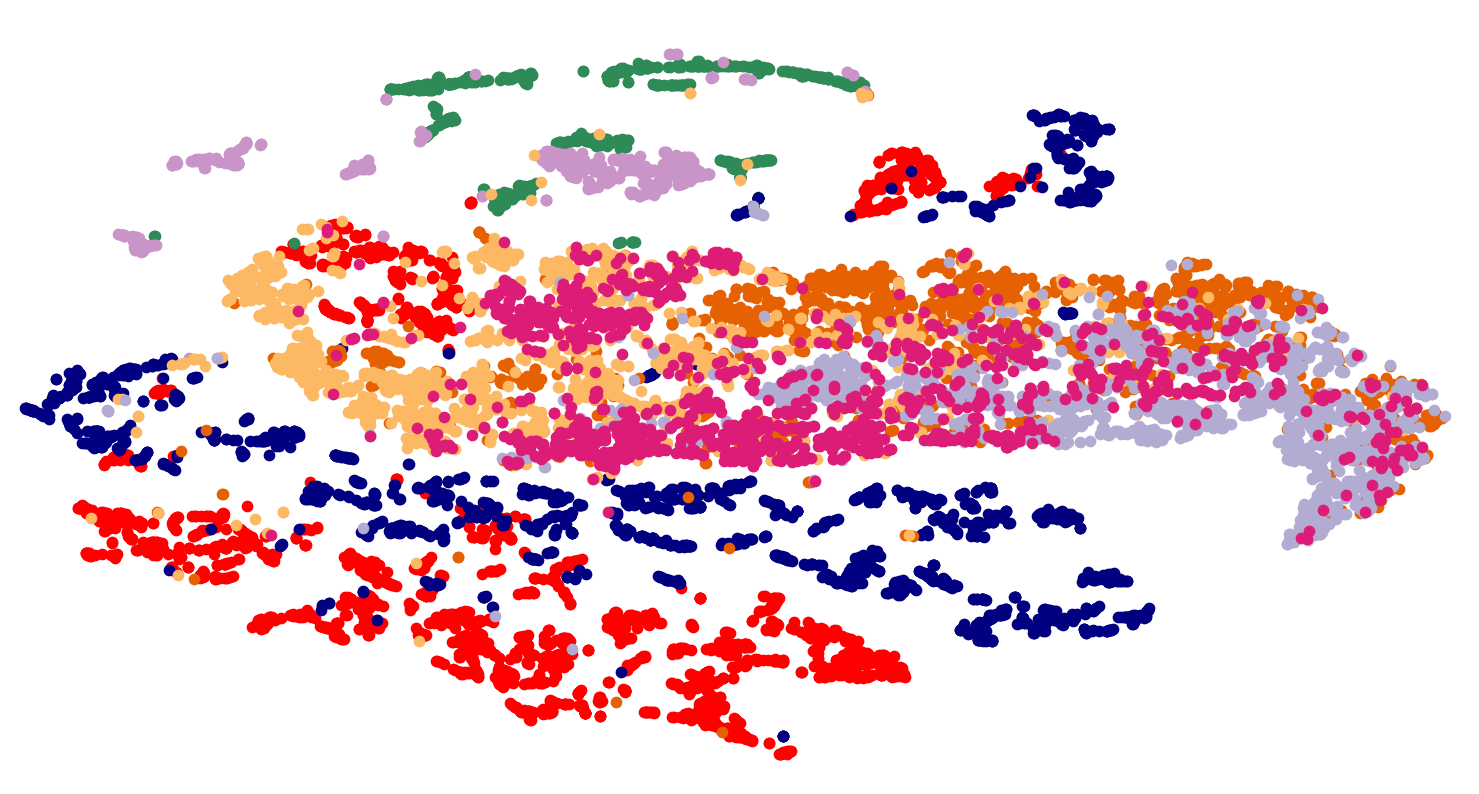}
        }
       \hfill
     \subfloat[\small Iteration-3K\label{fig:y equals x13}]{%
       \includegraphics[width=0.3\textwidth]{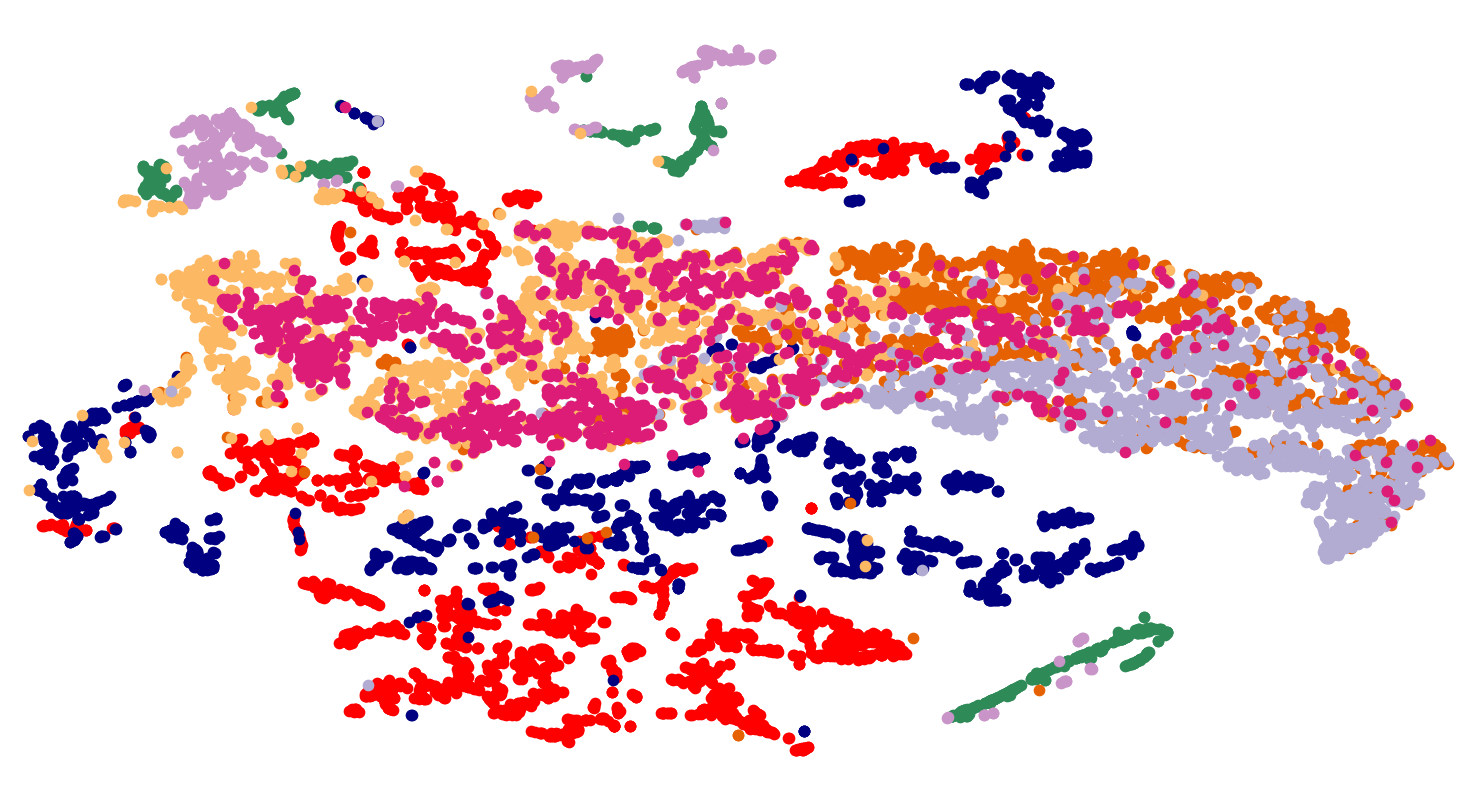}
        }
   
      \subfloat[\small Iteration-4K\label{fig:y equals x14}]{%
       \includegraphics[width=0.32\textwidth]{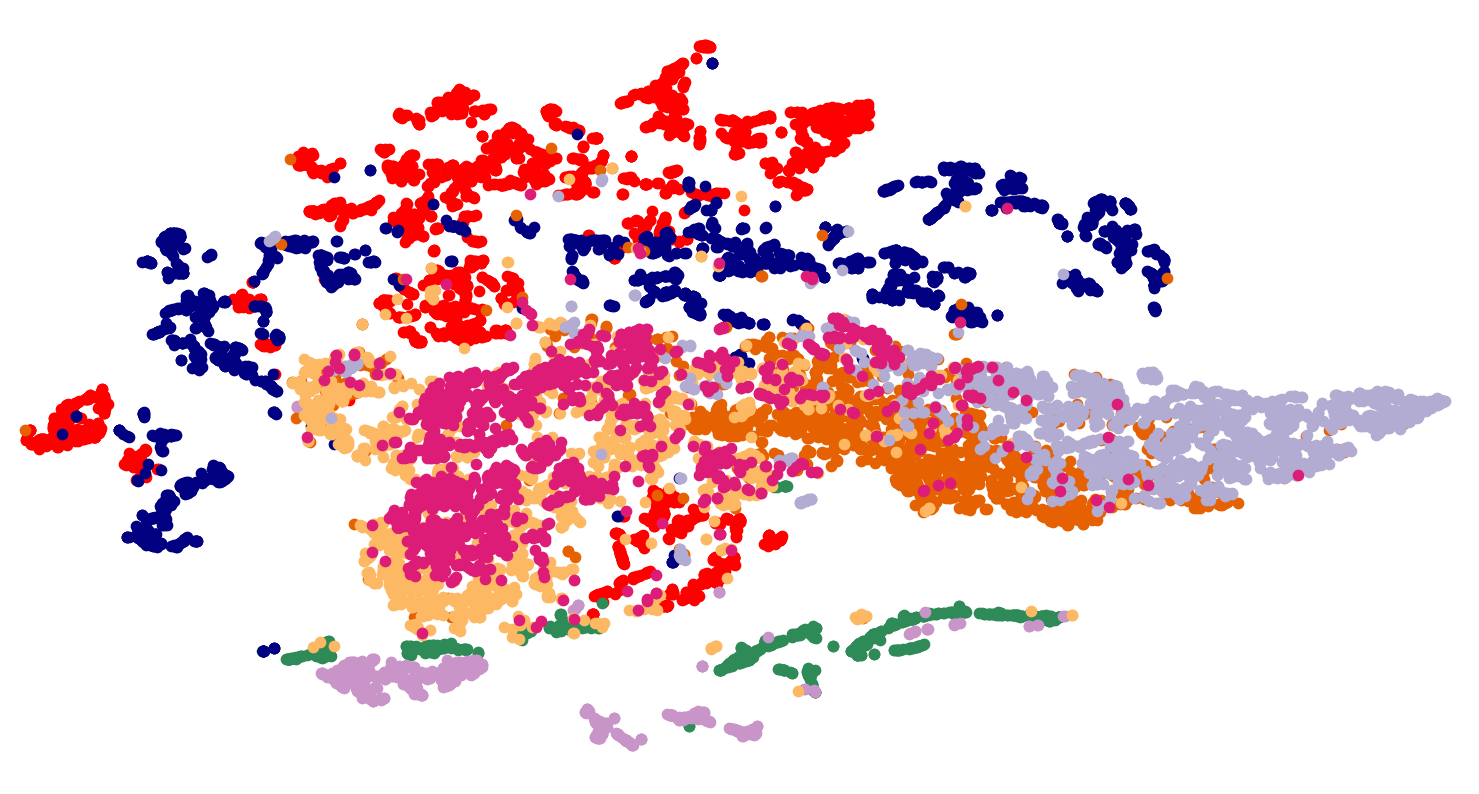}
        }
     \hfill
     \subfloat[\small Iteration-5K\label{fig:y equals x15}]{%
       \includegraphics[width=0.32\textwidth]{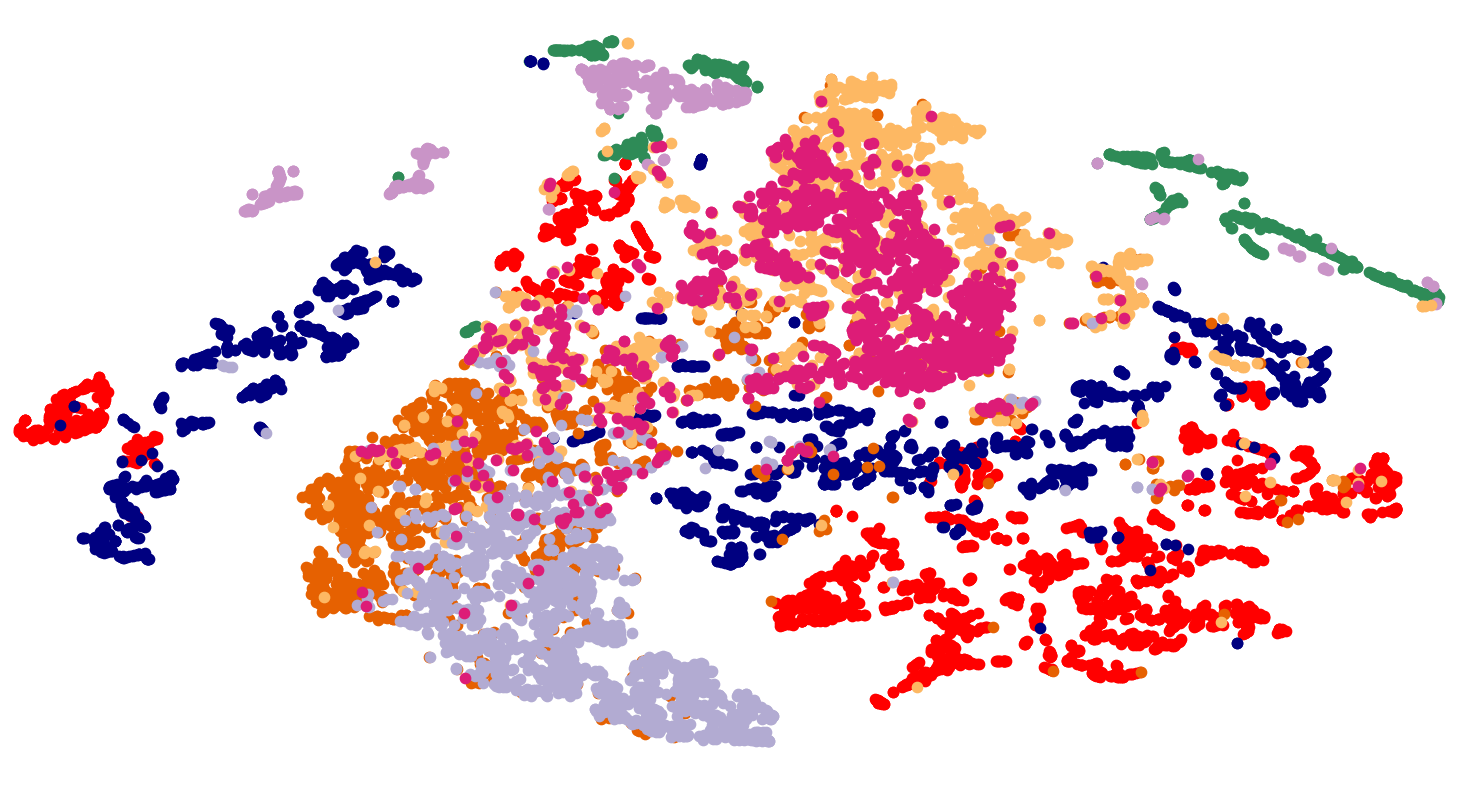}
        }    
      \hfill
       \subfloat[\small Iteration-6K\label{fig:y equals x16}]{%
       \includegraphics[width=0.32\textwidth]{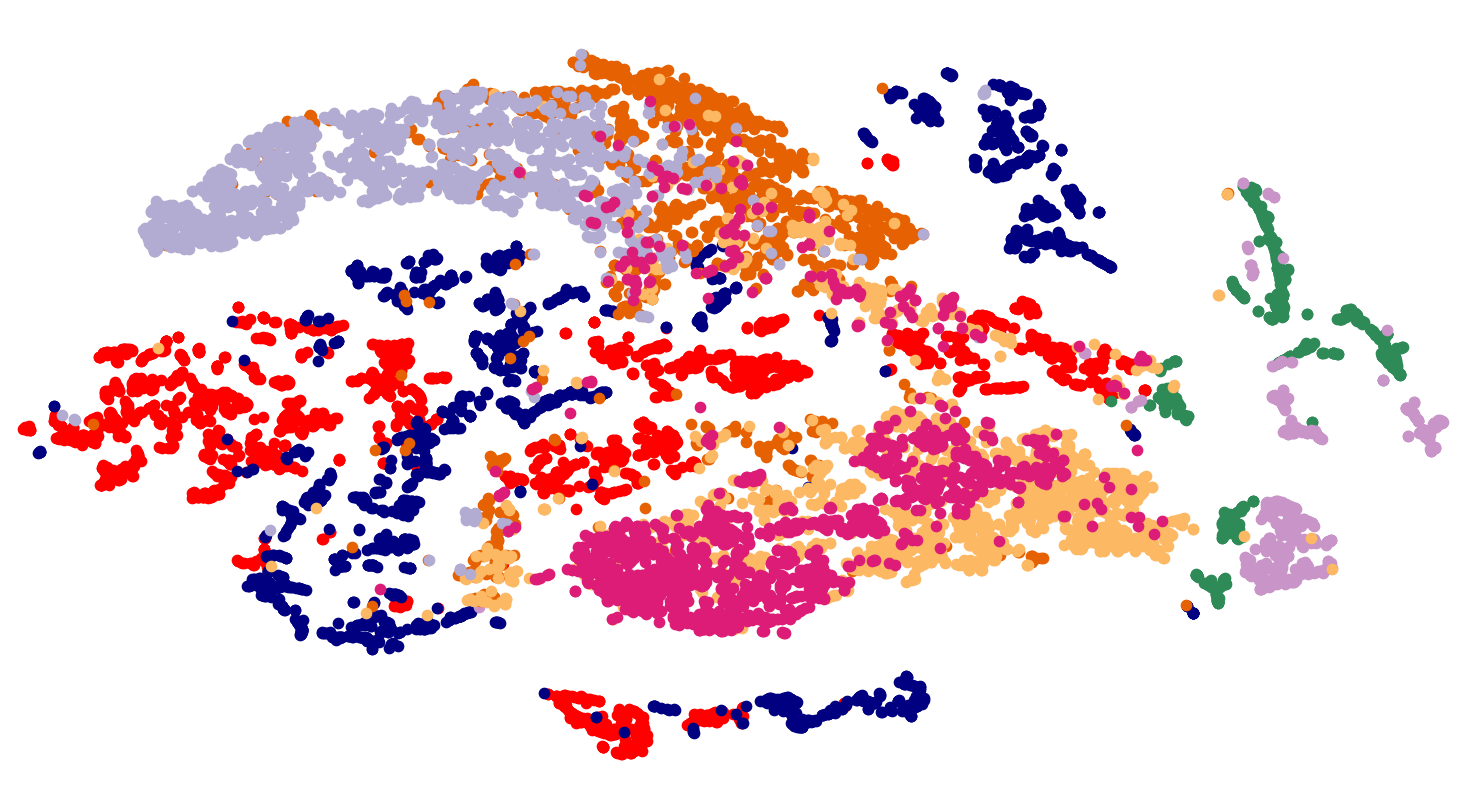}
        }
     
      \subfloat[\small Iteration-7K\label{fig:y equals x17}]{%
       \includegraphics[width=0.32\textwidth]{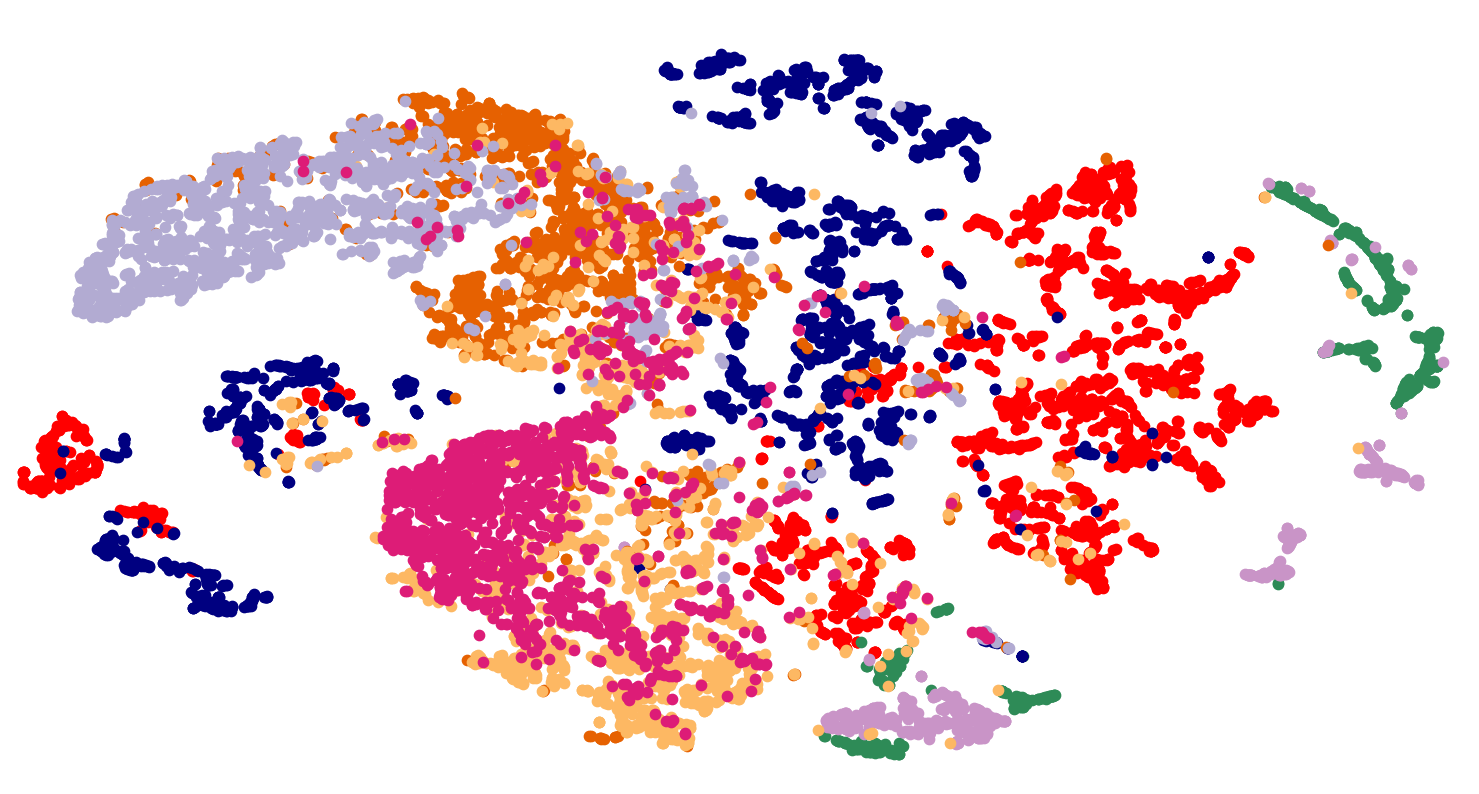}
        } 
     \hfill
      \subfloat[\small Iteration-8K\label{fig:y equals x18}]{%
       \includegraphics[width=0.32\textwidth]{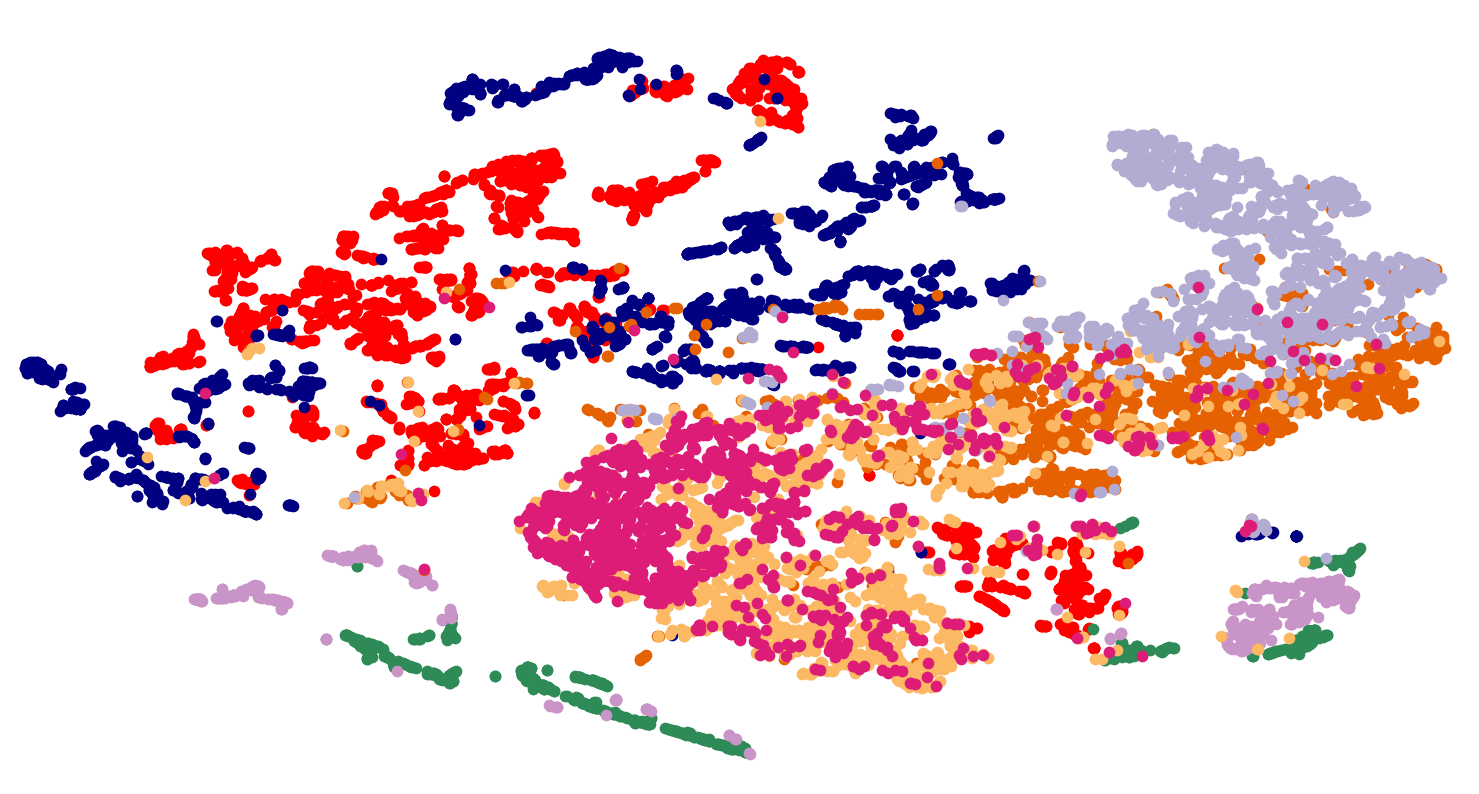}
        }
        \hfill  
    \subfloat[\small Iteration-9K\label{fig:y equals x19}]{%
       \includegraphics[width=0.32\textwidth]{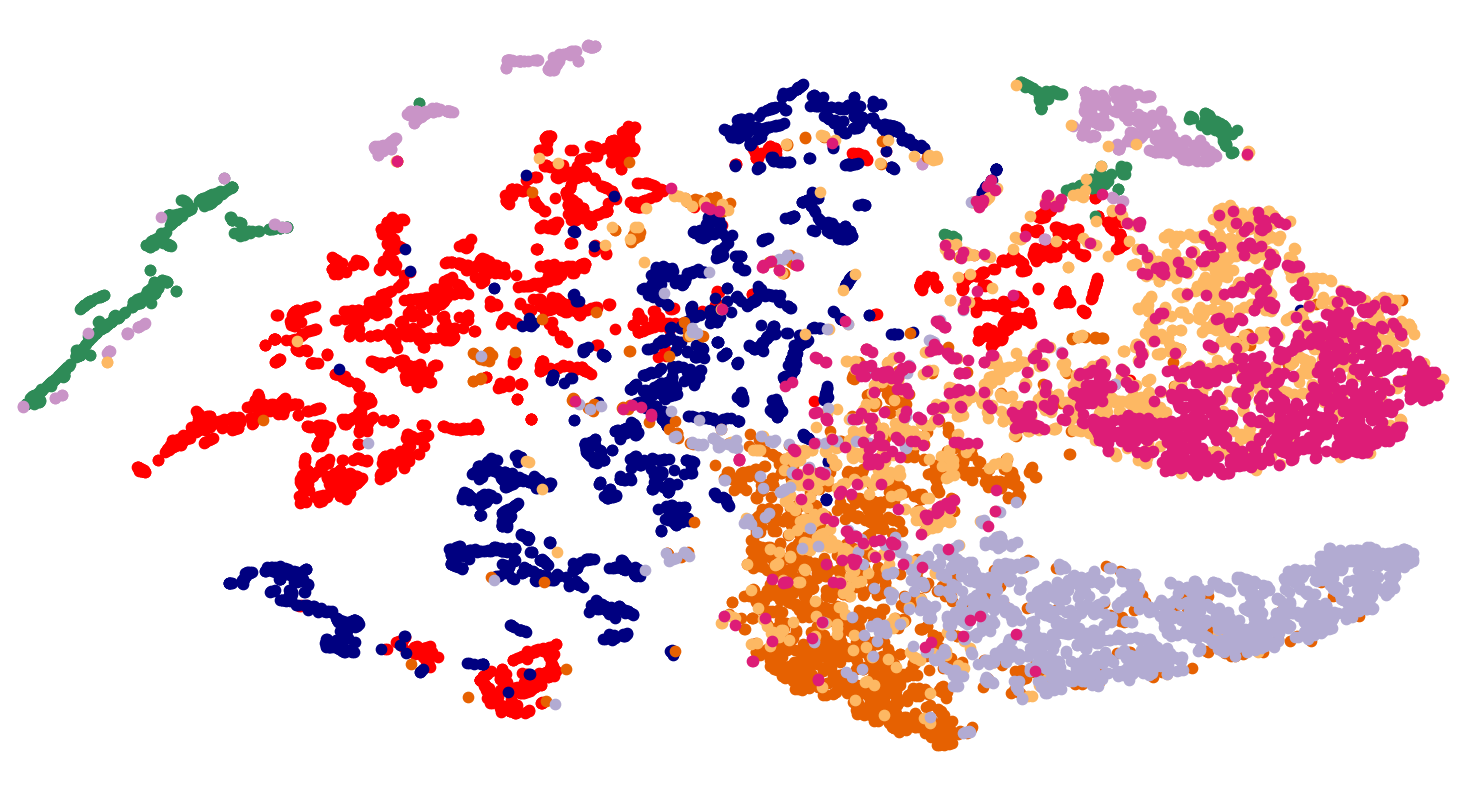}
        } 
     
      \subfloat[\small Iteration-10K\label{fig:y equals x20}]{%
       \includegraphics[width=0.32\textwidth]{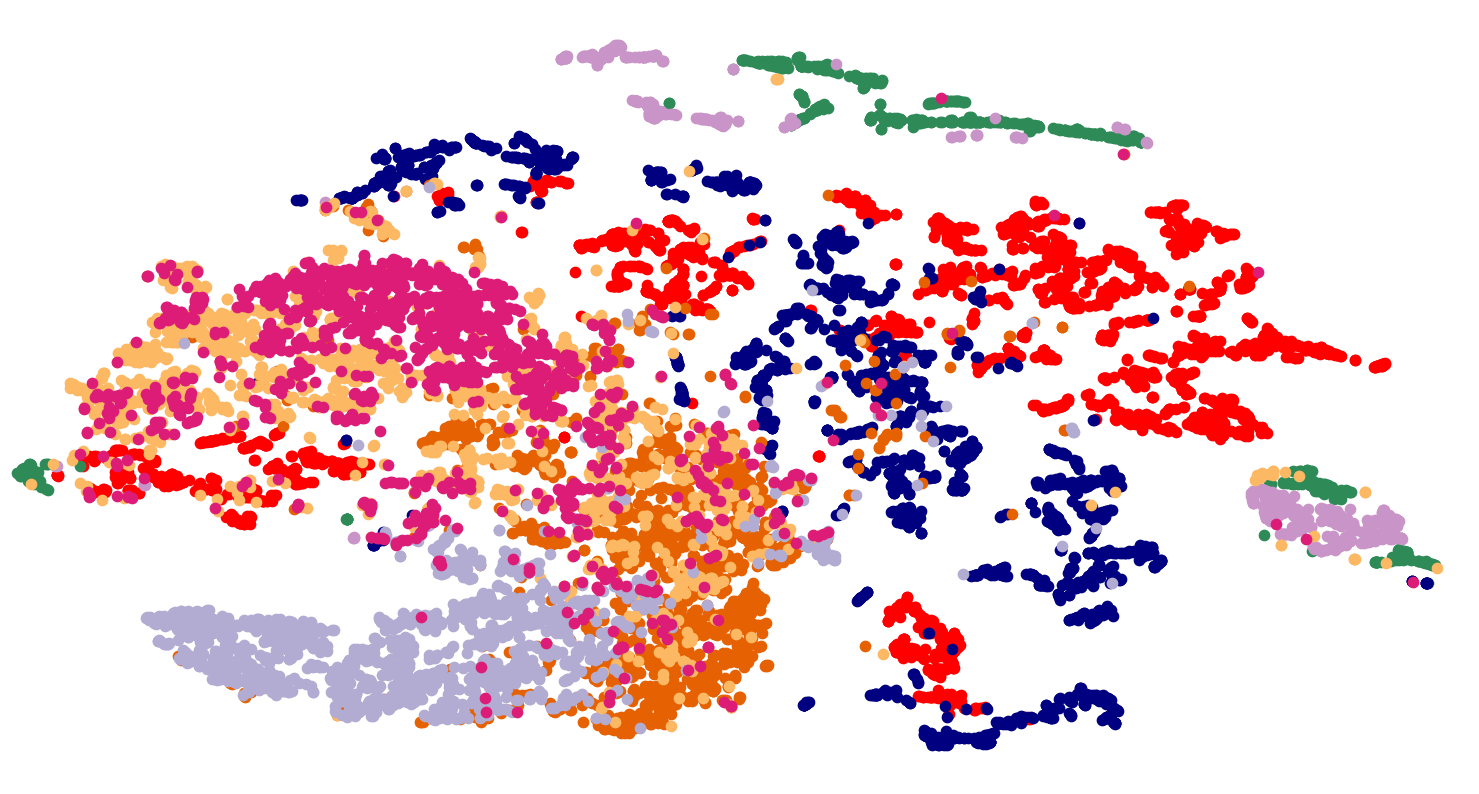}
        }
       \hfill
        \subfloat{%
       \includegraphics[width=0.3\textwidth]{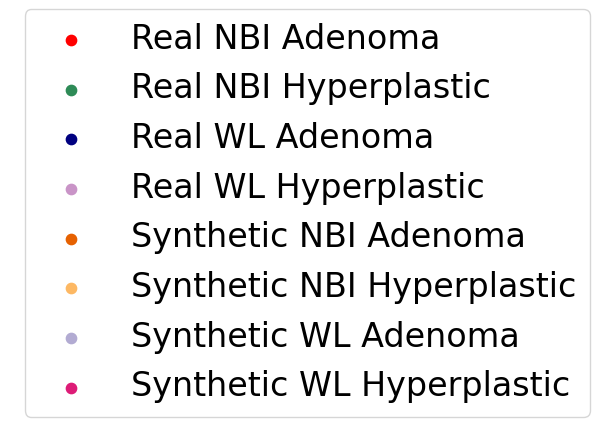}
        } 
    
        \caption{\small  Iteration-wise two-dimensional t-SNE embeddings to visualize the data points pertaining to synthetic and real adenomatous/hyperplastic images involving NBI/WLI imaging modalities. }
        \label{fig:step2}
\end{figure*}

\begin{figure}
    \centering
    \includegraphics[width=0.5\linewidth]{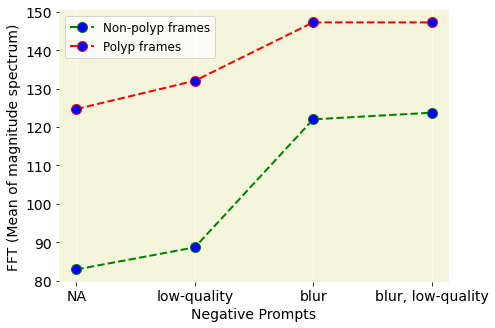}
    \caption{Quality assessment for validating the impact of negative prompt using FFT.}
    \label{fig:fft}
\end{figure}

\noindent \textbf{Stage-II:} To assess the quality of images generated in Stage-II, specifically in terms of pathological characteristics, we performed two different binary classifications. The corresponding purposes of these classifications are (a) To select the best model among different iterations (validation similar to Stage-I) and (b) To verify how the synthetic images impact the classification performance if used for augmentation. In the first case, we used a pre-trained DenseNet-121 model intended to classify adenomatous and hyperplastic classes with WLI and NBI modalities. The pre-training is conducted on the same training dataset that is later used to train the EfficientNet-B0 classifier, however, with all videos and frames. Also, note that the training data for the diffusion model and the classifier overlap, while the classifier's test set remains entirely separate from the training data of both models. This overlap in the training dataset does not affect the evaluation, as the synthetic images generated by the diffusion model are used solely to augment the classifier's training data, ensuring fair evaluation without bias. The synthetic images generated from every $1000^{th}$ epoch are evaluated using this model and the related results are shown in Table \ref{tab:step2}. Similar to Stage-I, we plotted t-SNE feature embeddings to analyze the overlay and proximity of synthetic features to real images. As illustrated in Fig. \ref{fig:step2}, the pathology-specific features are not learnt in the initial epochs. Hence, a significant overlap is observed among the generated images from different classes. It can also be inferred from the findings presented in Table \ref{tab:step2}, where DenseNet-121 exhibited significant challenges in effectively distinguishing between the two classes until $3000^{th}$ iteration. These outcomes are further supported by the confusion matrices in Fig. \ref{fig:conf}. These matrices demonstrate the biased shift of the model towards the adenomatous class, which gradually improves with increasing epochs and, after some epochs, again shows the same biased performance. This trend is observed due to the deviation of synthetic data from the expected pathological behavior as we train the model after a certain number of iterations. This analysis is based on the last few plots in Fig. \ref{fig:step2}. Considering the results in Table \ref{tab:step2}, Fig. \ref{fig:conf}, and Fig. \ref{fig:step2}, we selected the model at $6000^{th}$ iteration which reports the highest F1-score for both the classes (adenoma: 0.82, hyperplastic: 0.74).  

In the second case (i.e. case (b)), we used the synthetic images to augment the real data (using ISIT-UMR dataset). The effectiveness of using synthetic images is validated using a binary classifier, EfficientNet-B0 \cite{tan2019efficientnet}. In addition to validating the synthetic data inclusion, we also compared the quality of synthetic images obtained using two different text prompts (with and without cross-class label learning). Table \ref{tab:framecomp} shows the associated frame-wise results with various data proportions. Starting with real data samples of 16, 32 and 64 frames per video, we subsequently increased the sample count by adding synthetic images in $ix$ proportion, where $i=\{1,2,3\}$ and $x$ is the real frame count. This procedure is followed for A and B text prompts, which can be defined as \textit{"colonoscopy image with `y' polyp, `z'"} and \textit{"colonoscopy image with `y' polyp, `z', good-quality, clear"}, respectively where `y' denote adenomatous/hyperplastic and `z' denote NBI/WLI. It is noteworthy that the label \textit{"good-quality, clear"} used in the text prompt B is not directly related to the training samples provided to PathoPolyp-Diff during Stage-II; instead, they are learnt from a different dataset through cross-class label learning during Stage-I. The comparative analysis of synthetic images pertaining to the two text prompts aims to assess the effectiveness of cross-class label learning. As shown in Table \ref{tab:framecomp}, with 16 real images per video, adding an equal number of synthetic images improves the frame-wise balanced accuracy by 1.14\% to 4.75\% with NBI and WLI. A similar increase of 2.66\% to 3.93\% is reported when the ratio of synthetic data is doubled. With a further increase, i.e., when the proportion of synthetic data is three times, the results are enhanced by up to 7.91\%. However, with 32 or 64 real images per video, the performance increase is relatively less significant and limited to about 3.5\%. Therefore, it can be inferred that a relatively significant performance gain is achieved when a substantially small real dataset is merged with synthetic data. Further increasing the real dataset shows comparatively less improvement (without a monotonic trend) with a similar data augmentation approach.  The trend shown in Table \ref{tab:framecomp} also signifies that over-training the model with synthetic images degrades or saturates the results after a certain proportion, which aligns with the investigation conducted in literature in a similar domain \cite{sharma2024controlpolypnet,sagers2023augmenting}. 

\textbf{Impact of Cross-class Label Learning:} We further performed validation and analysis for the cross-label learning approach. It can be observed that in most cases, the balanced accuracy and class-wise F1-scores using text prompt B are comparatively higher than that of text prompt A. Some notable improvements in balanced accuracy include 0.6192 to 0.6614 (difference of +4.22\%, 95\% CI: 1.5\%, 6.91\% , p-value = 0.0068) using 64 real samples per video with an equal proportion of synthetic data, 0.5561 to 0.6254 (difference of +6.93\%, 95\% CI: 5.01\%, 8.85\%, p-value < 0.0001 ) using 16 real samples per video with two times samples of synthetic data, and 0.6228 to 0.6983 (difference of +7.55\%, 95\% CI: 2.88\%, 12.22\%, p-value = 0.0058) using 16 real samples per video with three times samples of synthetic data. These outcomes signify that the quality of generated data can be improved with variations in the input text prompts. Moreover, the labels used in these text prompts can be indirectly inferred from other classes, thus reducing the requirements of annotated data for each scenario.     

\begin{table}[h!]
\resizebox{\linewidth}{!}{
\begin{tabular}{ccccrrrrrrc}
\multicolumn{1}{l}{\cellcolor[HTML]{C0C0C0}}                                                                              & \multicolumn{2}{l}{\cellcolor[HTML]{C0C0C0}}                                                                                                             & \cellcolor[HTML]{C0C0C0}                             & \multicolumn{3}{c}{\cellcolor[HTML]{C0C0C0}Adenoma}                                                                                                               & \multicolumn{3}{c}{\cellcolor[HTML]{C0C0C0}Hyperplastic}                                                                                                          & \multicolumn{1}{c}{\cellcolor[HTML]{C0C0C0}}                                     \\
\multicolumn{1}{l}{\multirow{-2}{*}{\cellcolor[HTML]{C0C0C0}\begin{tabular}[c]{@{}l@{}}Imaging\\  Modality\end{tabular}}} & \multicolumn{2}{l}{\multirow{-2}{*}{\cellcolor[HTML]{C0C0C0}\begin{tabular}[c]{@{}l@{}}Training sample count\\ (Real + Synthetic)\end{tabular}}}         & \multirow{-2}{*}{\cellcolor[HTML]{C0C0C0}Text Prompt} & \multicolumn{1}{c}{\cellcolor[HTML]{C0C0C0}Precision} & \multicolumn{1}{l}{\cellcolor[HTML]{C0C0C0}Recall} & \multicolumn{1}{l}{\cellcolor[HTML]{C0C0C0}F1-score} & \multicolumn{1}{c}{\cellcolor[HTML]{C0C0C0}Precision} & \multicolumn{1}{l}{\cellcolor[HTML]{C0C0C0}Recall} & \multicolumn{1}{l}{\cellcolor[HTML]{C0C0C0}F1-score} & \multicolumn{1}{c}{\multirow{-2}{*}{\cellcolor[HTML]{C0C0C0}Balanced Accuracy}}  \\
\cellcolor[HTML]{EFEFEF}                                                                                                  &                                                                                      & x                    & - (baseline)                                                    & \underline{0.8173$\pm$0.026} &	0.2897$\pm$0.035 &	0.4266$\pm$0.037 &	0.4757$\pm$0.009	& \textbf{0.9079$\pm$0.020} &	0.6242$\pm$0.007 & 0.5988$\pm$0.012                 					                                                       \\

\cellcolor[HTML]{EFEFEF}                                                                                                  &                                                                                      &                       & A                                                    & 0.7876$\pm$0.122                        & 0.3194$\pm$0.111                                             & 0.4360$\pm$0.099                       & 0.4689$\pm$0.006                                                & 0.8449$\pm$0.121                     & 0.6005$\pm$0.026                                              & 0.5821$\pm$0.007                   					                                                       \\
\cellcolor[HTML]{EFEFEF}                                                                                                  &                                                                                      & \multicolumn{1}{c}{\multirow{-2}{*}{x+x}} & \cellcolor[HTML]{EFEFEF}B                                                    & 0.8146$\pm$0.080                                                & 0.3295$\pm$0.050                                             & 0.4671$\pm$0.056                                               & 0.4853$\pm$0.023                                                & 0.8910$\pm$0.059                                             & \underline{0.6281$\pm$0.031}                                               & \underline{0.6102$\pm$0.037}   					                                                                         \\
\cellcolor[HTML]{EFEFEF}                                                                                                  &                                                                                      &                                                                   & A                                                    & 0.6596$\pm$0.020                                                & \textbf{0.4131$\pm$0.025}                                             & \textbf{0.5077$\pm$0.021}                                               & 0.4580$\pm$0.012                                                & 0.6991$\pm$0.029                                             & 0.5533$\pm$0.016                                               & 0.5561$\pm$0.015                                				                                          \\
\cellcolor[HTML]{EFEFEF}                                                                                                  &                                                                                      & \multirow{-2}{*}{x+2x}                                            & \cellcolor[HTML]{EFEFEF}B                                                    & \textbf{0.8299$\pm$0.042}                        & \underline{0.3569$\pm$0.032}                     & \underline{0.4977$\pm$0.027}                       & \textbf{0.4965$\pm$0.007}                        & \underline{0.8939$\pm$0.040}                     & \textbf{0.6382$\pm$0.012}                       & \textbf{0.6254$\pm$0.011}                                           					                                 \\
\cellcolor[HTML]{EFEFEF}                                                                                                  &                                                                                      &                                                                   & A                                                    & 0.7359$\pm$0.059                                                & 0.3529$\pm$0.055                                             & 0.4730$\pm$0.046                                               & 0.4712$\pm$0.014                                                & 0.8133$\pm$0.076                                             & 0.5959$\pm$0.028                                               & 0.5831$\pm$0.022                              					                                            \\
\cellcolor[HTML]{EFEFEF}                                                                                                  & \multirow{-6}{*}{\begin{tabular}[c]{@{}l@{}}x=16 images\\  per video\end{tabular}} & \multicolumn{1}{c}{\multirow{-2}{*}{x+3x}}                                            & \cellcolor[HTML]{EFEFEF}B                                                    & 0.8133$\pm$0.113                                                & 0.3524$\pm$0.047                                             & 0.4857$\pm$0.027                                               & \underline{0.4867$\pm$0.010}                                                & 0.8680$\pm$0.097                                             & 0.6227$\pm$0.033                                 & \underline{0.6102$\pm$0.025}                                                                        					    \\ \cline{2-11}
\cellcolor[HTML]{EFEFEF}                                                                                                  &                                                                                      & x                    & - (baseline)                                                    & 0.8099$\pm$0.083	& 0.3320$\pm$0.036 &	0.4690$\pm$0.040	&  0.4840$\pm$0.020	&   \underline{0.8843$\pm$0.068}	   & 
 0.6253$\pm$0.032 & 0.6082$\pm$0.034                 					                                                       \\
\cellcolor[HTML]{EFEFEF}                                                                                                  &                                                                                      &                & A                                                    & 0.7713$\pm$0.078                                                & 0.3970$\pm$0.073                                             & 0.5170$\pm$0.051                                               & 0.4909$\pm$0.007                                                & 0.8192$\pm$0.093                                             & 0.6124$\pm$0.025                                               & 0.6081$\pm$0.014                       					                                                    \\
\cellcolor[HTML]{EFEFEF}                                                                                                  &                                                                                      & \multicolumn{1}{c}{\multirow{-2}{*}{x+x}} & \cellcolor[HTML]{EFEFEF}B                                                    & 0.7634$\pm$0.055                                                & 0.3980$\pm$0.015                                             & 0.5222$\pm$0.008                                               & 0.4917$\pm$0.011                                                & 0.8221$\pm$0.054                                             & 0.6151$\pm$0.023                                               & 0.6101$\pm$0.021   									                                                        \\
\cellcolor[HTML]{EFEFEF}                                                                                                  &                                                                                      &                                                                   & A                                                    & 0.7562$\pm$0.029                                                & \underline{0.4155$\pm$0.024}                                             & \underline{0.5358$\pm$0.022}                                               & \underline{0.4958$\pm$0.012}                                                & 0.8101$\pm$0.033                                             & 0.6150$\pm$0.017                                               & 0.6128$\pm$0.017                                 					                                          \\
\cellcolor[HTML]{EFEFEF}                                                                                                  &                                                                                      & \multicolumn{1}{c}{\multirow{-2}{*}{x+2x}}                                          & \cellcolor[HTML]{EFEFEF}B                            & \underline{0.8240$\pm$0.066}                                                & \textbf{0.4160$\pm$0.041}                                             & \textbf{0.5510$\pm$0.035}                                               & \textbf{0.5137$\pm$0.018}                                                & 0.8700$\pm$0.064                                             & \underline{0.6455$\pm$0.028}                                               & \textbf{0.6430$\pm$0.029}                        					                                                   \\
\cellcolor[HTML]{EFEFEF}                                                                                                  &                                                                                      &                                                                   & A                                                    & 0.7797$\pm$0.083                                                & 0.3841$\pm$0.036                                             & 0.5115$\pm$0.024                                               & 0.4905$\pm$0.013                                                & 0.8371$\pm$0.074                                             & 0.6180$\pm$0.028                                               & 0.6106$\pm$0.024                                                \\
\cellcolor[HTML]{EFEFEF}                                                                                                  & \multirow{-6}{*}{\begin{tabular}[c]{@{}l@{}}x=32 images\\  per video\end{tabular}} & \multirow{-2}{*}{x+3x}                                            & \cellcolor[HTML]{EFEFEF}B                                                    & \textbf{0.8464$\pm$0.090}                                                & 0.3805$\pm$0.043                                             & 0.5221$\pm$0.040                                               & 0.5060$\pm$0.020                                                & \textbf{0.8948$\pm$0.068}                                             & \textbf{0.6460$\pm$0.030}                                               & \underline{0.6377$\pm$0.031}                 					                                                          \\ \cline{2-11}

\cellcolor[HTML]{EFEFEF}                                                                                                  &                                                                                      & x                    & - (baseline)                                                    & \textbf{0.8889$\pm$0.035}	& 0.3698$\pm$0.022	& 0.5216$\pm$0.019  &	0.5124$\pm$0.005  &	\textbf{0.9338$\pm$0.025}	& \textbf{0.6615$\pm$0.007} & \underline{0.6516$\pm$0.008}                					                                                       \\
\cellcolor[HTML]{EFEFEF}                                                                                                  &                                                                                      &   & A                                                    & 0.7704$\pm$0.026                                                & \underline{0.4129$\pm$0.032}                                             & 0.5368$\pm$0.027                                               & 0.4995$\pm$0.011                                                & 0.8255$\pm$0.033                                             & 0.6221$\pm$0.014                                               & 0.6192$\pm$0.014                		                                                 \\
\cellcolor[HTML]{EFEFEF}                                                                                                  &                                                                                      & \multicolumn{1}{c}{\multirow{-2}{*}{x+x}} & \cellcolor[HTML]{EFEFEF}B                            & \underline{0.8463$\pm$0.062}                                                & \textbf{0.4405$\pm$0.021}                                             & \textbf{0.5780$\pm$0.011}                                               & \textbf{0.5277$\pm$0.010}                                                & \underline{0.8823$\pm$0.060}                                             & \underline{0.6601$\pm$0.024}                                               & \textbf{0.6614$\pm$0.022}       	                                             \\
\cellcolor[HTML]{EFEFEF}                                                                                                  &                                                                                      &                                                                   & A                                                    & 0.7964$\pm$0.031                                                & 0.4120$\pm$0.016                                             & 0.5426$\pm$0.011                                               & 0.5062$\pm$0.007                                                & 0.8501$\pm$0.034                                             & 0.6344$\pm$0.014                                               & 0.6310$\pm$0.013                                    					                                      \\
\cellcolor[HTML]{EFEFEF}                                                                                                  &                                                                                      & \multirow{-2}{*}{x+2x}                                            & \cellcolor[HTML]{EFEFEF}B                                                    & 0.7818$\pm$0.056                                                & 0.4033$\pm$0.013                                             & 0.5314$\pm$0.014                                               & 0.4988$\pm$0.013                                                & 0.8381$\pm$0.051                                             & 0.6252$\pm$0.024                                               & 0.6207$\pm$0.023  
					\\
\cellcolor[HTML]{EFEFEF}                                                                                                  &                                                                                      &                                                                   & A                                                    & 0.8348$\pm$0.080                                                & 0.4086$\pm$0.012                                             & \underline{0.5477$\pm$0.023}                                               & \underline{0.5132$\pm$0.021}                                                & 0.8810$\pm$0.069                                             & 0.6483$\pm$0.035                                               & 0.6448$\pm$0.036                                            \\

\multirow{-18}{*}{\cellcolor[HTML]{EFEFEF}NBI}                                                                            & \multirow{-6}{*}{\begin{tabular}[c]{@{}l@{}}x=64 images\\  per video\end{tabular}} & \multirow{-2}{*}{x+3x}                                            & \cellcolor[HTML]{EFEFEF}B                                                    & 0.7900$\pm$0.078                                                & 0.3922$\pm$0.028                                             & 0.5218$\pm$0.014                                               & 0.4959$\pm$0.013                                                & 0.8447$\pm$0.073                                             & 0.6245$\pm$0.029                                               & 0.6185$\pm$0.025                 					                                                         \\  \hline
\cellcolor[HTML]{EFEFEF}                                                                                                  &                                                                                      & x                    & - (baseline)                                                   & 0.7856$\pm$0.064 & 0.5689$\pm$0.084 &	0.6583$\pm$0.072 &	0.4237$\pm$0.068 &	0.6695$\pm$0.096 &	0.5176$\pm$0.074 & 0.6192$\pm$0.073                					                                                       \\
\cellcolor[HTML]{EFEFEF}                                                                                                  &                                                                                      &                       & A                                                    & 0.7869$\pm$0.057                                                & 0.6247$\pm$0.058                                             & 0.6937$\pm$0.031                                               & 0.4373$\pm$0.041                                                & 0.6268$\pm$0.129                                             & 0.5126$\pm$0.062                                               & 0.6257$\pm$0.051                          					                                                 \\
\cellcolor[HTML]{EFEFEF}                                                                                                  &                                                                                      & \multicolumn{1}{c}{\multirow{-2}{*}{x+x}} & \cellcolor[HTML]{EFEFEF}B                                                    & \underline{0.8563$\pm$0.039}                                                & 0.5296$\pm$0.051                                             & 0.6521$\pm$0.031                                               & 0.4444$\pm$0.010                                                & \textbf{0.8037$\pm$0.070}                                             & \underline{0.5714$\pm$0.015}                                               & \underline{0.6667$\pm$0.013}                         				                                                  \\ 
\cellcolor[HTML]{EFEFEF}                                                                                                  &                                                                                      &                                                                   & A                                                    & 0.7709$\pm$0.037                                                & \textbf{0.7551$\pm$0.049}                                             & \textbf{0.7616$\pm$0.028}                                               & \underline{0.4944$\pm$0.054}                                                & 0.5144$\pm$0.111                                             & 0.5013$\pm$0.073                                               & 0.6348$\pm$0.048                                				                                           \\
\cellcolor[HTML]{EFEFEF}                                                                                                  &                                                                                      & \multirow{-2}{*}{x+2x}                                            & \cellcolor[HTML]{EFEFEF}B                            & 0.8385$\pm$0.044                                                & 0.5635$\pm$0.113                                             & 0.6658$\pm$0.072                                                 & 0.4504$\pm$0.028                                              & 0.7534$\pm$0.119                                            & 0.5593$\pm$0.026                                                & 0.6585$\pm$0.020                         			                                                 \\
\cellcolor[HTML]{EFEFEF}                                                                                                  &                                                                                      &                                                                   & A                                                    & 0.7652$\pm$0.017                                                & \underline{0.7196$\pm$0.052}                                             & \underline{0.7407$\pm$0.028}                                               & 0.4687$\pm$0.029                                                & 0.5260$\pm$0.066                                             & 0.4937$\pm$0.035                                               & 0.6228$\pm$0.023                                	                                          \\ 
\cellcolor[HTML]{EFEFEF}                                                                                                  & \multirow{-6}{*}{\begin{tabular}[c]{@{}l@{}}x=16 images\\  per video\end{tabular}} & \multirow{-2}{*}{x+3x}                                            & \cellcolor[HTML]{EFEFEF}B                                                    & \textbf{0.8621$\pm$0.053}                                                & 0.6302$\pm$0.156                                             & 0.7154$\pm$0.099                                               & \textbf{0.5079$\pm$0.079}                                                & \underline{0.7665$\pm$0.130}                                             & \textbf{0.6010$\pm$0.040}                                               & \textbf{0.6983$\pm$0.039}                                                             \\ \cline{2-11}
\cellcolor[HTML]{EFEFEF}                                                                                                  &                                                                                      & x                    & - (baseline)                                                   & \underline{0.8460$\pm$0.033} &	0.6760$\pm$0.089 &	0.7475$\pm$0.046 &	0.5188$\pm$0.047 &	0.7291$\pm$0.089 &	\textbf{0.6018$\pm$0.019} & \textbf{0.7026$\pm$0.016}                					                                                       \\
\cellcolor[HTML]{EFEFEF}                                            &                                                                                      &                   & A                                                    & 0.7983$\pm$0.030                                                & \underline{0.7695$\pm$0.089}                                            & \underline{0.7801$\pm$0.034}                                               & \underline{0.5493$\pm$0.050}                                                & 0.5751$\pm$0.126                                             & 0.5514$\pm$0.046                                               & 0.6723$\pm$0.024                                           					          \\
\cellcolor[HTML]{EFEFEF}                                                                                                  &                                                                                      & \multicolumn{1}{c}{\multirow{-2}{*}{x+x}} & \cellcolor[HTML]{EFEFEF}B                                        & 0.8328$\pm$0.034 &	0.7009$\pm$0.023 &	0.7605$\pm$0.020 &	0.5199$\pm$0.017 &	0.6952$\pm$0.080 &	\underline{0.5939$\pm$0.039}                         & \underline{0.6981$\pm$0.032}                                                                                                           \\
\cellcolor[HTML]{EFEFEF}                                                                                                  &                                                                                      &                                                                   & A                                                    & 0.7868$\pm$0.024                                                & 0.7011$\pm$0.093                                             & 0.7395$\pm$0.059                                               & 0.4920$\pm$0.075                                                & 0.5957$\pm$0.051                                             & 0.5351$\pm$0.047                                               & 0.6484$\pm$0.044                                  	                                         \\
\cellcolor[HTML]{EFEFEF}                                                                                                  &                                                                                      & \multirow{-2}{*}{x+2x}                                            & \cellcolor[HTML]{EFEFEF}B                           & 0.8454$\pm$0.040                                                & 0.5983$\pm$0.092                                             & 0.6963$\pm$0.057                                               & 0.4728$\pm$0.037                                                & \underline{0.7591$\pm$0.097}                                             & 0.5793$\pm$0.032      &                                     0.6787$\pm$0.032                                                                      				                 \\
\cellcolor[HTML]{EFEFEF}                                                                                                  &                                                                                      &                                                                   & A                                                    & 0.7960$\pm$0.015                                                & \textbf{0.7799$\pm$0.089}                                             & \textbf{0.7857$\pm$0.045}                                               & \textbf{0.5628$\pm$0.079}                                                & 0.5714$\pm$0.068                                             & 0.5602$\pm$0.029                                               & 0.6756$\pm$0.025                                                         					                  \\
\cellcolor[HTML]{EFEFEF}                                                                                                  & \multirow{-6}{*}{\begin{tabular}[c]{@{}l@{}}x=32 images\\  per video\end{tabular}} & \multirow{-2}{*}{x+3x}                                            & \cellcolor[HTML]{EFEFEF}B                                                    & \textbf{0.8542$\pm$0.056}                                                & 0.5664$\pm$0.186                                             & 0.6626$\pm$0.113                                               & 0.4701$\pm$0.072                                                & \textbf{0.7675$\pm$0.157}                                             & 0.5696$\pm$0.020                                               & 0.6670$\pm$0.023   					                     \\ \cline{2-11}
\cellcolor[HTML]{EFEFEF}                                                                                                  &                                                                                      & x                    & - (baseline)                                                   & 0.8007$\pm$0.053 &	0.5690$\pm$0.149 &	0.6574$\pm$0.120 &	0.4432$\pm$0.084 &	0.7023$\pm$0.095 &	0.5389$\pm$0.073 & \underline{0.6357$\pm$0.075}                					                                                       \\
\cellcolor[HTML]{EFEFEF}                                                                                          &                   &                       & A                                                    & 0.7657$\pm$0.035                                                & \textbf{0.7225$\pm$0.066}                                             & \textbf{0.7408$\pm$0.019}                                              & \textbf{0.4634$\pm$0.018}                                                & 0.5171$\pm$0.139                                             & 0.4824$\pm$0.073                                               & 0.6198$\pm$0.038   
					\\
\cellcolor[HTML]{EFEFEF}                                                                                                  &                                                                                      & \multicolumn{1}{c}{\multirow{-2}{*}{x+x}} & \cellcolor[HTML]{EFEFEF}B                            & \underline{0.8445$\pm$0.029}                                                & 0.4436$\pm$0.062                                             & 0.5792$\pm$0.052                                               & 0.4096$\pm$0.019                                                & \textbf{0.8229$\pm$0.050}                                             & \textbf{0.5463$\pm$0.019}                                               & 0.6333$\pm$0.022    
					\\
\cellcolor[HTML]{EFEFEF}                                                                                                  &                                                                                      &                                                                   & A                                                    & 0.7767$\pm$0.025                                                & 0.6606$\pm$0.100                                             & 0.7107$\pm$0.060                                               & \underline{0.4570$\pm$0.054}                                                & 0.5928$\pm$0.083                                             & 0.5112$\pm$0.037                                               & 0.6267$\pm$0.035                                                         					                \\
\cellcolor[HTML]{EFEFEF}                                                                                                  &                                                                                      & \multirow{-2}{*}{x+2x}                                            & \cellcolor[HTML]{EFEFEF}B                                                    & 0.8183$\pm$0.062                                                & 0.4888$\pm$0.120                                             & 0.6016$\pm$0.086                                               & 0.4086$\pm$0.027                                                & \underline{0.7499$\pm$0.143}                                             & 0.5247$\pm$0.042                                               & 0.6194$\pm$0.036  
		\\
\multirow{-17}{*}{\cellcolor[HTML]{EFEFEF}WLI}                                                                             &                                                                                      &                                                                   & A                                                    & 0.7559$\pm$0.011                                                & \underline{0.6762$\pm$0.023}                                             & \underline{0.7135$\pm$0.012}                                               & 0.4343$\pm$0.013                                                & 0.5320$\pm$0.038                                             & 0.4778$\pm$0.020                                             & 0.6041$\pm$0.013                                 				                                          \\
\cellcolor[HTML]{EFEFEF}                                                                                                      &      \multirow{-6}{*}{\begin{tabular}[c]{@{}l@{}}x=64 images\\  per video\end{tabular}}                                                                                 & \multirow{-2}{*}{x+3x}                                            & \cellcolor[HTML]{EFEFEF}B                                                    & \textbf{0.8230$\pm$0.043}                                                & 0.5547$\pm$0.075                                             & 0.6589$\pm$0.044                                              & 0.4362$\pm$0.021                                                & 0.7361$\pm$0.113                                             & \underline{0.5451$\pm$0.042}                                               & \textbf{0.6454$\pm$0.033}                               				                                           
\end{tabular}}
\caption{Classification results using different proportions of real and synthetic images. Text prompt A and B stand for \textit{"colonoscopy image with `y' polyp, `z'"} and \textit{"colonoscopy image with `y' polyp, `z', good-quality, clear"}, respectively where `y' denote adenomatous/hyperplastic and `z' denote NBI/WLI. The values represent mean $\pm$ standard deviation. In `x+ix', the first term denotes the real frame count and the second term represents the synthetic image count. i is the ratio in which synthetic images are added.} 
\label{tab:framecomp}
\end{table}

\begin{table}[h!]
    \centering
    \begin{subtable}{0.4\linewidth}
        \centering
        \resizebox{0.8\linewidth}{!}{
       \begin{tabular}{cccc}
\multicolumn{4}{l}{\cellcolor[HTML]{C0C0C0}}              
                                            \\
\multicolumn{2}{l}{\multirow{-2}{*}{\cellcolor[HTML]{C0C0C0}\begin{tabular}[c]{@{}l@{}}Training sample count\\ (Real + Synthetic)\end{tabular}}}         & \multirow{-2}{*}{\cellcolor[HTML]{C0C0C0}Text Prompt} & \multirow{-2}{*}{\cellcolor[HTML]{C0C0C0}Balanced Accuracy} \\
     & x                     & -                                                     & 0.6333$\pm$0.126                                                             \\                               &                      & A                                                     & 0.6$\pm$0.137                                                                      \\                              & \multicolumn{1}{c}{\multirow{-2}{*}{x+x}} & \cellcolor[HTML]{EFEFEF}B                                                     & \textbf{0.7833$\pm$0.046}                                                             \\                      &                               & A                                                     & 0.6167$\pm$0.046                                                             \\                             & \multirow{-2}{*}{x+2x}                    & \cellcolor[HTML]{EFEFEF}B                                                     & 0.7167$\pm$0.046                                                             \\
 &                                                                   & A                                                     & 0.6$\pm$0.109                                                                      \\    \multirow{-6}{*}{\begin{tabular}[c]{@{}l@{}}x=16 images\\  per video\end{tabular}} & \multirow{-2}{*}{x+3x}                                            & \cellcolor[HTML]{EFEFEF}B                                                     & 0.75$\pm$0.083                                                                     \\ \hline
 & x                     & -                                                     & \textbf{0.7833$\pm$0.046}                                                             \\
  &                      & A                                                     & 0.7167$\pm$0.119                                                             \\
   & \multicolumn{1}{c}{\multirow{-2}{*}{x+x}} & \cellcolor[HTML]{EFEFEF}B                                                     & \textbf{0.7833$\pm$0.047}                                                             \\
  &   & A                                                     & 0.6333$\pm$0.095                                                             \\
 & \multirow{-2}{*}{x+2x}                    & \cellcolor[HTML]{EFEFEF}B                                                     & 0.7333$\pm$0.037                                                             \\
       &                                                                   & A                                                     & 0.6333$\pm$0.046                                                             \\
\multirow{-6}{*}{\begin{tabular}[c]{@{}l@{}}x=32 images\\  per video\end{tabular}} & \multirow{-2}{*}{x+3x}                                            & \cellcolor[HTML]{EFEFEF}B                                                     & 0.7167$\pm$0.046                                                             \\ \hline                                                                                    & x                     & -                                                     & 0.6833$\pm$0.109                                                             \\                              &                       & A                                                     & 0.6$\pm$0.070                                                                      \\                                & \multicolumn{1}{c}{\multirow{-2}{*}{x+x}} & \cellcolor[HTML]{EFEFEF}B                                                     & \textbf{0.7167$\pm$0.075}                                                             \\                             &  & A                                                     & 0.5667$\pm$0.037                                                             \\                                & \multirow{-2}{*}{x+2x}                    & \cellcolor[HTML]{EFEFEF}B                                                     & 0.6167$\pm$0.095                                                             \\                                                                                                 &                                                                   & A                                                     & 0.5333$\pm$0.046                                                             \\                             \multirow{-6}{*}{\begin{tabular}[c]{@{}l@{}}x=64 images\\  per video\end{tabular}} & \multirow{-2}{*}{x+3x}                                            & \cellcolor[HTML]{EFEFEF}B                                                     & 0.6333$\pm$0.095 \\
            \bottomrule
        \end{tabular}}
        \caption{}\label{tab:patient_wli}
    \end{subtable}
    \hfill 
    \begin{subtable}{0.4\linewidth}
        \centering
         \resizebox{0.8\linewidth}{!}{
       \begin{tabular}{cccc}
\multicolumn{4}{l}{\cellcolor[HTML]{C0C0C0}}                                                \\
\multicolumn{2}{l}{\multirow{-2}{*}{\cellcolor[HTML]{C0C0C0}\begin{tabular}[c]{@{}l@{}}Training sample count\\ (Real + Synthetic)\end{tabular}}}         & \multirow{-2}{*}{\cellcolor[HTML]{C0C0C0}Text Prompt} & \multirow{-2}{*}{\cellcolor[HTML]{C0C0C0}Balanced Accuracy} \\                                                                                     & x                     & -                                                     & 0.6667                                                             \\
 &              & A                                                     & 0.7167$\pm$0.095                                                             \\
      & \multicolumn{1}{c}{\multirow{-2}{*}{x+x}} & \cellcolor[HTML]{EFEFEF}B                                                     & 0.75$\pm$0.059                                                                     \\                        &                                                                   & A                                                     & 0.6833$\pm$0.037                                                             \\
     & \multirow{-2}{*}{x+2x}                                            & \cellcolor[HTML]{EFEFEF}B                                                     & 0.7667$\pm$0.037                                                            \\
 &  & A                                                     & \textbf{0.7833$\pm$0.046}                                                             \\
\multirow{-6}{*}{\begin{tabular}[c]{@{}l@{}}x=16 images\\  per video\end{tabular}} & \multirow{-2}{*}{x+3x}                    & \cellcolor[HTML]{EFEFEF}B                                                     & 0.75                                              \\ \hline
  & x                     & -                                                     & 0.7333$\pm$0.070                                                          \\
  &                                                                           & A                                                     & 0.7333$\pm$0.070                                                             \\
 & \multicolumn{1}{c}{\multirow{-2}{*}{x+x}} & \cellcolor[HTML]{EFEFEF}B                                                     & 0.7667$\pm$0.037                                                             \\
  &  & A                                                     & \textbf{0.8$\pm$0.046}                                                                      \\
 & \multirow{-2}{*}{x+2x}                    & \cellcolor[HTML]{EFEFEF}B                                                     & 0.7833$\pm$0.046                                                             \\
   &   & A                                                     & 0.7667$\pm$0.037                                                             \\
\multirow{-6}{*}{\begin{tabular}[c]{@{}l@{}}x=32 images\\  per video\end{tabular}} & \multirow{-2}{*}{x+3x}                    & \cellcolor[HTML]{EFEFEF}B                                                     & 0.7833$\pm$0.046                                                             \\ \hline
 & x                     & -                                                     & \textbf{0.8167$\pm$0.037}                                                             \\
 &  & A                                                     & \textbf{0.8167$\pm$0.037}                                                             \\
     & \multicolumn{1}{c}{\multirow{-2}{*}{x+x}} & \cellcolor[HTML]{EFEFEF}B                                                     & 0.8$\pm$0.046                                                                      \\
   
 &                                          & A                                                     & \textbf{0.8167$\pm$0.037}                                                             \\
  & \multirow{-2}{*}{x+2x}                    & \cellcolor[HTML]{EFEFEF}B                                                     & 0.7667$\pm$0.037                                                             \\
    &                                        & A                                                     & \textbf{0.8167$\pm$0.037}                                                             \\
\multirow{-6}{*}{\begin{tabular}[c]{@{}l@{}}x=64 images\\  per video\end{tabular}} & \multirow{-2}{*}{x+3x}                    & \cellcolor[HTML]{EFEFEF}B                                                     & 0.7667$\pm$0.037  \\   
            \bottomrule
        \end{tabular}}
         \caption{}\label{tab:patient_nbi}
    \end{subtable}
    \caption{Video-wise results using (a) WLI modality and (b) NBI modality.}
    \label{tab:side_by_side_tables}
\end{table}

\begin{figure}[h!]
    \centering
   \begin{tabular}{cc}
     \subfloat[\small p-value obtained using two-tailed t-test for statistical significance analysis. The values are rounded off to 3 decimal places. The label names used in rows and columns can be read as ``Text Prompt\_Sample Count per Video (Real Image Count + Synthetic Image Count)". \label{fig:pvalue_wli}]{%
       \includegraphics[width=0.61\textwidth]{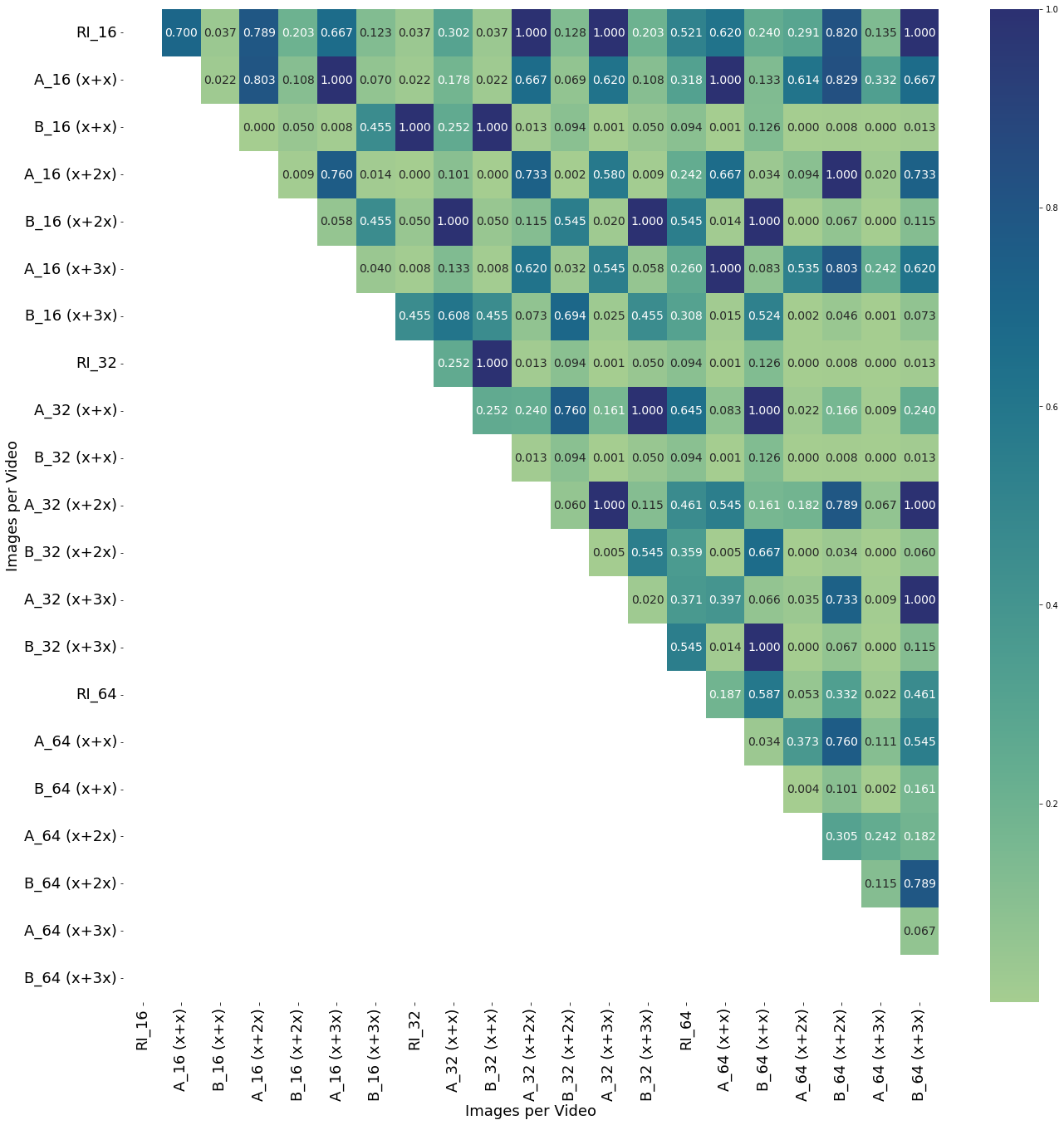}
        } &  \subfloat[\small Sample heatmaps for both real and augmented data. \label{fig:heatmap_wli}]{%
       \includegraphics[width=0.25\textwidth]{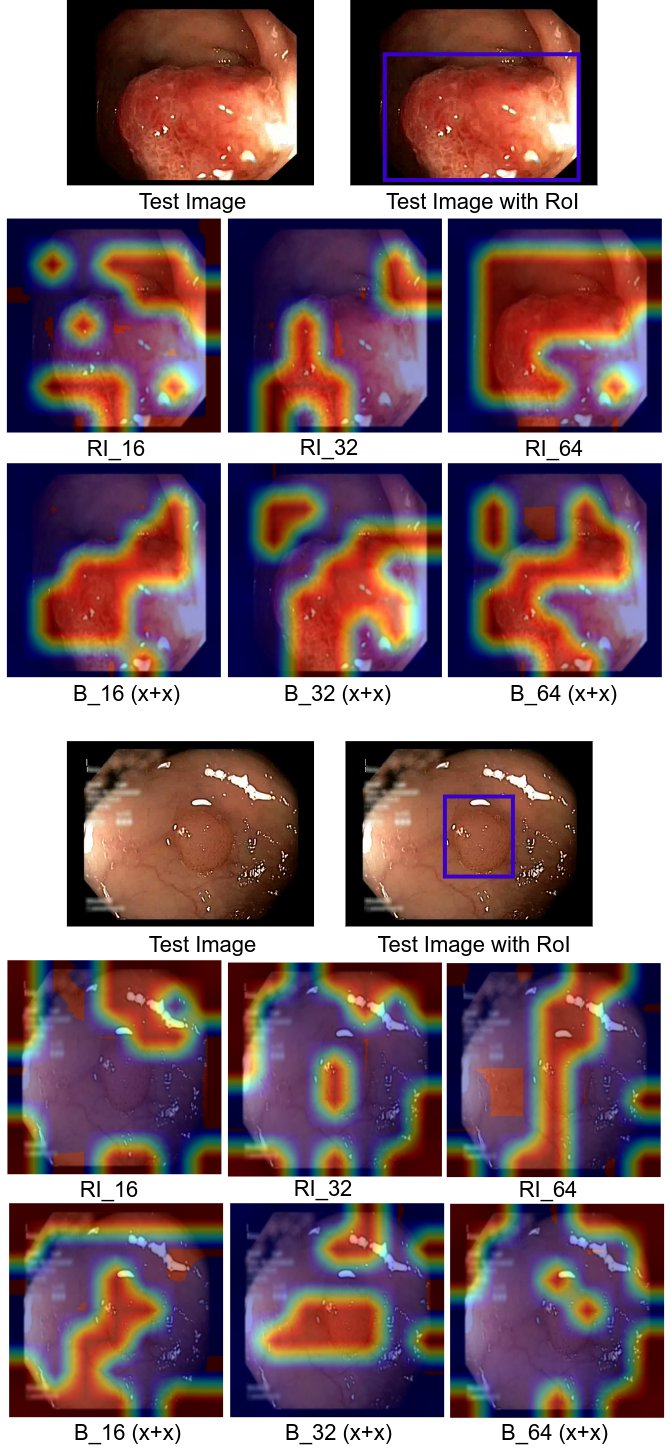}
        }

   \end{tabular}
    \caption{Video-wise outcomes obtained using WLI images and the associated p-values and heatmaps. RI stands for real images.}
    \label{fig:patient_wli}
\end{figure}

\begin{figure}[h!]
    \centering
   \begin{tabular}{cc}
     \subfloat[\small p-value obtained using two-tailed t-test for statistical significance analysis. The values are rounded off to 3 decimal places. The label names used in rows and columns can be read as ``Text Prompt\_Sample Count per Video (Real Image Count + Synthetic Image Count)". \label{fig:pvalue_nbi}]{%
       \includegraphics[width=0.61\textwidth]{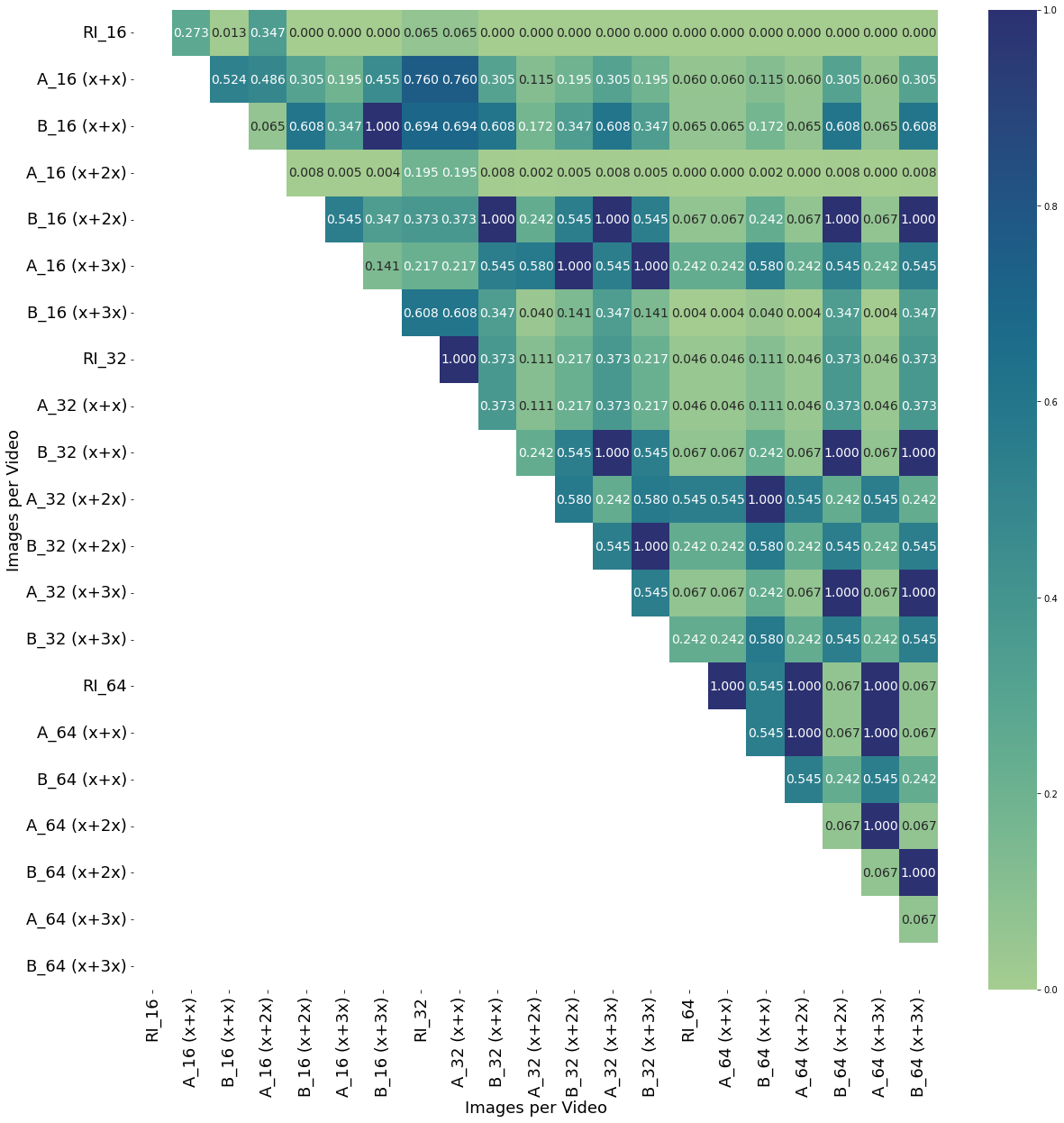}
        }
        & \subfloat[\small Sample heatmaps for both real and augmented data.  \label{fig:heatmap_nbi}]{%
       \includegraphics[width=0.25\textwidth]{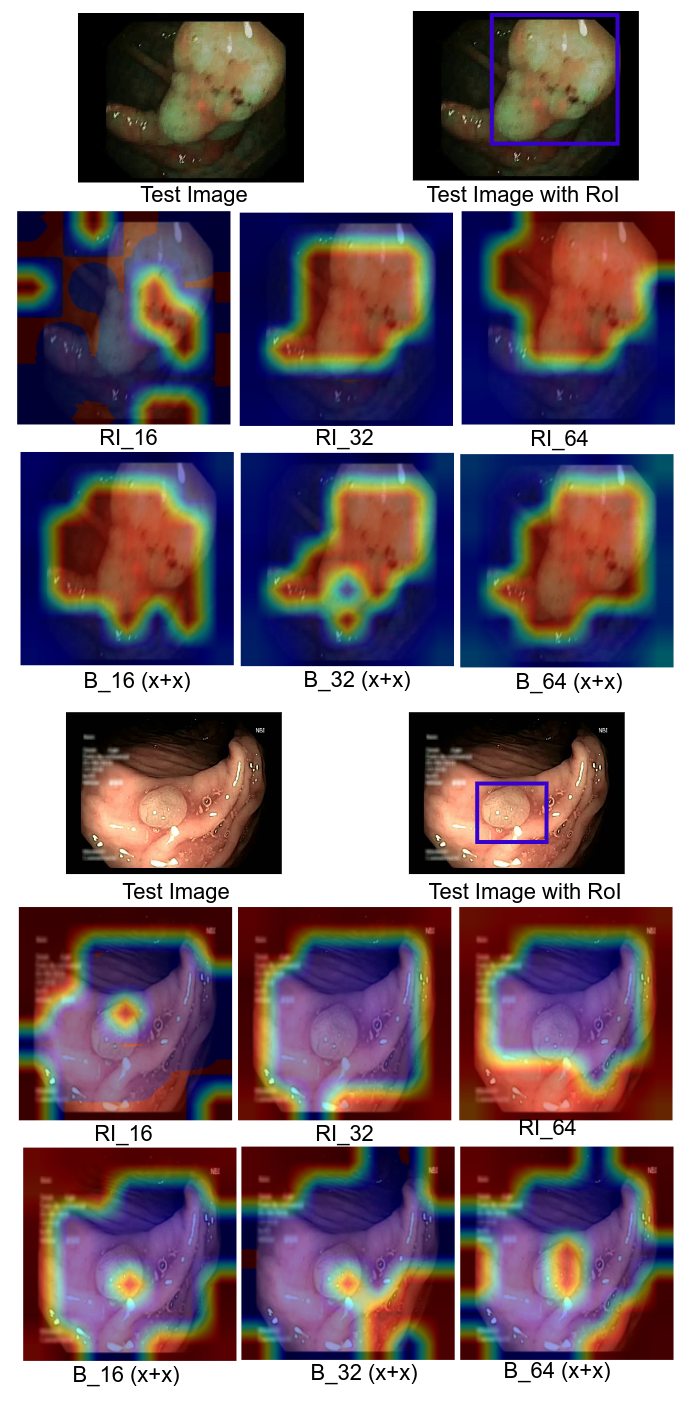}
        } \\
  
   \end{tabular}
    \caption{Video-wise outcomes obtained using NBI images and the associated p-values and heatmaps. RI stands for real images.}
    \label{fig:patient_nbi}
\end{figure}

\begin{figure*}[h!]
     \centering
     \subfloat[\small \label{fig:ad_gd_wli1}]{%
       \includegraphics[width=0.15\textwidth]{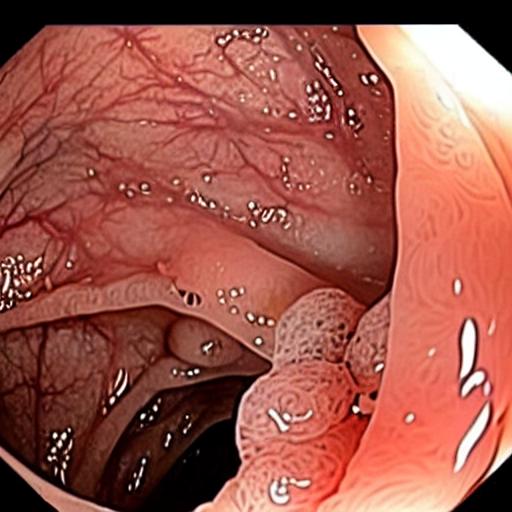}
        }  
      \hfill
     \subfloat[\small \label{fig:ad_gd_wli2}]{%
       \includegraphics[width=0.15\textwidth]{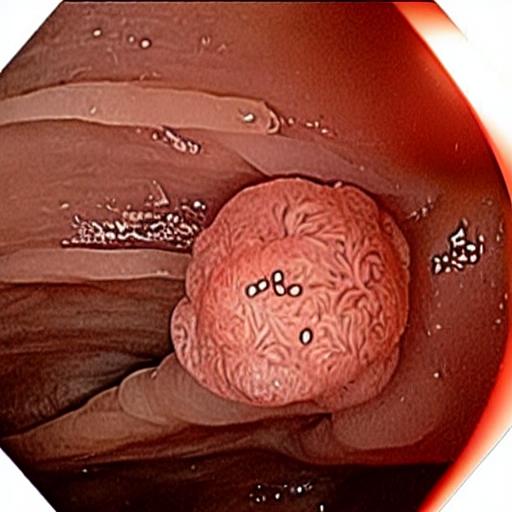}
        }
     \hfill
     \subfloat[\small \label{fig:hp_gd_wli3}]{%
       \includegraphics[width=0.15\textwidth]{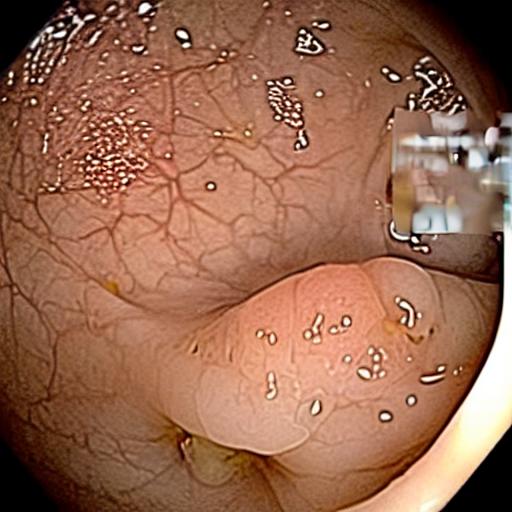}
        }
      \hfill
      \subfloat[\small \label{fig:hp_gd_wli4}]{%
       \includegraphics[width=0.15\textwidth]{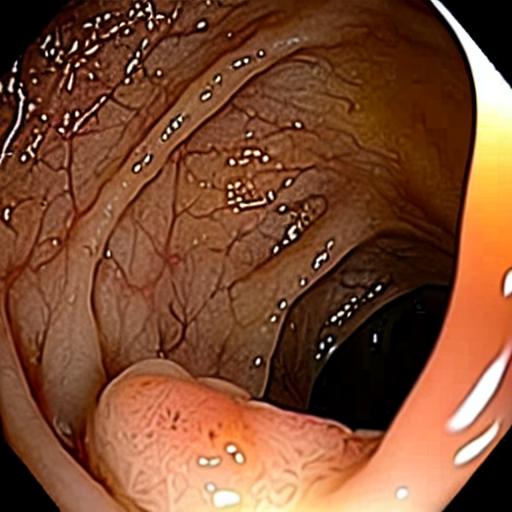}
        }
     \hfill
     \subfloat[\small \label{fig:ad_gd_nbi1}]{%
       \includegraphics[width=0.15\textwidth]{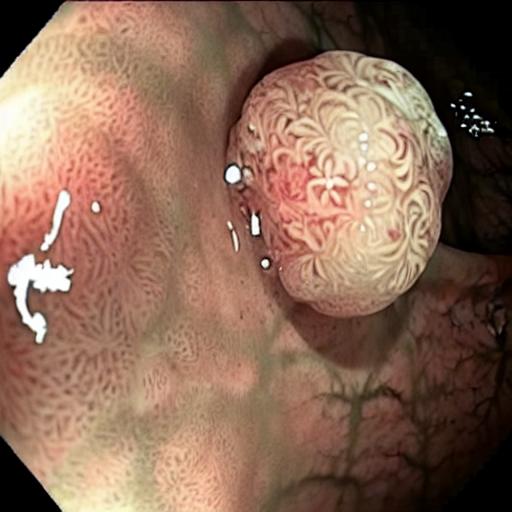}
        }  
      \hfill
     \subfloat[\small \label{fig:ad_gd_nbi2}]{%
       \includegraphics[width=0.15\textwidth]{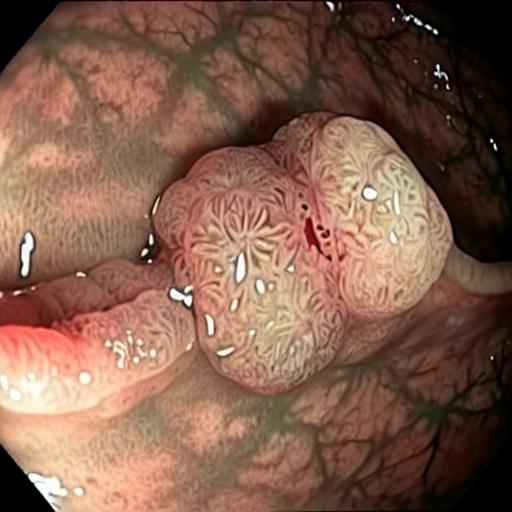}
        }
     \hfill
     \subfloat[\small \label{fig:hp_gd_nbi1}]{%
       \includegraphics[width=0.15\textwidth]{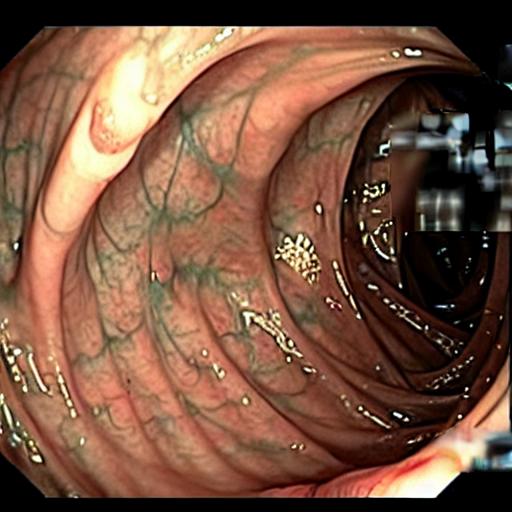}
        }
      \hfill
      \subfloat[\small \label{fig:hp_gd_nbi2}]{%
       \includegraphics[width=0.15\textwidth]{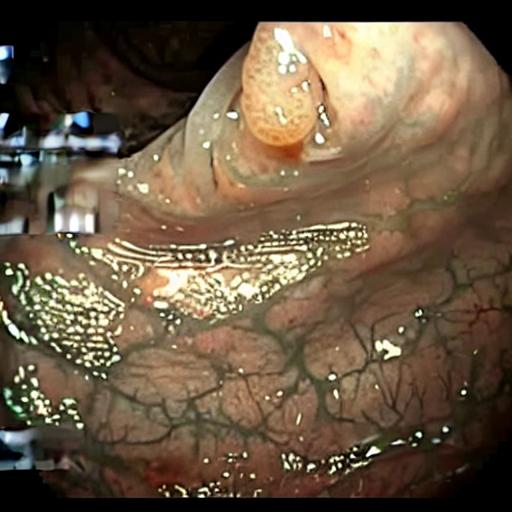}
        }
     \hfill
     \subfloat[\small \label{fig:ad_wli1}]{%
       \includegraphics[width=0.15\textwidth]{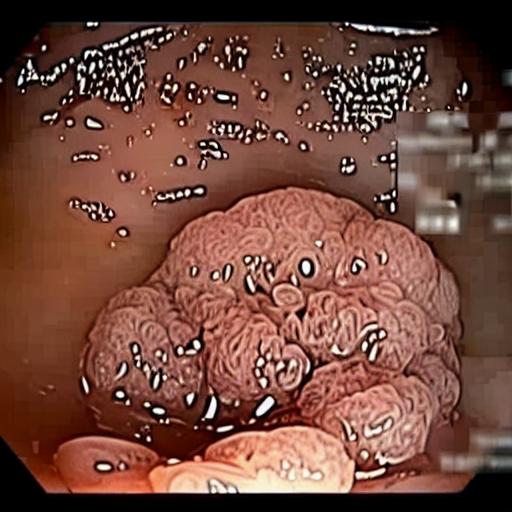}
        }  
      \hfill
     \subfloat[\small \label{fig:ad_wli2}]{%
       \includegraphics[width=0.15\textwidth]{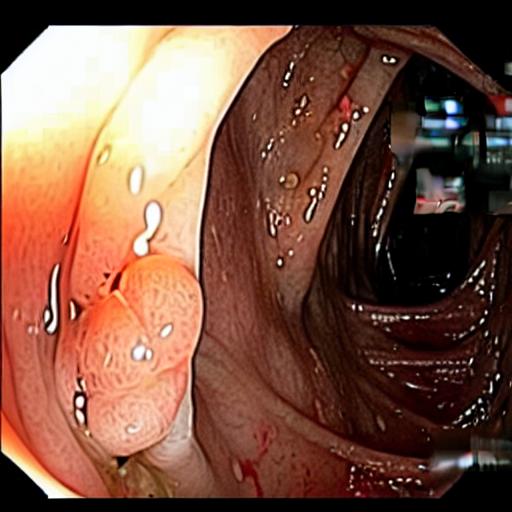}
        }
     \hfill
     \subfloat[\small \label{fig:hp_wli1}]{%
       \includegraphics[width=0.15\textwidth]{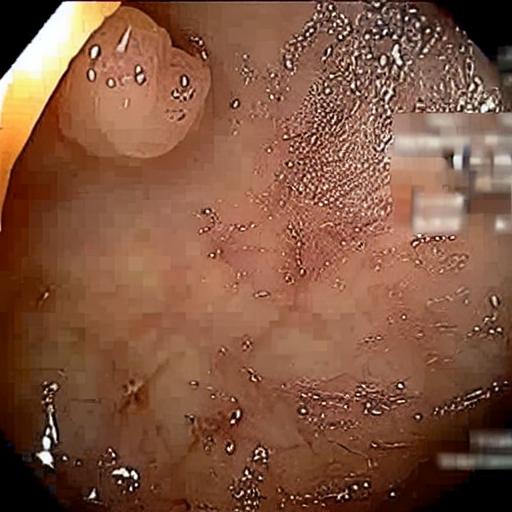}
        }
      \hfill
      \subfloat[\small \label{fig:hp_wli2}]{%
       \includegraphics[width=0.15\textwidth]{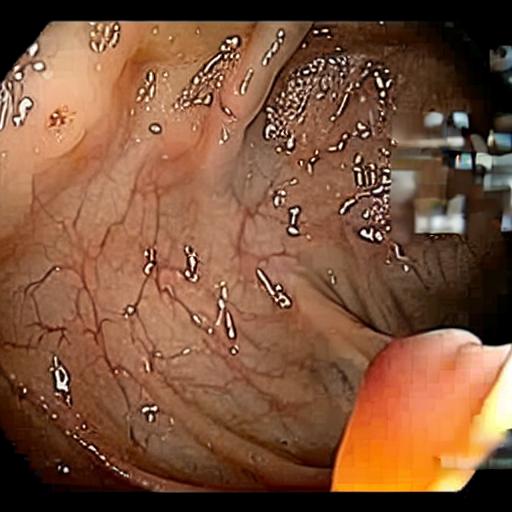}
        }
     \hfill
      \subfloat[\small \label{fig:ad_nbi1}]{%
       \includegraphics[width=0.15\textwidth]{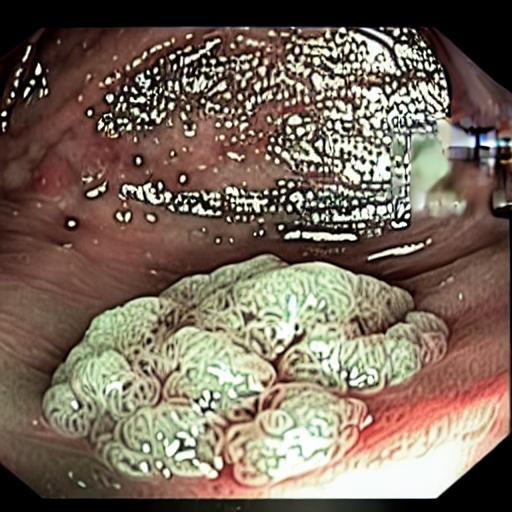}
        }  
      \hfill
     \subfloat[\small \label{fig:ad_nbi2}]{%
       \includegraphics[width=0.15\textwidth]{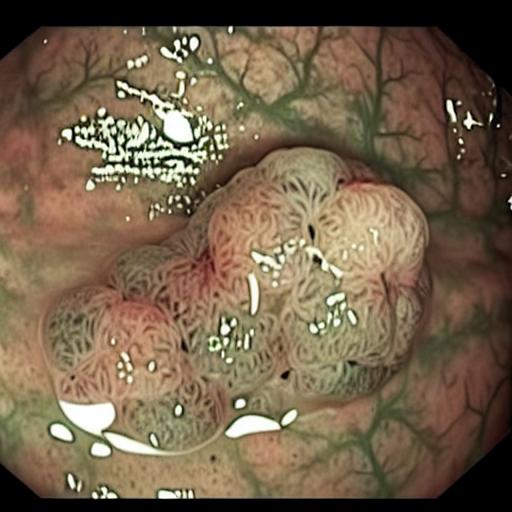}
        }
     \hfill
     \subfloat[\small \label{fig:hp_nbi1}]{%
       \includegraphics[width=0.15\textwidth]{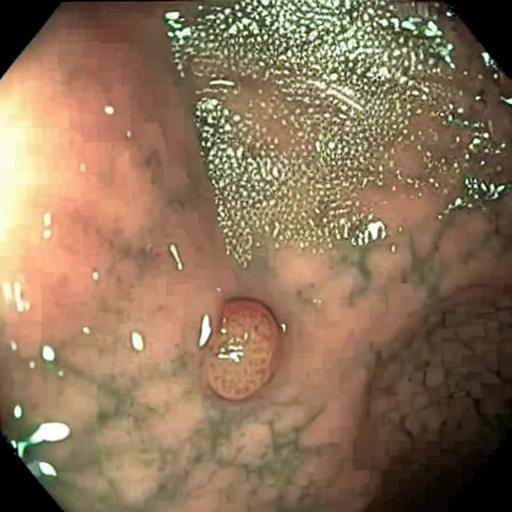}
        }
      \hfill
      \subfloat[\small \label{fig:hp_nbi2}]{%
       \includegraphics[width=0.15\textwidth]{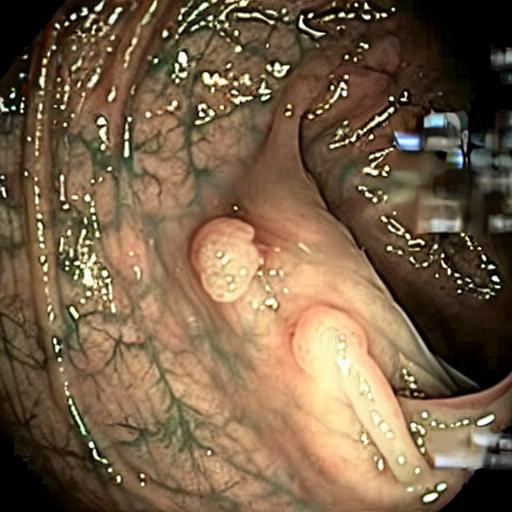}
        }
     \hfill
     \subfloat[\small \label{fig:nbi_gd1}]{%
       \includegraphics[width=0.15\textwidth]{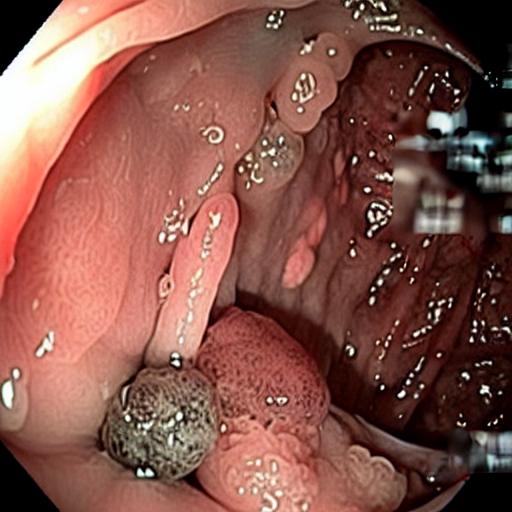}
        }  
      \hfill
     \subfloat[\small \label{fig:nbi_gd2}]{%
       \includegraphics[width=0.15\textwidth]{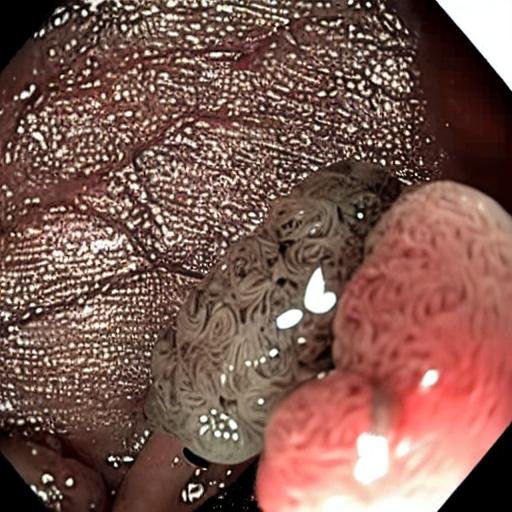}
        }
     %\hfill
     %\subfloat[\small \label{fig:nbi_gd3}]{%
     %  \includegraphics[width=0.15\textwidth]{pic/nbi_gd391.jpg}
     %   }
     % \hfill
     % \subfloat[\small \label{fig:nbi_gd4}]{%
     %  \includegraphics[width=0.15\textwidth]{pic/nbi_gd41.jpg}
     %   }
     
      %\subfloat[\small Iteration-10K\label{fig:y equals x}]{%
      % \includegraphics[width=0.32\textwidth]{pic/t-SNE_plot_10k.png}
      %  }
      %  \hfill
      % \subfloat{%
      % \includegraphics[width=0.25\textwidth]%{pic/legend1.png}
       %} 
        \caption{\small Sample generated images depicting (a)-(b) adenomatous polyp in WLI (using text prompt B) (c)-(d) hyperplastic polyp in WLI (using text prompt B), (e)-(f) adenomatous polyp in NBI (using text prompt B), (g)-(h) hyperplastic polyp in NBI (using text prompt B), (i)-(j) adenomatous polyp in WLI (using text prompt A), (k)-(l) hyperplastic polyp in WLI (using text prompt A), (m)-(n) adenomatous polyp in NBI (using text prompt A), (o)-(p) hyperplastic polyp in NBI (using text prompt A), and (q)-(r) shows some undesired images generated using text prompt B with NBI. }
        \label{fig:gensamples}
\end{figure*}

\subsubsection*{Video-wise Analysis with Statistical Significance Test}
\label{sec:patient-wise}
A comparative study has been conducted on patient-wise results. Although the training experiments are performed on the frame level, the inference is computed on both the frame and video levels. A majority voting scheme has been adopted for such computations. For instance, a video is labelled as the class `adenomatous' if the majority of frames (> mean number of total video frames) are predicted as adenomatous. These results are provided in the Table \ref{tab:patient_wli} and Table \ref{tab:patient_nbi}. Additionally, a statistical significance test is conducted using a two-tailed t-test, which indicates the significance of an increase or decrease in video-wise outcomes and reports if this change is insignificant. This test calculates the p-values between all possible combinations of the data proportions used in our work. The associated p-values are depicted in Fig. \ref{fig:pvalue_wli} and Fig. \ref{fig:pvalue_nbi}. Using the WLI modality, we observed that the best outcomes are obtained using 32 real images per video and 16 or 32 real images per video combined with synthetic images generated from text prompt B in equal proportion. The last two cases align with the second-best results in frame-level evaluation presented in Table \ref{tab:framecomp}. Moreover, considering the case with 16 real images per video and an equal proportion of synthetic samples, video-level analysis reports a significant improvement of 15\% (p-value = 0.037). Similar to frame-level analysis, the video-level validations signify that the performance improvement with synthetic data is reduced by increasing the real sample count. Although with 64 real images per video, an increase of 3.34\% is observed, this increase is not statistically significant (p-value = 0.587). We further examined the performance difference between the two text prompts, A and B, to evaluate the video-level cross-class label learning ability. Our analysis demonstrates that synthetic images generated using text prompt B are superior and statistically significant to those generated using text prompt A. This observation is supported by some of the notable improvements that include 0.6 to 0.7833 (difference of +18.33\%, 95\% CI: 3.43\%, 33.23\%, p-value = 0.022) using 16 real samples per video with an equal number of synthetic samples, 0.6 to 0.7167 (difference of +11.67\%, 95\% CI: 1.09\%, 22.25\%, p-value = 0.034) using 64 real samples per video with an equal number of synthetic samples, and 0.6333 to 0.7167 (difference of +8.34\%, 95\% CI: 1.63\%, 15.05\%, p-value = 0.020) using 32 real samples per video combined with three times as many synthetic samples. 

Using the NBI modality, the overall performance trend with data proportion is similar. Also, the comparison between the real data and a combination of real and synthetic data shows a similar trend as observed using the WLI modality. However, in the NBI modality, the synthetic data generated using text prompt A presented better outcomes than text prompt B in some cases. Although the best results are obtained using text prompt A, the average performance over all data proportions is the same for both text prompts. Moreover, it is noteworthy that the difference is not statistically significant in almost every case. Some such examples include the improvement from 0.7167 to 0.75 (difference of +3.33\%, 95\% CI: -8.2028\% to 14.86\%, p-value = 0.524) using 16 real images per video and an equal number of synthetic images, 0.7833 to 0.80 (difference of +1.67\%, 95\% CI: -5.04\%, 8.38\%, p-value = 0.580) using 32 real images per video and twice as many synthetic images, and 0.7667 to 0.8167 (difference of +5\%, 95\% CI: -0.40\%, 10.40\%, p-value = 0.067) using 64 real images per video and twice/thrice as many synthetic images. Despite such change in trend in video-wise analysis, it can be observed from Table \ref{tab:framecomp} that on a frame-wise level, most of the cases favored text prompt B over text prompt A. This inconsistent shift can be due to the fact that control over diffusion models is limited and also depends on the seed value. The original dataset used in Stage-I with annotations based on the quality (good-quality, clear/ low-quality) comprises WLI images, whereas the  WLI and NBI images used in Stage-II lack such annotations. It was a relatively simple task for the model to generate a combination of WLI and good-quality data. On the contrary, the constrained control over the generated data occasionally resulted in blending WLI characteristics into some NBI images. This inconsistency emerged from the association between WLI and quality learned during Stage-I training. Random seed initialization and limited control over diffusion models resulted in some arbitrary outcomes with text prompt B in the case of NBI images. This justification is supported by the qualitative outcomes (shown in Fig. \ref{fig:nbi_gd1}-\ref{fig:nbi_gd2}) discussed in the next section.          

\subsubsection*{Qualitative Results and Interpretability through Visualization}
In addition to the quantitative analysis, we examined the qualitative results and further studied the related heatmaps for visualization and interpretability. The heatmaps were obtained using GradCAM on EfficientNet-B0. After selecting a target layer, heatmaps were generated by attributing gradients to that layer, highlighting the most influential regions in the input images. The outcomes were then overlaid on the original images. The heatmaps pertaining to the scenarios involving real images and also those related to some cases involving synthetic images are shown in Fig. \ref{fig:heatmap_wli} and Fig. \ref{fig:heatmap_nbi}. These heatmaps illustrate the region the classification model focuses on before providing the final prediction scores. It can be observed that the classifier learns the complex polyp-specific features better when trained using a more diverse set of polyp images obtained using PathoPolyp-Diff. However, the classifier's performance drops in identifying polyp features when the count of synthetic images increases. This decline in performance could be because synthetic images might carry noise and can not exactly replicate real image characteristics; thus, added noise could deviate the model after a certain limit.

Further, we analyzed the generated images for qualitative analysis. It can be observed that the synthetic images obtained using text prompt B are visually more appealing than those produced with text prompt A. The texture is more prominent and clear in images shown in Fig. \ref{fig:ad_gd_wli1} to Fig. \ref{fig:hp_gd_nbi2} compared to those presented in Fig. \ref{fig:ad_wli1} to Fig. \ref{fig:hp_nbi2}. Moreover, in both scenarios, qualitatively, the generated images are close to real images in terms of structure, color and texture. The fundamental color criteria that differentiate NBI and WLI remain consistently evident in the images, making them easily distinguishable. However, as already discussed, text prompt B with NBI images fails for some samples, as can be inferred from Fig. \ref{fig:nbi_gd1} and Fig. \ref{fig:nbi_gd2}. This failure is attributed to the limited control over the image generation and dependency on seed initialization. Note that the pathology-focused data used in Stage-II has both WLI and NBI images, but none of the images have annotations based on quality. Still, the model learnt the relation between quality and WLI easily because of the relation developed earlier in Stage-I. Therefore, in Stage-II, our model more precisely established the relation between WLI, quality, and pathology. Due to the development of this direct relationship, the model sometimes recollects ``WLI'' characteristics when it assigns more weightage to the tokens \textit{"good-quality, clear"} in the given text prompt \textit{"colonoscopy image with adenomatous polyp, narrow band imaging, good-quality, clear"} or \textit{"colonoscopy image with hyperplastic polyp, narrow band imaging, good-quality, clear"}. Consequently, some samples present a combination of WLI and NBI images or complete WLI images.  

To overcome this issue, we assigned more weight to the tokens representing \textit{"colonoscopy image with `y' polyp, narrow band imaging"} where `y' can be hyperplastic/adenomatous. This modification resulted in better qualitative outcomes, as depicted in Fig. \ref{fig:ad_plus1} to Fig. \ref{fig:hp_plus2}. However, the related quantitative outcomes are reduced in most of the cases covered by Fig. \ref{fig:nbi_gd_1x} to Fig. \ref{fig:nbi_gd_3x}. One possible reason for such a decline in results could be due to a lack of pathology-specific features when images are generated using a weighted approach. Such outcomes signify that slight modifications in the text prompt could impact synthetic images' characteristics and visual properties.

\begin{figure*}[t]
     \subfloat[\small \label{fig:ad_plus1}]{%
       \includegraphics[width=0.24\textwidth]{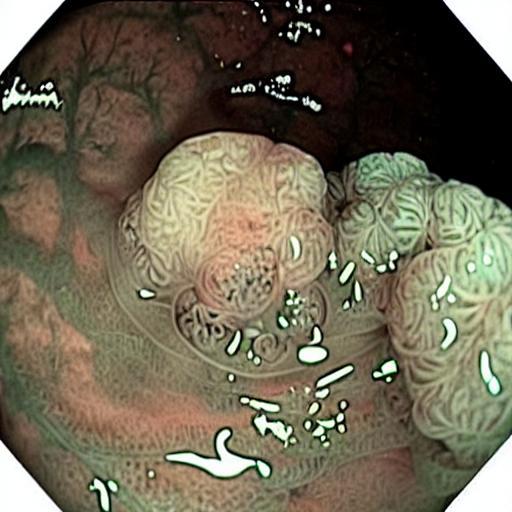}
        }  
      \hfill
     \subfloat[\small \label{fig:ad_plus2}]{%
       \includegraphics[width=0.24\textwidth]{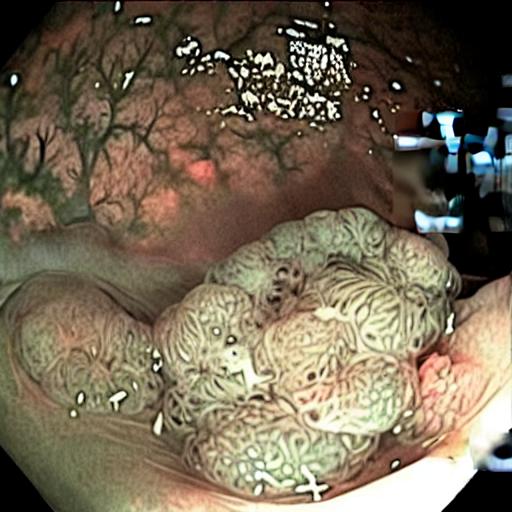}
        }
     \hfill
     \subfloat[\small \label{fig:hp_plus1}]{%
       \includegraphics[width=0.24\textwidth]{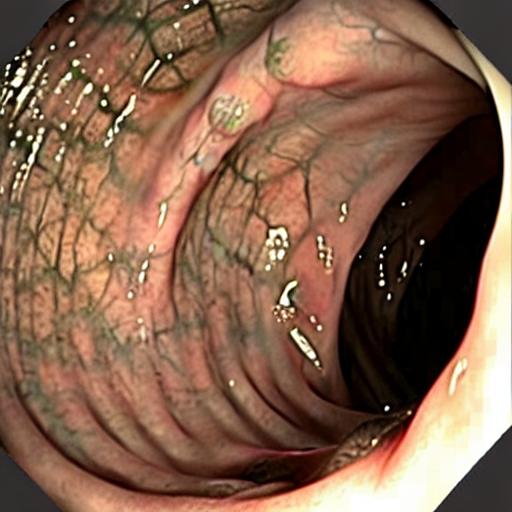}
        }
      \hfill
      \subfloat[\small \label{fig:hp_plus2}]{%
       \includegraphics[width=0.24\textwidth]{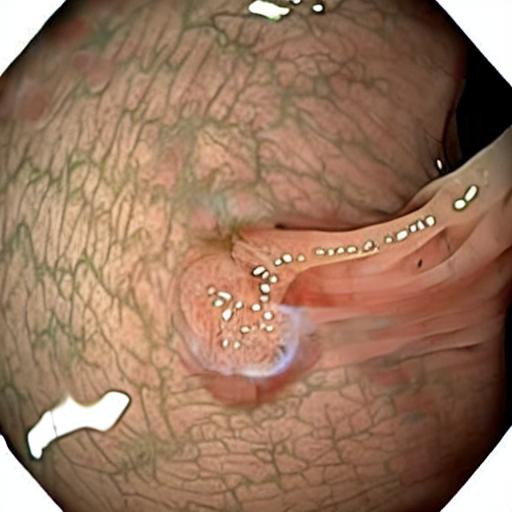}
        }  
        \hfill
      \subfloat[\small \label{fig:nbi_gd_1x}]{%
       \includegraphics[width=0.24\textwidth]{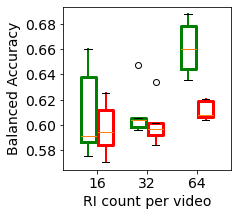}
        }  
        \hfill
      \subfloat[\small \label{fig:nbi_gd_2x}]{%
       \includegraphics[width=0.24\textwidth]{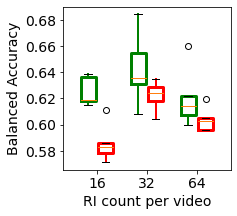}
        }  
        \hfill
      \subfloat[\small \label{fig:nbi_gd_3x}]{%
       \includegraphics[width=0.24\textwidth]{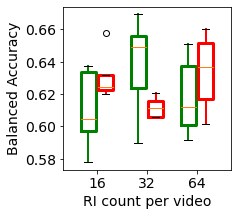}
        }  
        \caption{\small Sample images depicting (a)-(b) adenomatous polyps in NBI, and (c)-(d) hyperplastic polyps in NBI using weighted control mechanism. The boxplots (e) to (g) demonstrate the comparison between text prompt B with and without a weighted control mechanism when synthetic images are added in equal proportion or twice or thrice in proportion to the real images, respectively. The former and latter text prompts are denoted by green and red color, respectively. RI stands for Real Images. }
        \label{fig:plus_samples}
\end{figure*}

%\subsubsection*{Mode-II}
%During \textit{Mode-II}, the pre-trained weights of the \textit{Mode-I} model is used to transfer   

%\begin{equation}
%    y_c = \mathcal{H}(x,\alpha) + \mathcal{C}(\mathcal{H}(x + \mathcal{C}(c,\alpha_{c1});\alpha_c);\alpha_{c1})
%\end{equation}
%where $y_c$ is the output and $x$ is th
%e input feature map.
%The overall objective after including the downstream task can be modified as shown below:
%\vspace{-0.2cm}
%\begin{equation}
%    L_{CN} := \mathbb{E}_{E(a),b,b',\epsilon,t}[{\lVert \epsilon-\epsilon_\theta({a_l}_t,t, \mathcal{Z}_\theta(b), \mathcal{Z}_\theta(b'))\rVert}^2_2]
%\end{equation}
%where $\mathcal{Z}_\theta(b')$ is the intermediate representation of the task-specific conditional text prompt. 

%\subsection*{Training Algorithm and Protocols}

%\subsubsection{}

% \usepackage{multirow}

% \usepackage{multirow}

% \usepackage{multirow}

\section*{Discussion and Limitations}
 In this work, our proposed text-controlled polyp generation approach presented a novel technique to generate polyps with different pathologies, which is an underexplored research area in the literature. An exhaustive investigation is conducted to study the training process of diffusion models using iteration-wise t-SNE plots, confusion matrices and quality assessment metrics.  Considering the visual outcomes of t-SNE plots and the quantitative outcomes of the classification test, it can be inferred that during the training phase, our diffusion based model gradually performs well with increasing iterations; however, it starts to degrade after a certain extent. Such a trend indicates that over-training of a diffusion model could result in synthetic images with unacceptable characteristics.

Following this assessment, a standard classification test is conducted utilizing pre-trained (on real images) models, including DenseNet-201 and DenseNet-121. This aims to assess the clinical relevance of synthetic images obtained in different iterations by validating if a model trained using real images could identify the synthetic images during the inference phase. These results, along with the outcomes of KID and t-SNE plots assisted in selecting the best training iteration of our proposed diffusion model. In Stage-I, the F1-score improved with every iteration, contradicting both the qualitative results and the KID score. The t-SNE plots suggested that such a scenario occurred because, despite unsatisfactory image generation, the separation between the two classes (polyp and non-polyp) was sufficient to achieve a significantly high F1-score. However, the lack of overlap with their real counterparts resulted in unsatisfactory KID scores at higher iterations. Therefore, our study reveals that the clinical relevance of synthetic images needs to be evaluated using different standard metrics. A similar approach was followed in Stage-II, yielding more convincing outcomes with fewer contradictions. However, a common observation in both stages was that the synthetic image data points began to deviate from the real data points after certain training iterations.  
 
We further validated the pathological content of the synthetic images by using the generated images to augment the real images of a public dataset. This augmented dataset is then used for the downstream task of binary classification on adenomatous/hyperplastic polyps using EfficientNet-B0. We compared our augmentation approach with the baseline (only real data) using different proportions of the dataset. The best result reported an increase of 7.91\% (0.6983$\pm$0.039 vs. 0.6192$\pm$0.073). With a similar comparison approach (using different data proportions), we also examined the synthetic images generated using different variations of text prompts. It is observed that the text prompts formulated using the cross-class label concept outperformed those without such labels in most of the cases (for both NBI and WLI). In addition to frame-level analysis, we conducted video-level investigations. The associated quantitative results are supported by a statistical significance test (two-tailed t-test) and heatmaps. At video-level analysis, a statistically significant difference can be observed in favor of the cross-class label concept. Although the text prompts without cross-class labels achieved the best outcomes for NBI cases, the overall performance of both text prompts was similar, and the difference was not statistically significant. In addition, we explored the concept of weighted text prompts and presented both qualitative outcomes and quantitative analysis through box plots. It is noteworthy that, to the best of our knowledge, this is the first work that focused on text-controlled pathology-based polyp generation and also introduced the concept of cross-class label learning.

\textbf{Limitations and Future Work:} It is evident from the above-mentioned analysis that our approach achieved visually appealing outcomes followed by promising quantitative results. However, our model PathoPolyp-Diff also has some limitations. In a few cases, utilizing the cross-class labels resulted in undesired qualitative outcomes due to limited generation control. When the weighted mechanism was incorporated, it improved the qualitative results but at the expense of degraded quantitative outcomes. From a clinical perspective, such undesired outcomes could be critical, as pathologically unrealistic images, if used for training diagnostic models, can lead to inaccurate predictions and misdiagnosis. Further exploration is needed in such cases. For instance, an auxiliary classifier trained separately on good quality and clean samples can be used to guide the diffusion process. In addition, some parameters like guidance and temperature scale can be adjusted for improvement. Further, our method, based on diffusion models, inherently involves a complex architecture and computationally intensive processes. Given the current design and objectives of the model, it may not be well-suited for deployment on low-end devices at this time. In a clinical setting, this issue can lead to local adaptation challenges where fine-tuning on a local dataset could be desirable for a specific task. The other consequences can include slow inference speed during emergency scenarios and high memory demands in remote healthcare facilities. To address this issue, some techniques, such as model quantization, pruning, and knowledge distillation, can be explored. Despite the above limitations, we successfully developed a novel diffusion-based technique to generate a diverse set of polyps using cross-class labels; we also provide a roadmap for the research community to build upon our work, extending the synthetic polyp dataset and experimenting with text prompts to enhance overall outcomes. Additionally, as already discussed, other possible exploration directions could include model simplifications, quantization, or developing lightweight variants that maintain performance while being more resource-efficient. Further, clinical validation by experts would help to rigorously assess the quality and realism of the generated images, validating their acceptability in a clinical setting.

\section*{Conclusion}
In this paper, we developed a novel diffusion based model, PathoPolyp-Diff, which generates a visually appealing diverse set of polyp images in a two-stage process. This set of polyps covers multiple categories, including pathology (adenomatous/hyperplastic), imaging modality (NBI/WLI), and quality (informative/uninformative). In addition, we introduced the concept of cross-class label learning, which allows the generation of diversified polyps exhibiting features from other classes without requiring additional labels. This text-controlled image generation is validated by performing a downstream binary classification task on adenomatous and hyperplastic polyp classes with NBI and WLI modalities. Further, we studied the impact of text prompt variations on colonoscopy image generation by incorporating a weighted control mechanism. The results are analyzed at both frame and video level outcomes with a statistical significant test. Our research also reveals the significance of text prompt formulation on synthetic images' visual and clinical properties. The proposed approach achieved an improvement of 7.91\% and 18.33\% in balanced accuracy when used for augmentation and cross-class label learning, respectively.

\section*{Data Availability}
All data used in this study are publicly available or accessible on request. The SUN Database details are available at http://sundatabase.org/ (Ref. 26, 27), and it requires a prior request from the respective authors for access, while the ISIT-UMR Colonoscopy Dataset is directly available at http://www.depeca.uah.es/colonoscopy\_dataset/ (Ref.28). The quality-based annotations can be obtained on request (Ref. 29). This study did not involve humans or any living organism subjects. 
\bibliography{sample}

\section*{Acknowledgements}
V.S. is supported by the INSPIRE fellowship (IF190362), DST, Govt. of India. D.J. is supported by the NIH funding: R01-CA246704,  R01-CA240639, U01 DK127384-02S1, and U01-CA268808. V.S. would like to thank Dr. Abhishek for reviewing the article. 

\section*{Author contributions statement}

V.S. and D.J. conceived the study. M.K.B, P.K.D and U.B designed the article. V.S. conducted experiments, designed illustrations and substantially contributed to the writing of the article. D.J. analyzed the results and participated in manuscript writing. All authors reviewed the manuscript and provided critical feedback. 

\section*{Competing interests}
The authors declare no competing interests.

\section*{Additional information}
\textbf{Correspondence} and requests for materials should be addressed to V.S.

%To include, in this order: \textbf{Accession codes} (where applicable); \textbf{Competing interests} (mandatory statement). 

%The corresponding author is responsible for submitting a \href{http://www.nature.com/srep/policies/index.html#competing}{competing interests statement} on behalf of all authors of the paper. This statement must be included in the submitted article file.

\end{document}